\documentclass[longauth]{aa} 
\usepackage{amsmath}
\usepackage{lscape}
\usepackage{graphicx}
\usepackage{stfloats}
\usepackage[dvips]{color}
\usepackage{color}
\usepackage{natbib}
\bibpunct{(}{)}{;}{a}{}{,}                           
\usepackage{txfonts}
\def \aj {AJ}
\def \mnras {MNRAS}
\def \apj {ApJ}
\def \apjl {ApJL}
\def \aap {A\&A}
\def \nat {Nature}
\def \araa {ARAA}

\def \pasp {PASP}

\def \apjs {ApJS}
\def \aaps {A\&AS}
\def\lesssim{\mathrel{\hbox{\rlap{\hbox{\lower4pt\hbox{$\sim$}}}\hbox{$<$}}}}
\def\gtrsim{\mathrel{\hbox{\rlap{\hbox{\lower4pt\hbox{$\sim$}}}\hbox{$>$}}}}

\DeclareMathAlphabet{\mathsc}{OT1}{a}{m}{sc}
\def\testbx{bx}%
\DeclareRobustCommand{\ion}[2]{%
\relax\ifmmode
\ifx\testbx\f@series
{\mathbf{#1\,\mathsc{#2}}}\else
{\Mathrm{#1\,\mathsc{#2}}}\fi
\else\textup{#1\,{\mdseries\textsc{#2}}}%
\fi}

\begin{document}

\title{PESSTO : survey description and products from the first data release by the Public ESO Spectroscopic Survey
of Transient Objects  
\thanks{Based on observations collected at the European Organisation for
Astronomical Research in the Southern Hemisphere, Chile, as part of 
programme 188.D-3003 (PESSTO)}\fnmsep\thanks{www.pessto.org}}

   \subtitle{}
  \author{S. J. Smartt\inst{1}
      \and S. Valenti \inst{2,3}
      \and M. Fraser \inst{4}
      \and C. Inserra \inst{1}
      \and D. R. Young  \inst{1}
\and M. Sullivan \inst{5}  
\and A. Pastorello \inst{6}
\and  S. Benetti \inst{6}
\and  A. Gal-Yam \inst{7}
\and  C. Knapic \inst{8}
% ...
}
     \institute{Astrophysics Research Centre, School of Mathematics
       and Physics, Queen's University Belfast, Belfast BT7 1NN, UK. \email{s.smartt@qub.ac.uk}
\and 
Las Cumbres Observatory Global Telescope Network, 6740 Cortona Dr.,
Suite 102, Goleta, California 93117, USA 
\and 
Department of Physics, University of California Santa Barbara, Santa
Barbara, CA 93106, USA
\and 
Institute of Astronomy, University of Cambridge, Madingley Road, Cambridge CB3 0HA, UK
\and  
School of Physics and Astronomy, University of Southampton, Southampton SO17 1BJ, UK
\and 
 INAF - Osservatorio Astronomico di Padova, Vicolo del l'Osservatorio 5, 35122 Padova, Italy
\and 
Benoziyo Center for Astrophysics, Weizmann Institute of Science, 76100 Rehovot, Israel
\and 
INAF - Osservatorio Astronomico di Trieste, Via G.B. Tiepolo 11, I-34143 Trieste, Italy
% ...
}

   \date{}
      
   \abstract
    {The Public European Southern Observatory 
      Spectroscopic Survey of Transient Objects (PESSTO) began  as a public
      spectroscopic survey  in April 2012. PESSTO classifies
      transients from publicly available sources and wide-field
      surveys, and selects science targets for
      detailed spectroscopic and photometric follow-up. PESSTO runs
      for nine months of the year, January - April and August -
      December inclusive, and typically has allocations of 10 nights
      per month. }
   {We describe the data reduction strategy and data products that are publicly available
     through the ESO archive as the Spectroscopic Survey Data
     Release 1 (SSDR1). } 
   {PESSTO uses the New Technology Telescope
     with the instruments EFOSC2 and SOFI to provide optical and NIR
     spectroscopy and imaging. We  target supernovae and
     optical transients brighter than 20.5$^{m}$ for classification. Science targets are selected for 
     follow-up based on the PESSTO science goal of extending knowledge
     of the extremes of the supernova population.  We use standard EFOSC2 set-ups providing
     spectra with resolutions of 13-18\AA\ between
     3345-9995\AA.  A subset of the brighter science targets are selected for SOFI 
     spectroscopy with the blue and red grisms (0.935-2.53$\mu$m and resolutions 23-33\AA) and imaging
   with broadband $JHK_{\rm s}$ filters.} 
   {This first data release  (SSDR1)  contains flux calibrated spectra from the  first year
    (April 2012  - 2013). A total of 221 confirmed supernovae 
      were classified, and we 
     released calibrated optical spectra and classifications publicly
     within 24hrs of the data being taken (via WISeREP).  The data in SSDR1 replace 
     those released spectra. They have more reliable and quantifiable
     flux calibrations, correction for telluric absorption, and are
     made available in standard ESO Phase 3 formats. We estimate the absolute
     accuracy of the flux calibrations for EFOSC2 across the whole survey in
     SSDR1 to be typically $\sim$15\%, although a number of spectra will have less reliable absolute flux calibration because of weather and slit losses. Acquisition images for each spectrum are available which, in principle, can allow the user to refine the absolute flux calibration.  The standard NIR reduction process does not    produce high accuracy absolute spectrophotometry but synthetic   photometry with accompanying $JHK_{s}$ imaging can improve  this. 
Whenever possible, reduced SOFI images are provided to allow this. }
 {Future data releases will focus on improving the automated 
    flux calibration of the data products. The rapid turnaround
    between discovery and classification and access to reliable
    pipeline processed data products has allowed early science papers
    in the first few months of the survey. }

\titlerunning{PESSTO: survey description and data products}
\maketitle

\section{Introduction} 
\label{sec:intro}
The search for transient phenomena in the Universe has entered a new
era, with the construction and operation of dedicated wide-field
optical telescopes,
coupled with large format digital cameras and rapid data analysis pipelines. The
current surveys in operation with 0.7 - 1.8m aperture wide-field telescopes
combined with cameras covering 5--10 square degrees that are actively
searching for transients are 
the Palomar Transient Factory \citep[PTF; ][]{2009PASP..121.1395L},
the Pan-STARRS1 survey  \citep[PS1;][]{2010SPIE.7733E..12K}, 
the Catalina Real Time Survey \citep[CRTS;][]{2009ApJ...696..870D},  the La Silla QUEST survey \citep[LSQ;][]{2013PASP..125..683B}, and the SkyMapper survey \citep{2007PASA...24....1K}. With exposure times of 30-60 seconds, these surveys can reach magnitudes between 19-21, and can each cover around 1000 - 6000 square degrees
per night depending on the telescope.  Each survey runs transient object
detection pipelines, with differing methodology, but a common factor  amongst the 
surveys is the requirement for rapid spectroscopic observations of new transients. 
The simplest initial characteristic of a transient that is immediately required is 
distance, to provide an estimate of emitted energy, albeit initially in the narrow
wavelength region of the optical domain. 

The PTF built dedicated fast follow-up 
into the factory element of the project from the outset, successfully combining transient detection
on the Palomar 1.2m Schmidt with spectroscopic follow-up on accessible 2-4m telescopes
within their consortium. 
\cite[e.g.][]{2009PASP..121.1334R,2011ApJ...736..159G,2011Natur.480..344N}. 
PS1 has concentrated efforts on higher redshift regimes, covering
similar volume to PTF's imaging survey with its 10 Medium Deep fields, daily cadence, 
and follow-up on 4m, 6m, and 8m telescopes \cite[e.g.][]{2010ApJ...717L..52B,2011ApJ...743..114C,2012ApJ...755L..29B,2012Natur.485..217G}. Other projects are using smaller aperture ($\sim$0.1-0.4\,m) telescopes or cameras to cover
wider fields to shallower depths. The MASTER 
project employs a number of 40cm telescopes and 7cm cameras 
in the northern hemisphere, finding optical transients down to 
$\sim$20\,mag \citep{2010AdAst2010E..30L}. The ASAS-SN project is now running two 14cm telescopes 
in the northern and southern hemispheres, successfully 
finding transients brighter than $V\sim$17 \citep{2014ApJ...788...48S} with the aim of being all sky to approximately 
this flux limit. The very successful OGLE project is  now producing extragalactic transients in its 
$\sim$700 square degree footprint of the OGLE-IV survey with a 1.3m telescope and 1.4 square degree field of view \citep{2013AcA....63....1K, 2014AcA....64..197W}. 

Between them, these synoptic surveys are discovering new classes of transients that
challenge our ideas of the physics of stellar explosions. The long running and very successful nearby
supernova (SN) searches of 
LOSS \citep{2011MNRAS.412.1441L,2011MNRAS.412.1419L} and 
CHASE \citep{arxiv.org.0812.4923}, 
are  aided by the large community of well equipped and experienced
amateur astronomers throughout the world who have also increased their
detection limits to provide some critical scientific data \cite[see e.g.
K. Itagaki's contribution to][]{2007Natur.447..829P}. These have targeted bright
galaxies (within about $\sim$100 Mpc, $z\lesssim0.025$) for the
obvious reasons that they host much of the mass and star formation in
the local Universe.  The SN population in these
galaxies are well studied \citep{2011MNRAS.412.1441L}, and the progenitor
stars of many core-collapse SNe have been discovered
\citep{2009ARA&A..47...63S}
leading to physical insights into the explosions and the
progenitor population. However, surprises still appear in these 
galaxy focused surveys such as the faint hydrogen poor SNe  
\citep{2009Natur.459..674V,2010ApJ...723L..98K}. The origins of some of
these are disputed and it is not clear if they are
thermonuclear explosions of white dwarfs or related to core-collapse.
The nature of faint transients such as that in M85, which are between 1-2mags brighter than classical novae, have been suggested as potential stellar mergers rather than SNe
 \citep{2007Natur.447..458K}. 

The new wide-field transient searches discover transients with no
galaxy bias and  fainter limiting magnitudes, and probe shorter
timescales. This has opened up a new window on the transient Universe
-- and the physical diversity discovered thus far is challenging the
paradigms we hold for stellar deaths. It is likely that we are witnessing
the diversity in the transient Universe that depends on stellar mass,
metallicity, binarity, mass-loss rates, and rotation rates of the
progenitor systems. However the biggest challenge in the field is now
linking the discoveries to rapid spectroscopic and multi-wavelength
follow-up. PTF  has discovered and spectroscopically classified 2288
transients over the years 2010-2012, which made up 25-50\% of all 
SNe found and classified in this period \citep{2013A&G....54f6.17S}. 
PS1 has discovered
over 4000 transients, with spectroscopy of 10\%. Still many transients
go unclassified  
and wide-field searches in the south are only just
beginning. Furthermore, we will soon enter the era of
multi-messenger astronomy, which aims to link electromagnetic
detections to gravitational wave, neutrino, and high-energy cosmic ray
sources \citep{2013RSPTA.37120498O}. 

In response to the first call by the European Southern Observatory
(ESO) for public spectroscopic surveys, and particularly prompted by
the opportunities provided by the La Silla QUEST and SkyMapper
surveys, we proposed PESSTO (the Public ESO Spectroscopic Survey of
Transient Objects).  This built on the work and broad european consortium gathered together in the 
ESO Large Programme "Supernova Variety and Nucleosynthesis Yields" lead by S. Benetti 
\citep[ESO 184.D-1140, 30 nights/yr allocated at ESO- NTT; e.g. see][]{2013ApJ...767....1P}.  PESSTO 
 was accepted by ESO as one of two public
surveys, the other being the GAIA-ESO survey using FLAMES on the Very
Large Telescope. PESSTO was awarded 90 nights per year on the New
Technology Telescope (NTT) initially for two  years, which has been renewed to four years (2012-2016).  
The science goal of
PESSTO is to provide a public spectroscopic survey to deliver
detailed, high-quality, time series optical+NIR spectroscopy of about 150
optical transients covering the full range of parameter space that the
surveys now deliver : luminosity, host metallicity, explosion
mechanisms. The PESSTO team is composed of the major supernovae
research teams in the ESO community and rapid access to the reduced
data is an integral part of the project. 
To date, ten papers based primarily on PESSTO data have been accepted in refereed journals
\citep{2013MNRAS.433.1312F,
2013MNRAS.431L.102M,
2013ApJ...770...29C,
2014MNRAS.438L.101V,
2014MNRAS.441..289B,
2014MNRAS.437L..51I,
2014MNRAS.437.1519V,
2013MNRAS.436..222M,
2014MNRAS.445...30S,
2014MNRAS.444.2096N}

\section{Description of the survey and data reduction pipeline } 

PESSTO is allocated 90 nights per year, in visitor mode, on the
ESO NTT. There are no observations planned during the months of May,
June, and July because  the Galactic centre is at optimal right
ascension. These three months make it more difficult to search for
extragalactic SNe and there is large time pressure from the ESO 
community for Milky Way stellar science. PESSTO is typically  allocated 
10 nights per month split into three sub-runs of 4N, 3N and 3N. 
The middle sub-run is usually dark time, while the two others are
grey/bright with the moon up for around 50\% of the time. 
The instruments used are EFOSC2 and SOFI and both spectroscopy and 
imaging modes are employed.  The PESSTO collaboration host public
webpages which includes information on night reports, observing
conditions,  observing with the NTT and instructions for downloading the data reduction pipeline. This information is udpated during the survey and users should read
this document with the information on www.pessto.org and the wiki pages that the homepage points to. 

\subsection{Target selection and strategy}\label{sec:target}

We have built a web-based data aggregator that works to pool various
institutional and transient survey websites, alongside astronomical
transient alert resources such as The 
Astronomers Telegram (ATel) \footnote{http://www.astronomerstelegram.org} and the IAU
Central Bureau for Electronic Telegrams (CBET) \footnote{http://www.cbat.eps.harvard.edu/cbet/RecentCBETs.html}
services. This aggregator (the \emph{PESSTO Marshall}),
cross-correlates all transient event metadata coming from these
various sources, grouping duplicate objects together, and presenting
the user with a detailed overview of what is currently known about each
transient event.

The PESSTO Marshall also provides a structured workflow which allows
users to both promote objects they wish to be classified with the NTT 
and to track observations of objects they have chosen for detailed
follow-up. The Marshall also works well as a collaboration and
communication platform for the PESSTO community, allowing users to
comment on objects they are interested in, to append useful object
metadata, to state their intentions about individual objects and to
provide observers at the NTT with detailed instructions as to what
observations they require. 

The major science goal of PESSTO is to extend the detailed time series
data for unusual and rare transient events that challenge our
understanding of the two standard physical mechanisms for SNe : core-collapse and thermonuclear. Furthermore, nearby supernovae allow
for more detailed study such as progenitor detections and
multi-wavelength  (x-ray to radio) follow-up that probes both the
explosion mechanisms and shock physics. To this end, the PESSTO
collaboration has formed Science groups who study the classification
spectra and promote targets to ``PESSTO Key Science Targets'' on the
basis of  them falling into these areas. 
PESSTO takes input sources for
classification from  all publicly available sources, and has 
partnerships in particular with the La Silla QUEST,  SkyMapper and OGLE-IV surveys for access to its targets as early as possible. The public 
feed of CRTS targets has also proved to be a very valuable 
source of targets.  PESSTO employs a magnitude limit for classification of 20.5 in $BVR$ or unfiltered CCD magnitudes
and particularly looks to prioritise targets according to the following criteria

\begin{itemize}
\item  Distance ($d<40$\,Mpc) and very high priority for candidates with $d<25$\,Mpc
\item Young phase :  non-detection at $<$ 7 days, or fast rise time
  ($>0.5$mag/day) or sharp drops in magnitude over short timescales
\item  Luminosity extremes  : objects with expected $M < -19.5$ and $M
  > -15.5$. This is difficult to implement as the distance or redshift
  of the host is required, but when possible it is used. 
\item  Fast declining light curves ($\Delta$mag$> 1$\,mag/5days)  or
  very slow-rising light curves $(t_{\rm rise} > 30$\,days)
\item Variability history - for example pre-discovery outburst such as
  SN2006jc, SN2009ip, SN2010mc 
\citep{2007Natur.447..829P,2013ApJ...767....1P,2013MNRAS.433.1312F,2013Natur.494...65O}
\item  XRF \& GRB alerts : these have not yet been observed, because of a
  lack of targets at the right magnitude and availability 
\item Peculiar host environments  : for example low-luminosity
  galaxies: $M_{B} > -18$ or hostless transients;  remote locations in
  E/S0 galaxies and in the halos of spirals ($d > 20$\,kpc) from the
  nucleus; enhanced star formation environments such as Arp galaxies
  -interacting systems or tidal galaxy tails or galaxies which have hosted multiple SNe. 
\end{itemize}

\begin{figure*}
\centering
\parbox[t]{4cm}{
\vspace*{1cm}
\includegraphics[scale=0.3,angle=0]{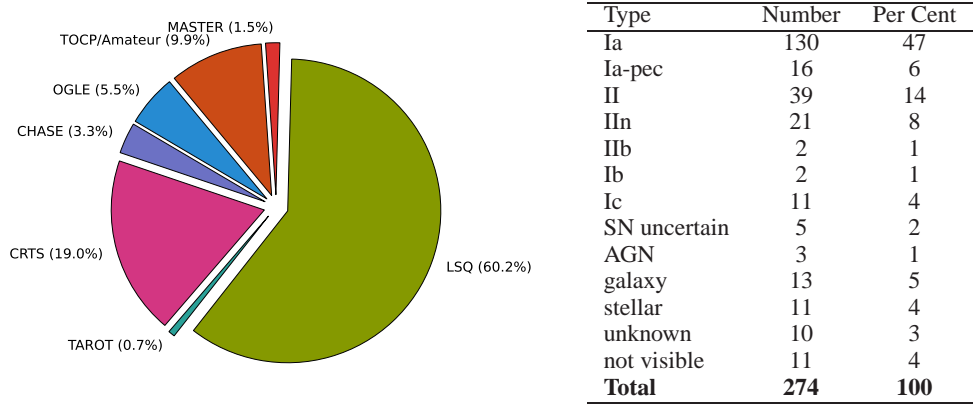}
}
\hfill
\parbox[t]{10cm}{
\vspace{1cm}
\begin{tabular}{lcc}\hline
Type & Number & Per Cent \\\hline
Ia                      &    130       &    47 \\      
Ia-pec                &      16     &    6 \\            
II                       &    39               & 14    \\
IIn                     &     21               &   8  \\
IIb                     &    2                 &    1 \\
Ib                      &       2                  &  1   \\
Ic                      &      11                 &   4  \\
SN uncertain       &      5                       & 2    \\
AGN                    &      3                     &   1  \\
galaxy                 &      13                    &  5   \\
stellar                   &      11                 &   4  \\
unknown             &     10                   &    3 \\
not visible           &     11                     &   4  \\
{\bf Total}                   &      {\bf 274}                         & {\bf 100} \\\hline
\end{tabular}
}
\caption{{\em Left :} Breakdown of the source of targets for PESSTO
  classifications from April 2012 to April 2013. The details of the
  survey names are in the text in Sect.\ref{sec:target}, {\em Right :}
classification types.}
\label{fig:survey-stats}
\end{figure*}

The breakdown of the targets from the various feeder surveys is shown
in Fig.\ref{fig:survey-stats} for the first year, which covers the
first public data release. 
PESSTO formally collaborates with the La Silla-QUEST (LSQ) Low Redshift
Supernova Survey \citep{2013PASP..125..683B} which operates 
as part of the La Silla-QUEST Southern Hemisphere Variability Survey
\citep{2012Msngr.150...34B}. The 160-megapixel QUEST camera is installed on the ESO Schmidt telescope,
providing a sensitive pixel area of 8.7 square degrees and it uses almost all of the 
Schmidt telescope time. The pixel scale is $0\farcs88$  and the system gets to a
depth of around 20$^m$ in 60s through a wide-band filter that covers
the SDSS $g$+$r$ bands. This magnitude limit is well matched to the
capability of the NTT+EFOSC2 for obtaining spectra with signal-to-noise
of greater than 10 in typically 20 min exposures. The LSQ 
survey repeats a sky area twice per night to remove bogus
objects and asteroids, and in this way 1500 square degrees is
typically covered each night. Fig.\ref{fig:survey-stats} shows the source of targets
for PESSTO classification from April 2012-2013 illustrating the 
extensive use of the the LSQ  survey targets. The Catalina
Real-Time Transient Survey \citep[CRTS][]{2009ApJ...696..870D} is another very useful source of
targets for PESSTO, as again the survey depths are well matched to
PESSTO spectroscopic capabilities and the wide-field searching
produces large numbers of transients that can be filtered for
objects at the extreme end of the supernova luminosity distribution
\citep[e.g.][]{2010ApJ...718L.127D,2010ApJ...724L..16P,2013ApJ...770..128I} 
and luminous transient events in the cores of galaxies 
\citep{2011ApJ...735..106D}.  PESSTO also parses the OGLE-IV transient
list which is very useful as the declination of the fields are around
$-60$ to $-80$ degrees, allowing the NTT to point in this direction
during wind restrictions and providing targets available for long
observational seasons. The other sources of targets as seen in 
Fig.\ref{fig:survey-stats} are the amateur reports which are posted on the 
Central Bureau's ``Transient Objects Confirmation Page"
\footnote{http://www.cbat.eps.harvard.edu/unconf/tocp.html}, 
the Chilean Automatic Supernova Search \citep[CHASE][]{arxiv.org.0812.4923} 
and a small number of targets from MASTER 
\citep{2010AdAst2010E..30L}
and TAROT
\citep{2008AN....329..275K}.  
Targets from the 3$\pi$ survey from Pan-STARRS1 survey
have also been classified, mostly during the second year of operations
\citep{2014ATel.5850....1S}. 
 In the future, PESSTO plans further exploitation of targets from the
 SkyMapper survey \citep{2007PASA...24....1K}. Although we have a 
formal partnership and agreement to access the SkyMapper targets as soon as they
are discovered, this survey was not  functioning in full science survey mode during
the period covered by the data in this paper. We also intend to exploit the  transient
stream likely to be produced by the ESA GAIA mission 
\cite[``GAIA alerts'', see][]{2013RSPTA.37120239H}.

\begin{figure*}
\centering
\parbox[t]{8cm}{
\includegraphics[scale=0.45,angle=0]{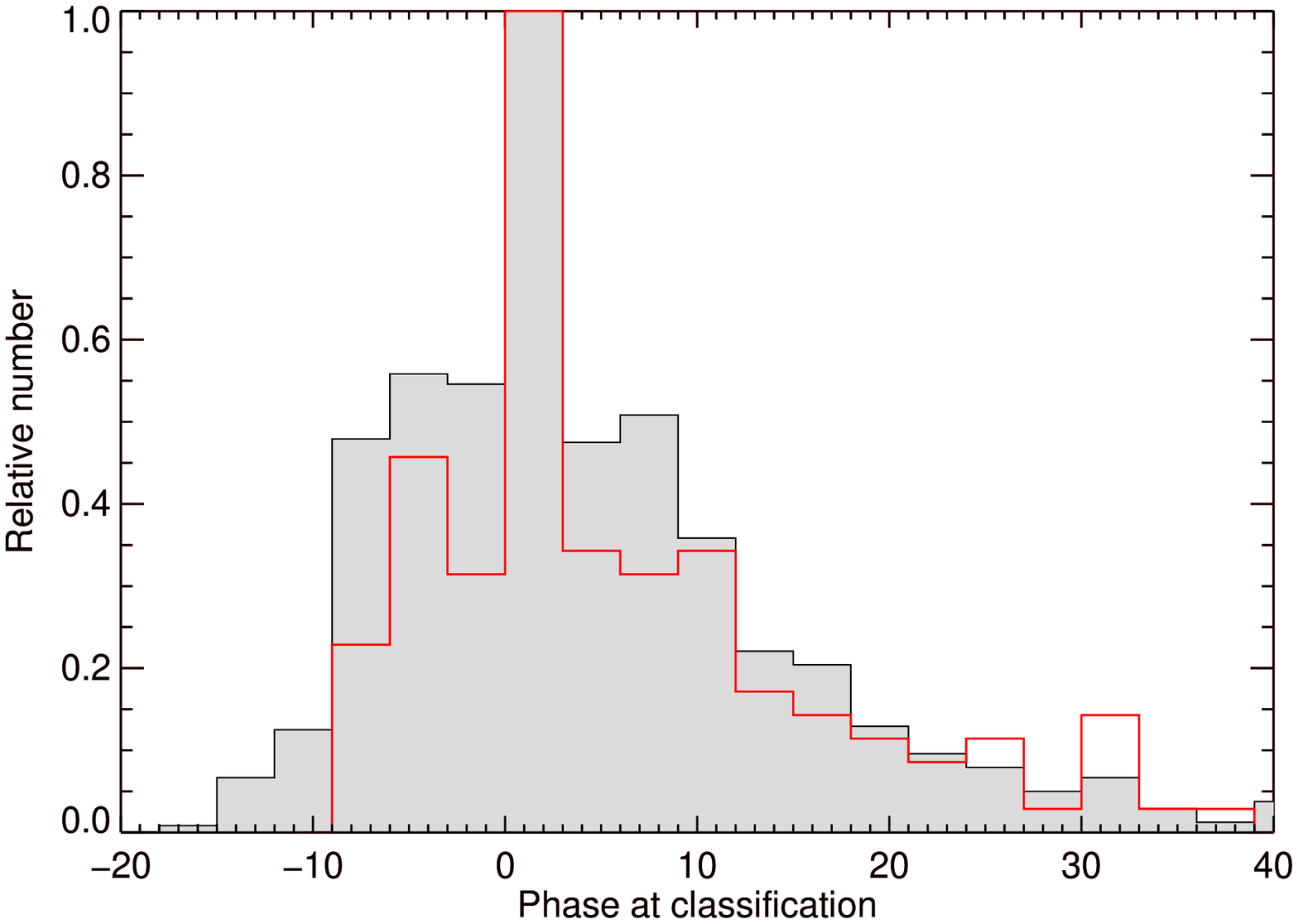}
}
\hfill
\parbox[t]{8cm}{
\includegraphics[scale=0.35, angle=0]{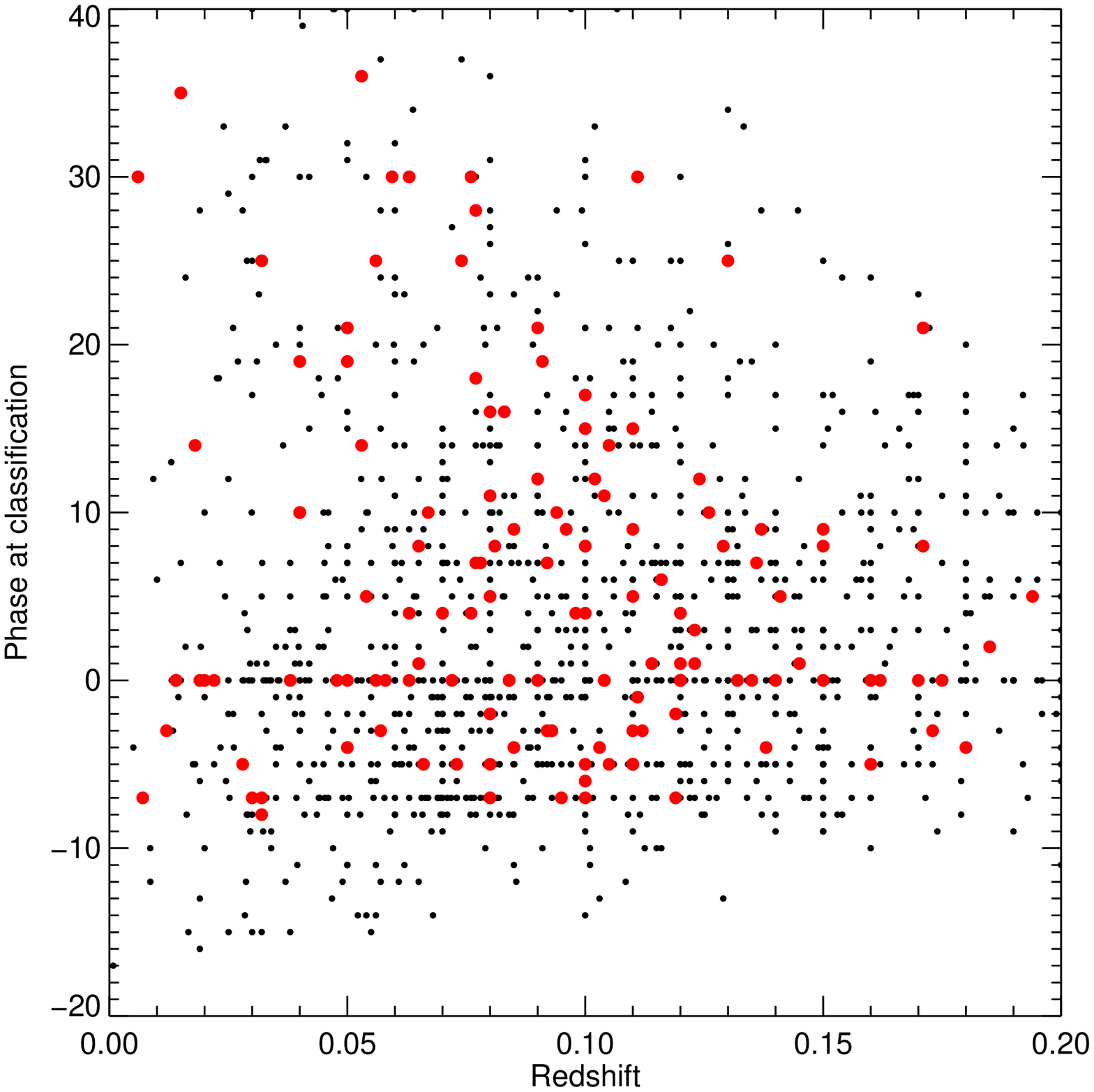}
}
\hfill
\caption{The phase of type Ia SNe at first spectrum taken with
  PESSTO. Results from the Palomar Transient Factory (over 2009-2012) are in grey
  or black \citep[data from][]{2014MNRAS.444.3258M}, with  PESSTO and La Silla QUEST targets in red. 
}
\label{fig:phase-Ia}
\end{figure*}

One of the goals of PESSTO is to provide early spectra for both fast
classification and for probing the early explosions of supernovae. The
earlier
an object can be classified, the more opportunity there is  for the
community to observe it with multi-wavelength facilities in the
interesting early phases of a few days after explosion
\citep[e.g.][]{2008Natur.453..469S,2013ApJ...775L...7C}
To gauge how
the first year of PESSTO progressed, we compare the phases of 
the first classification spectra taken by PESSTO of La Silla-QUEST
targets, with the Palomar Transient Factory 
in Fig.\ref{fig:phase-Ia} \citep{2014MNRAS.444.3258M}. 
 We chose type Ia SNe for this comparison as
they have well defined rise times, are the most common events found in 
magnitude limited surveys and can be typed to within a few days (type 
II SNe are hard to date from explosion, since the rise time takes hours to days
if the progenitor is an extended red supergiant, and the peak is not a
well  defined epoch in this case). This shows that we recover type Ia
SNe
down to around 10 days before peak. There is some uncertainty in  the dating of spectra as this epoch, but it is encouraging to see that
we can recover SNe competitively at these epochs by combining immediate
exchange of information between La Silla QUEST and PESSTO. 
There is an obvious peak in both surveys at $phase=0$ which is simply a 
labelling effect.  Observers tend to label SNe which are close to peak 
magnitude as being ``at peak" and therefore having $phase=0$. Spectroscopically  
dating type Ia SNe which are within a few days of peak magnitude is not 
accurate to within about $\pm$3-5 days. Hence there is a human tendency to label these with $phase=0$. With the hindsight of lightcurves
one can of course pin down the maximum light phase and then the true date 
of the first spectrum, but the phases plotted here typically come from 
spectroscopic dating only. In reality this peak should be smeared out within 
about $\pm$5 days around the $phase=0$ epoch. 
Where 
PESSTO could do better, and where PTF has excelled, is in very low redshift
early discoveries i.e. the bottom left corner of the right hand
panel in Fig\,\ref{fig:phase-Ia}. This is an area for rich scientific
exploitation \citep[e.g.][]{2011Natur.480..344N,2014Natur.509..471G}. 
PESSTO publicly releases the first spectrum of each newly classified transient
object it observes in reduced, flux calibrated form via WISeREP\footnote{http:// http://wiserep.weizmann.ac.il}
\citep{2012PASP..124..668Y}. The raw data for every PESSTO observation are immediately available to the public once they are taken, via the ESO Archive.  This includes
the raw data for all classification spectra, imaging (including acquisition images), follow-up spectra and  all calibration frames. The only delay is the time taken to copy data from La Silla and  ingest into the ESO Garching archive which is typically minutes to hours. We 
release final reduced data of the science follow-up targets during
yearly official data releases, the first of which is SSDR1.

\subsection{PESSTO Data reduction pipeline}

PESSTO uses fixed set-ups for EFOSC2 and SOFI which allow reduced data
products to be provided rapidly and uniformly to the PESSTO
survey members and the public
(see Table\,\ref{tab:efosc2} and \ref{tab:sofispec} for the fixed set-ups). 
The PESSTO consortium has developed a
data reduction pipeline, written in $python$ by Stefano Valenti, with
input from other members and based on the basic python packages 
$numpy, pylab, pyraf, pyfits$.  The pipeline produces fully reduced, flux
calibrated spectra (1D and 2D images) for both EFOSC2 and SOFI and
reduced images for EFOSC2 and SOFI on which photometric measurements
can be made (a nominal zeropoint is provided as discussed below).
This pipeline is publicly available on the PESSTO wiki, with
instructions for installation and use. All the data released in SSDR1
have been processed through the pipeline and indeed all subsequent 
data releases will be similarly processed. The following sections
describe the instrument set-ups, calibrations and data products
delivered including details on the header keywords employed that
assist the user in interpreting the calibrated data. 

In the first year of PESSTO and for SSDR1, we have not focused on high
accuracy absolute spectrophotometry. Reliable relative flux
calibration is essential for supernovae and transient science, but for
general classification and screening absolute spectrophotometry
is not critical. We placed more importance on screening and
classifying as many targets as possible rather than higher accuracy
calibration of fewer objects. For follow-up targets, the standard methods of
improving spectrophotometry involve correcting the spectra with
photometric measurements from time series lightcurves. Since most of
the PESSTO science targets do not yet have a fully calibrated
lightcurve, we have not corrected the spectra in bulk with this
method. Discussion of the accuracy and reliability of the absolute
flux scales is presented in Sect.\,\ref{sec:efosc-specphot} and future
data releases will focus on improving this.

\section{PESSTO EFOSC2  spectroscopic observations and calibrations}
\label{sec:efosc-spec}


The ESO Faint Object Spectrograph and Camera 2 (EFOSC2)\footnote{http://www.eso.org/sci/facilities/lasilla/instruments/efosc.html}, has been mounted at the $f/11$ nasmyth focus on the NTT
since 2008. It is a focal reducer which uses multi-layer coated,
all-transmission optics (i.e. no reflecting surfaces). The F/49 camera
has a focal length of 200mm, which provides a pixel scale of
$0\farcs12$\,pix$^{-1}$, for detector pixels of $15\mu$m physical size. As
described in the EFOSC2 manual, the filter and grism wheels are
located in the parallel beam, between the collimator and the camera
which means that the EFOSC2 focus is quite stable and does not drift
significantly as a function of temperature, rotator angle, wavelength
or observing mode. 

As an all-transmission optical instrument, the dispersing elements are
grisms, providing a fixed wavelength coverage dependent on detector
size and camera beam width. Despite the large number of grisms
available, PESSTO uses only three default settings for EFOSC2, listed
in Table\,\ref{tab:efosc2}. The CCD on EFOSC2 is ESO\#40 which is a
Loral Lesser Thinned AR coated, UV flooded detector with
$2048\times2048$ 15$\mu$m pixels and driven with an ESO-FIERA 
controller. PESSTO uses the CCD in Normal
readout mode with $2\times2$ pixel binning giving 2-dimensional science images with $1024\times1024$ physical pixels, which have a plate scale of $0\farcs24$ pixel$^{-1}$.
 The CCD is never windowed for PESSTO observations,
hence the field size is $4.1\times4.1$\,arcmin. The lowest resolution
setting (Gr\#13) of 17.7\AA\ has the broadest wavelength coverage and
is the default setting for classification spectra
(Table\,\ref{tab:efosc2}). 
 All 
classification spectra which are reduced and released immediately
(within 24\,hrs of the data being
taken) employ this set-up. The other two grism settings
are used for some of the PESSTO Key Science Targets, in which
either the higher resolution or full wavelength coverage (or both) are required. 
All PESSTO observations with Gr\#16 use the order blocking filter OG530 
to remove second-order effects in this red setting. For science targets PESSTO
does not use an order blocking filter for Gr\#13. Hence for blue objects there may
be second-order contamination at wavelengths greater than $>7400$\AA. 
In order to correct for this,  spectrophotometric flux standards are taken with 
and without a blocking filter; further details on how this correction is applied 
are in Sect.\,\ref{sec:efosc-specphot}. 
The EFOSC2 aperture wheel has fixed width slits. When the seeing  is $\leq 1\farcs4$ then the $1\farcs0$ slit is used and when it is $\geq1\farcs5$ then then the $1\farcs5$ slit is
employed. Flat field and flux calibrations are then taken with the
appropriate spectrograph configuration and matched to the 
science frames within the data reduction pipeline. 
The spectrograph set-ups are summarised in 
Table\,\ref{tab:efosc2}. 

\begin{table*}
\caption[]{PESSTO settings for EFOSC2 spectroscopy. The number of pixels and wavelength ranges are those in the final trimmed spectra for public release.  The blocking filter OG530 is used 
only (and always) for Gr\#16. The spectral resolution
$R$ is given at the central wavelength, as is the velocity resolution $V$. The FWHM is the full-width-half-maximum of a Gaussian fit to either the [O\,{\sc i}] 5577\AA\ or 6300\AA\  night sky line, for a  
$1\farcs$0 slit. The respective values when a  
$1\farcs$0 slit is used can be determined by simply multiplying the values by 1.5. 
The column headed Arclines indicates the number of
lines used. The RMS is the typical residual for the wavelength calibration solution.}
\label{tab:efosc2}
\begin{tabular}{rllllllllll}
\hline\hline
Grism    &  Wavelength    & Filter         & $n_{\rm pix}$ & Dispersion    &  FWHM & Resolution  &   R  &  $V$ resolution  & Arclines & RMS              \\
              &  (\AA)              & (blocking) &  (pixels) & (\AA\,pix$^{-1}$)     & (pixels) & (\AA)            &  ($\lambda_c/\Delta\lambda$) &  km\,s$^{-1}$ & (number)  &     (\AA) \\ \hline
\#13    &  3650 - 9250   &  none     & 1015 & 5.5 &  3.3 & 18.2 & 355 & 845 & 13-15 & 0.10-0.15             \\
\#11    &  3345 - 7470   &  none     & 1011 & 4.1 &  3.4  & 13.8 &  390 & 765 & 9        & 0.10-0.15		\\
\#16    &  6000 - 9995   &  OG530 & 950  & 4.2  &  3.2  & 13.4 &  595 & 504 & 11-14  & 0.05-0.10	  \\\hline\hline
\end{tabular}
\end{table*}

\subsection{Detector characteristics : bias level, gain and readnoise}
\label{sec:efosc-ccd}

The EFOSC2 chip, CCD\#40, is always used by PESSTO in Normal readout
mode and $2\times2$ binning (Mode 32 as defined by 
ESO.\footnote{\scriptsize{http://www.eso.org/sci/facilities/lasilla/instruments/efosc/inst/Ccd40.html}})
All acquisition images are also taken in this mode.  At the beginning
of the PESSTO survey and during April 2012, we began with the EFOSC2
default acquisition Observation Blocks (OBs) which use Fast readout mode for acquisition
images. This is immediately visible to the
user as Fast readout mode uses two amplifiers and the frame shows a
split appearance with two halves of the chip having different bias
levels.  This readout mode was never used for PESSTO science frames and from
August 2012 onwards, PESSTO has uniformly used Normal readout mode for
all  acquisition frames 
(except in a few occasions because of OB selection error).  In this
section we present characterisation checks of the CCD only in Mode 32 :
Normal readout mode and 2$\times$2 binning. 
The CCD has a physical size of 2048$\times$2048 pixels. With 12 pixels
of prescan and overscan this results in a 2060$\times$2060 pixel FITS
image, or a 1030$\times$1030 pixel FITS file after binning. 
Hereafter, all pixel coordinates referred to will be on this 1030$\times$1030
reference  frame. 

\begin{figure}
\centering
\includegraphics[scale=0.7,angle=270]{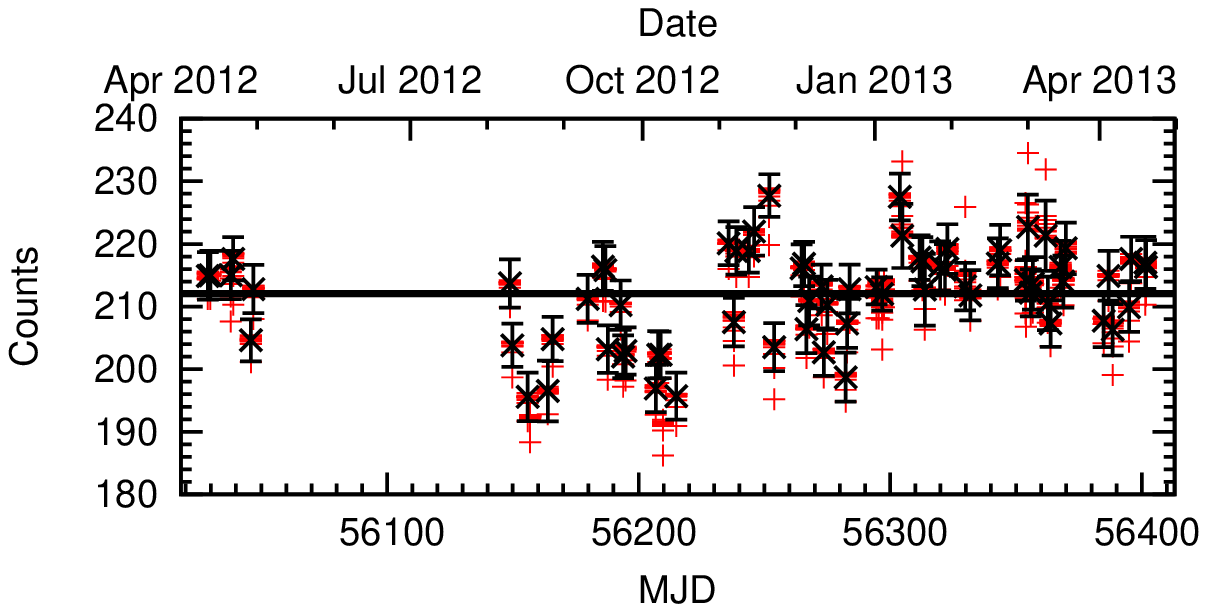}
\caption{The bias level of CCD\#40 over the first year of PESSTO
  survey operations. Red pluses are the mean count as measured over the central 
200$\times$200 binned pixels in each raw bias frame. Black crosses are 
the mean of each combined masterbias frame produced by the PESSTO 
pipeline from a set of raw biases, and measured over the same region. 
Error bars correspond to the standard deviation of the measured pixels, while the solid line is the average over the year.}
\label{fig:efbias}
\hspace{0.5cm}
\centering
\includegraphics[scale=0.65, angle=270]{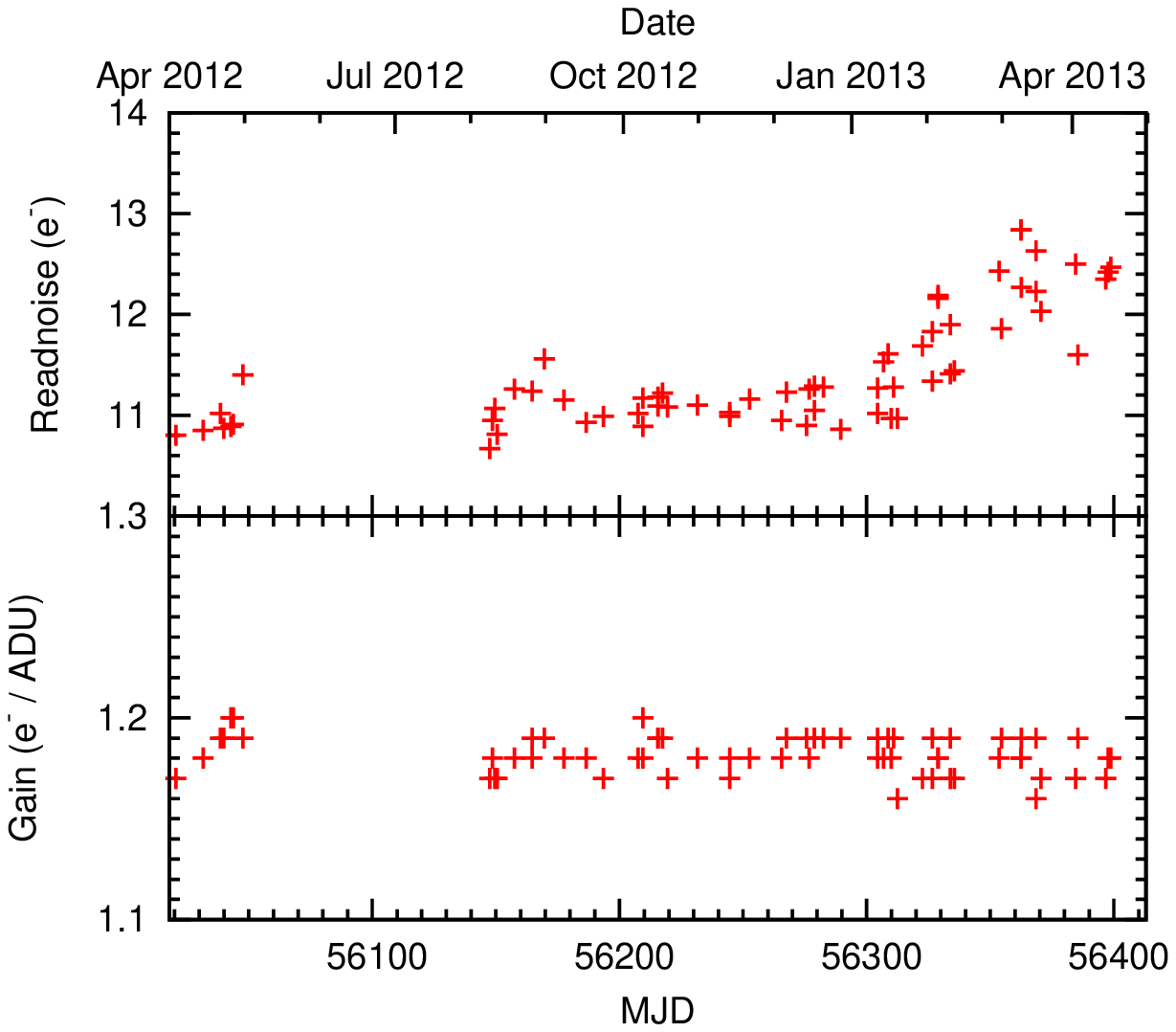}
\caption{The readnoise and gain of CCD\#40 over the period over the
    first year of PESSTO survey observations. The method to determine
    the values is described in the text. }
\label{fig:efgain}
\end{figure}

The bias level of CCD\#40 on EFOSC2 appears to be stable to within
5-10\% of the mean level (212 ADU)  across a period of a year. In
Fig.\,\ref{fig:efbias} we plot the bias level from all PESSTO nights
during the first year of survey operations. This was measured in the
central 200$\times$200 binned pixel region. The bias frames are 
always flat and uniform and no evidence of gradients have been found. 
Hence no overscan region corrections are ever applied in the PESSTO
pipeline processing.  The EFOSC2 Users Manual
 \citep{efosc2manual} notes that overscan correction should not be used owing to
 their  small size (6 and 12 pixels only)  and bleeding effects of charge from the
 science and calibration frames into these sections. 

To measure the gain and readout noise of CCD\#40, we used the {\sc
  findgain} task within {\sc iraf}. {\sc findgain} uses Janesick's
algorithm to determine the gain and readout noise of a CCD from a set
of two bias frames and two frames with uniformly high signal levels.
We used pairs of EFOSC2 dome flats from the ESO archive (in the V\#641
filter and Mode 32) during the months of PESSTO observations. We
selected two exposures from each set of domeflats, and used the
closest bias frames available (usually from the same night, but
occasionally from the preceding or subsequent night) and ensured that
all were taken in Mode 32 (i.e. Normal readout mode and 2$\times$2
binning).  The gain and
readnoise were measured over the region of the CCD from
[601:800, 401:600] (in binned pixels) to minimise the effects of any
slope in the flat field. Sixty five such measurements were made over
the period of PESSTO observations from April 2012 to April 2013 and
the results are plotted in Fig.  \ref{fig:efgain}.  The
gain was found to be stable, with a mean
value of 1.18 $\pm$ 0.01 e$^{-}$/ADU. The
readout noise appears to be constant until January 2013, when it
increased from an average of $\sim$11 e$^{-}$ to $\sim$12.5
e$^{-}$. The cause of the readout noise increase is not known to us, 
but our measurements match those of the ESO La Silla Quality Control
programme which also records an increase in the noise in this readout
mode.

As described in the EFOSC2 instrument handbook the bad pixel map of 
CCD\#40 (see also the EFOSC2 BADPIXMASK
\footnote{\scriptsize{http://www.eso.org/sci/facilities/lasilla/instruments/efosc/inst/BADPIXMASK.html}})
indicates a bad column at X=486 (in binned image coordinates). The
dispersion direction runs along the Y-direction, and hence as is
standard practice for EFOSC2 observing, PESSTO targets are positioned at
X=550. To be clear, these X-positions refer to the CCD pixel coordinates in the
full  raw 1030$\times$1030  frame (i.e. including the prescan and overscan
sections). After processing, the PESSTO data products are trimmed to 
851 pixels in the spatial direction and the targets are typically at pixel X=450.

\subsection{Spectroscopic calibration data and reduction}

PESSTO observations are carried out in visitor mode and the project
aims to homogenise all calibration frames and tie these directly to
what is required in the data reduction pipeline. To achieve that,
standard sets of OBs for calibration purposes have been created, 
and are available on the PESSTO public wiki. 
Here we describe the calibration data that are taken during routine
PESSTO observing and how they are applied in the data reduction 
process. 

\subsubsection{Bias calibration}
\label{sec:efosc-bias}

As discussed above in Sect.\,\ref{sec:efosc-ccd} the bias level has
been measured to be quite stable over a one year period.  A set of 11
bias frames are typically taken each afternoon of PESSTO EFOSC2
observations and are used to create a nightly master bias. This 
nightly master bias frame is applied to all EFOSC2 data taken, 
including the spectroscopic frames, the acquisition images and any 
photometric imaging.  The frame used for the bias subtraction can be
tracked in the header keyword 

\begin{small}
\begin{verbatim}
ZEROCOR = 'bias_20130402_Gr11_Free_56448.fits'
\end{verbatim}
\end{small}

The file name gives the date the bias frames were taken, the Grism and filter
combinations for which it is applicable (of course for biases this is not relevant
but the pipeline keeps track with this nomenclature) and the MJD of
when the master bias was created. 
The dark current is less than 3.5 e$^{-}$\,pix$^{-1}$\,hr$^{-1}$,
hence with typical PESSTO exposures being 600-1800s, no dark frame
correction is made

\subsubsection{Flat field calibration}
\label{sec:efosc-flat}

Techniques for flatfielding spectroscopic data depend on the
particular instrument or detector response that is being targeted for
removal. The type of calibration data required needs to be tailored to
the process and the final science product requirement.

The PESSTO survey takes sets of spectroscopic flatfields 
in the afternoons at a typical frequency of once per sub-run of 
3-4 nights.  The illumination for these comes from outside the
instrument, hence they are referred to as ``spectroscopic dome flats''.  An
integrating sphere is illuminated with a ``flatfield''  lamp
(a tungsten halogen with a quartz bulb)
  which is directed towards the NTT focal plane and the
telescope pupil is approximately simulated. The EFSOC2 calibration OBs
allow the user to set a required level of counts. PESSTO takes five
exposures with maximum count levels of 40,000-50,000 ADU for each of the grism, 
order sorting filter, and slit width combinations that we use. 
There are 8
combinations in total: the three basic configurations as
listed in Table\,\ref{tab:efosc2}, which are used with 1 and 1.5 arcsec slits and in addition 
 Gr\#13 flats are taken with the GG495 filter, to allow for second-order correction as discussed
in Sect.\,\ref{sec:efosc-specphot}).  Each of these is combined and normalised to give a
masterflat which can be associated with the appropriate science
observations from the sub-run. This is recorded in the FITS header for each data product, for example : 

\begin{scriptsize}
\begin{verbatim}
FLATCOR = 'nflat_20130413_Gr11_Free_slit1.0_100325221_56448.fits'
\end{verbatim}
\end{scriptsize}

As these are primarily used for removal of pixel-to-pixel response then 
higher frequency observations are not necessary. In fact it is debatable as to whether 
pixel-to-pixel response removal is required at all in long-slit spectroscopy at moderate 
signal-to-noise.   PESSTO does not take, nor use, spectroscopic sky flats during twilight 
for correction for slit illumination patterns. As we are primarily concerned with flux 
calibration of point sources, and the spectroscopic standard is placed at the same 
position on the slit and detector as the science targets (at CCD
pixel position X=555) then application of a slit illumination
correction is not necessary.  

\begin{figure}
\centering
\includegraphics[scale=0.65,angle=270]{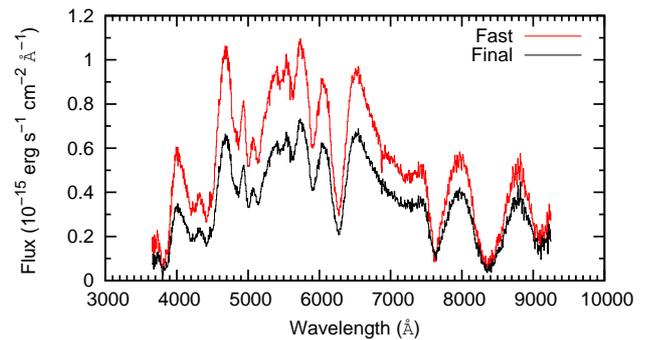}
\caption{Comparison of the rapid and final reductions for the classification spectrum of SN 2012fx, taken with Gr\#13 and a 1\arcsec\ slit on 2012 August 26. These spectra show that fringing is not as significant a problem with Gr\#13 as it is for Gr\#16.}
\label{fig:fast_full}
\end{figure}

The EFOSC2 CCD\#40 is a thinned chip, hence has significant fringing
beyond 7200\AA\ and the severity depends upon the grating used.  
To remove this fringing pattern (in spectroscopic mode) a
calibration flat field lamp exposure {\em immediately after or before}
the science image is required and is divided into the science spectrum
\citep[another method is to construct fringe frames directly from on-sky frames
during the night, but this requires that the targets are dithered along the slit; as discussed in][]{2008ApJ...674...51E}. 
Afternoon flats are not useful for this owing to wavelength drifts that 
alter the fringe pattern and hence do not allow for removal. PESSTO
always takes internal lamp flats (3 exposures of typically 40,000
ADU maximum count level) after taking any 
science spectra with Gr\#16. We do not, as default, take fringe
correcting flats when using Gr\#13, as the fringing does not affect this grism to a significant extent, as can be seen in Fig. \ref{fig:fast_full}. 
The internal lamp in EFOSC2 is a quartz lamp, which suffers from  water
vapour absorption that is not present in the afternoon dome flats. 

\begin{figure}
\centering
\includegraphics[scale=0.6,angle=270]{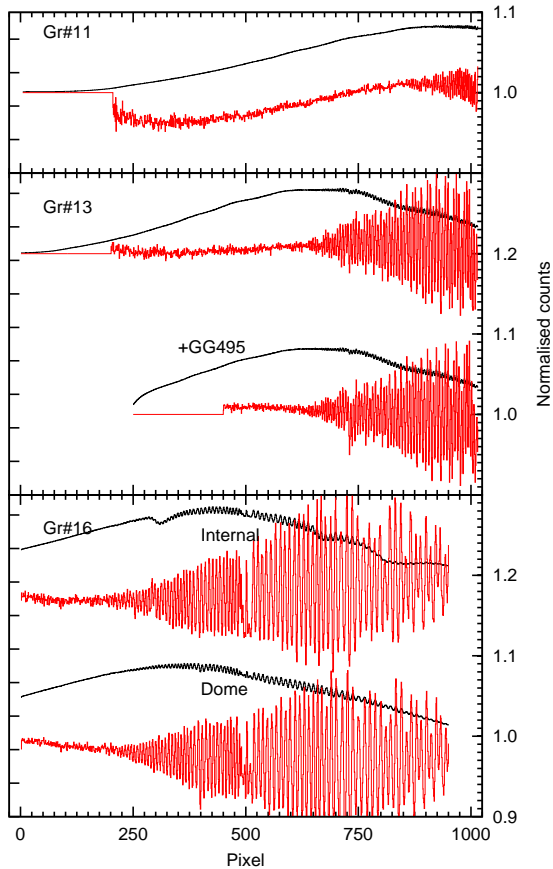}
\caption{Representative flat fields for Gr\#11, 13 and 16, taken in
  August 2013. Each panel shows a cut along column 450 of the
  two-dimensional flat-field, corresponding to the approximate pixel
  coordinates of the target. The black lines show the intrinsic shape
  of the flat fields, while the red lines show the normalised flat
  field after fitting with a high-order polynomial (all spectra are
  normalised to 1, and offset for clarity). For Gr\#13, the
  flat field with the GG495 order blocking filter (as used when
  correcting for second-order contamination) is also shown. For
  Gr\#16, both internal flats and dome flats are shown; in the former 
the absorptions due to H$_2$O vapour are visible in the un-normalised
flat. This absorption feature is fit by the high-order polynomial
during normalisation and hence is removed in the normalised flat. 
}
\label{fig:flat}
\end{figure}

In all cases, the combined flats are normalised to remove their
overall shape, while leaving the pixel to pixel response (and in the
case of Gr\#16, the fringing pattern) that we are attempting to
correct for. High-order splines are fitted to the flat fields, with
orders 90, 35 and 70 for Gr\#13, Gr\#11 and Gr\#16 respectively.
 In Fig. \ref{fig:flat} we show the shape of the flat fields used by
PESSTO, both before and after normalisation.
In the pixel regions below 200  for Gr\#11 
(which is approximately 4450 \AA), the response
of the dome flat field lamp is very poor, resulting in low
signal. Even after combining the frames, the count rate is typically
1700e$^{-}$ which would only increase the noise in the science frames
rather than improve it. Hence we set the flat level to unity between
pixels 1 and 200. The same effect occurs for Gr\#13 and we do not
flat field the first 200 pixels of each spectra. 
In summary, for Gr\#11 and Gr\#13 we employ masterflats constructed
using the dome lamps 
for each sub-run of 3-4 nights. For  Gr\#16 we do not use masterflats,
but instead use three flatfields taken with the internal
lamp immediately after the science spectrum is taken. 
Figure\,\ref{fig:flat} illustrates the profile of the normalised flats used in
the  flat-fielding process.  It is noticeable that Gr\#11 is not flat
at the unity level, but has variations of order 3-4\%. While this 
is not ideal, and will be fixed in future data releases (e.g. SSDR2), 
we have left it as shown, as the variation will not
affect the flux calibration since both the standards and science
objects are treated with the same flat-field. We also note that  the normalisation of the dome flats often is not continuous at the point where we set it to unity and discrete steps are 
induced at the level of 1-3\% (the examples shown in Fig\,\ref{fig:flat} for Gr\#11 is one of the worst cases). One might expect this to be propagated into the science spectra since the response functions derived from the spectrophotometric  standards
will not reproduce such a step function. 
We have checked the highest signal-to-noise science spectra and don't see obvious signatures of this flat-field feature. Nevertheless it is an undesirable
feature of the released dataset and we will aim to fix in the next data release.

\begin{figure}
\centering
\includegraphics[scale=0.65,angle=270]{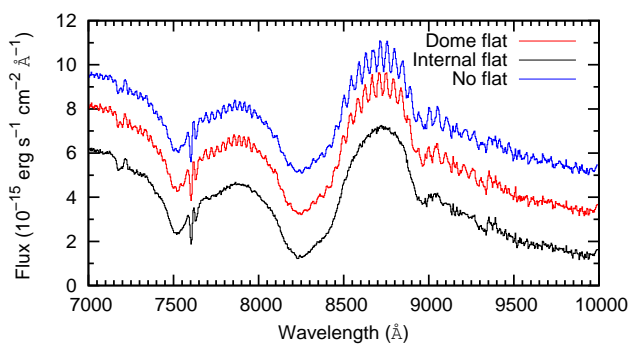}
\caption{Gr\#16 spectrum of SN 2011gr. Three reductions are shown, in the first there is no flat-fielding, in the second a dome flat from the start of the night is used, in the third an internal flat taken immediately after the science observations.}
\label{fig:gr16}
\end{figure}

Fig.\ref{fig:gr16} shows how fringing is almost completely removed in Gr\#16
spectra using   an internal contemporaneous flat-field. It also
illustrates that using a flat from the afternoon (not at the same telescope
position) has no impact on fringe removal, and in fact makes it
slightly worse.

\subsubsection{Cosmic ray removal} 
\label{sec:efosc-lacos}

\begin{figure}
\centering
	\includegraphics[width=0.9\linewidth]{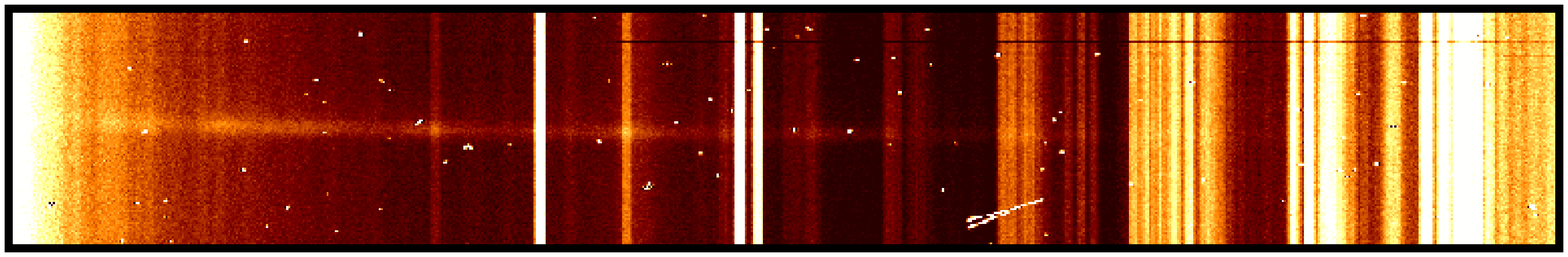}
	\includegraphics[width=0.9\linewidth]{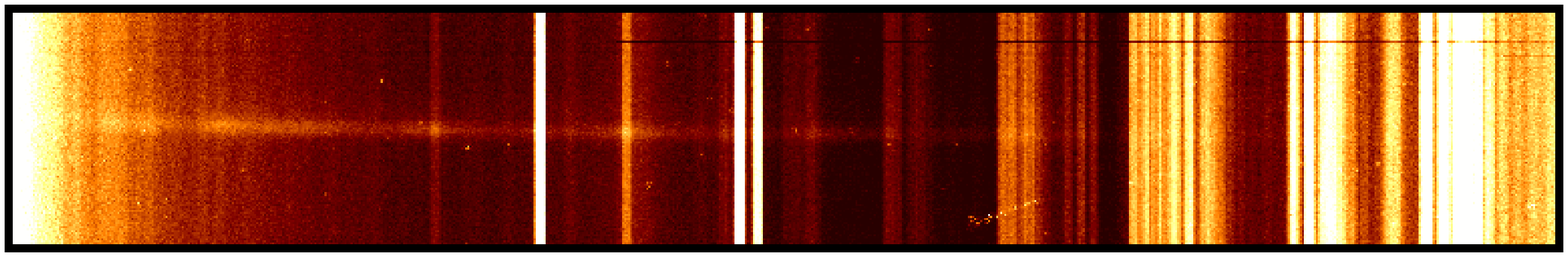}
\caption[]{2D spectrum of LSQ12drz, reduced with and without LACOSMIC
  applied. The PESSTO pipeline applies this, but only in the central
  $\pm$100 pixels around the science object. }
\label{fig:clean2d}
\end{figure}

\begin{figure}
\centering
\includegraphics[scale=0.65,angle=270]{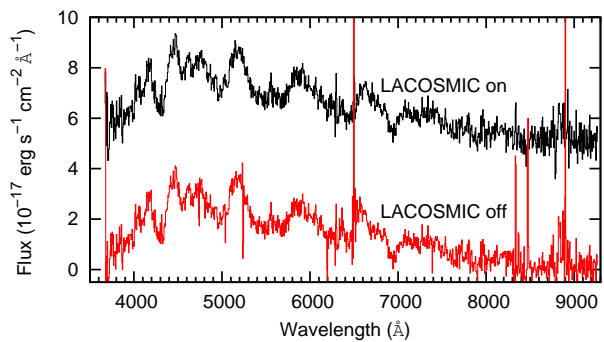}
\caption{The classification spectrum for LSQ12drz taken on 2012 August
  7, reduced with PESSTOFASTSPEC with and without cosmic ray
  rejection. The relatively long exposure time for this spectrum (2700
  s) results in a large number of cosmic ray hits, which are removed
  in the 2D image using the algorithm of 
\cite{2001PASP..113.1420V}. 
}
\label{fig:clean}
\end{figure}

EFOSC2 spectra with a typical exposure time of $\sim$1800 s will show numerous cosmic ray hits in the 2D frames, as can be seen in the upper panel of Fig. \ref{fig:clean2d}. After de-biasing and flat-fielding, we use the cosmic ray rejection algorithm {\sc lacosmic}\footnote{http://www.astro.yale.edu/dokkum/lacosmic/}
 presented in \cite{2001PASP..113.1420V} to remove these. The PESSTO pipeline incorporates a modified version of the {\em python}
 implementation\footnote{See http://obswww.unige.ch/$\sim$tewes/cosmics\_dot\_py/} of {\sc lacosmic} that
avoids the use of the {\em scipy} package. As it is a computationally expensive
process during the manual extraction, only the central 200 pixels around the object are cleaned (i.e. central pixel $\pm$100 pixels). 
Figure\,\ref{fig:clean} shows an
example of the extracted spectrum before and after cosmic ray cleaning. One concern of applying this cosmic ray rejection is whether it may erroneously remove real, narrow, emission features from spectra. However, we have tested this in the EFOSC2 data, and in particular for SN 2009ip which has narrow lines, and are confident that this is not the case. Whether cosmic rays were removed or not is recorded in the header as described in
Sect.\ref{Appdx:cosmic}. 

\subsubsection{Arc frames and wavelength calibrations} 
\label{sec:efosc-arc}

Arc frames are generally taken in the evening before observing and are never taken during dark time. 
EFOSC2 has helium and argon lamps and PESSTO uses both of these lamps
turned on together. No arc frames are taken during the night to reduce overheads. 
Although EFOSC2 suffers from significant flexure as the instrument rotates at
the nasmyth focus (which can be 4 pixels over 200 degrees in rotation), 
the flexure causes a rigid shift of the wavelength frame. Hence we apply 
the calibration determined from the evening arc frames and adjust this with a 
linear offset as measured from either the skylines or atmospheric
absorption lines. 

\begin{figure*}
\centering
\includegraphics[scale=0.65,angle=270]{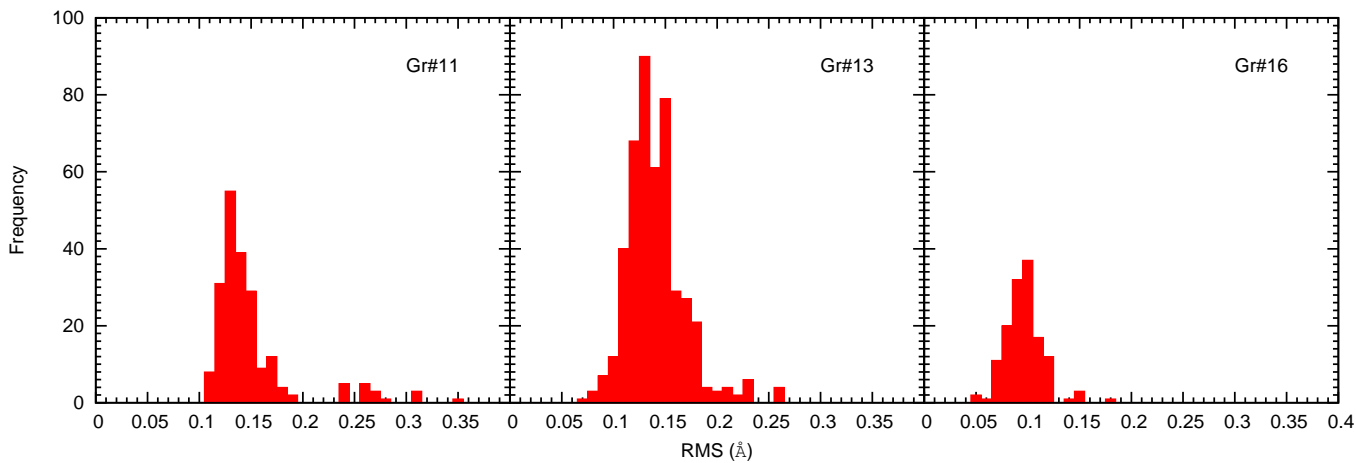}
\caption{Histogram of the root mean square error of the fit to the arc lines for each science spectrum. Each panel includes 1.0\arcsec\ and 1.5\arcsec\ slit spectra.}
\label{fig:wave_rms}
\end{figure*}

Relatively high-order Legendre polynomial fits are needed to fit the
EFOSC2  arc lines with a fit which produces no systematic residuals. 
For Gr\#13, 13-15 lines were used with a fifth- or sixth-order fit; when the
GG495 order blocking filter was also used the order of the fit was
reduced to 5 (due to the smaller wavelength range). For Gr\#11,
9 lines were used with a fifth-order fit, while for Gr\#16 11-14 lines
were used for a fifth or sixth-order fit. The root mean square (RMS) error
of the fit was typically found to lie between 0.1 and 0.2\AA\ as shown
in Fig. \ref{fig:wave_rms}.  The number of arc lines used for the
dispersion solution of each object, along with the RMS error, are
given in the header of the reduced spectra by the keywords
\texttt{LAMNLIN} and \texttt{LAMRMS} respectively. 
The formal  RMS values are probably too small to realistically represent
the uncertainty in the wavelength calibration at any particular
point, given the FWHM of the arclines is  13-17\AA. Hence this might 
suggest over-fitting of the sampled points. As a comparison,  Legendre polynomials with
order 4 produced obvious systematic residuals and RMS values of
between 0.4-1.0\AA\ for a $1\farcs0$ slit and 1-1.8\AA\ for a
$1\farcs5$ slit.

For exposures longer than 300\,s, the linear shift applied to the
dispersion solution is measured from the night sky emission lines. For
shorter exposures (brighter objects), the night sky
lines are weak or not visible, and the shift is instead measured from the
telluric absorptions in the extracted 1D spectrum.  The linear shifts
are calculated by cross-correlating the observed spectrum (sky or
standard) with a series of library restframe spectra which are offset
by 1\AA.  The library spectrum which produces the minimum in the
cross-correlation function is taken as the correct match and this
shift is applied. This method limits the precision of the shift to
1\AA, which is roughly $\frac{1}{4}$ of a pixel and less than $\frac{1}{10}$ of
a resolution element. This value of 1\AA\ is recorded in the header as the
systematic error in the wavelength calibration (\texttt{SPEC\_SYE}, see
Appendix\,\ref{Appdx:wavecal}). 

The value of the linear shifts applied are typically in the range of
6-13 \AA\ for Gr\#11 and Gr\#13. In the case of Gr\#16 spectra the
shifts were usually smaller, usually 4-9 \AA.  This value is recorded
in the header keyword \texttt{SHIFT}. Full details of the header
keywords applicable for the wavelength solution are in
Appendix\,\ref{Appdx:wavecal}.

\subsection{Spectrophotometric standards and flux calibration} 
\label{sec:efosc-specphot}

PESSTO uses a set of 9 spectrophotometric standard stars for EFOSC2, listed in
Table\,\ref{tab:specphotstandards}.  The data in this table are taken
directly from Simbad\footnote{http://simbad.u-strasbg.fr/simbad/}. The EFOSC2 finding charts, including proper
motion projections, are available for
users from the PESSTO website, and the data tables used for flux
calibration standards are available in the publicly accessible PESSTO
pipeline.  These standards provide year round coverage, have full
wavelength coverage from the atmospheric cut-off up to $1\mu$m and are
in a suitable magnitude range for a 3.5m aperture telescope. PESSTO
standard policy is to observe an EFOSC2 spectrophotometric standard
three times per night (start, middle and end), although if there are
significant SOFI observations or weather intervenes then this may be
reduced. Generally, the three observations will include 2 different
stars and a set of observations is taken with all grism, slit and
filter combinations used during the nights observing.
From September 2012 to November 2012  the standard EG21 
was frequently observed.  We later realised 
however that the photometric flux tables for this star did not cover
the full, telluric corrected regions for Gr\#16 and Gr\#13 and hence
stopped using it after 2012-11-21. We have
only used it to calibrate PESSTO data taken with Gr\#11 in SSDR1.

The wavelength coverage of GR\#13 is 3650 - 9250\AA, and for science
targets we do not use an order blocking filter. Hence second-order
contamination is possible for blue objects beyond around 7200\AA,
depending on their colour. This would also affect the flux standards
and hence the flux calibration of science targets. To remove any
second-order contamination in the flux standards, PESSTO always takes
Gr\#13 data for these stars with and without the filter GG495, to allow correction for the effect during pipeline reductions. 
The blocking filter GG490 has a transmission of 90\% from 5000\AA\ and upwards. The sensitivity function for the combination of Gr\#13+GG490 is scaled up to match the sensitivity function of Gr\#13+Free at the position of 5500\AA. 
To construct a final sensitivity function which is corrected for any second-order flux in the standards, we merge the sensitivity function Gr\#13+free (from 3650-5500\AA) and the scaled up sensitivity function of Gr\#13+GG490 (for wavelengths $>5500$\AA).  Flux
standards are always observed unless clouds, wind or humidity force
unexpected dome closure. Hence even during nights that are not
photometric, flux standards are taken and the spectra are flux
calibrated; we deal with the issue of the absolute flux reliability
below. 

\begin{figure*}
\centering
\includegraphics[scale=0.65,angle=270]{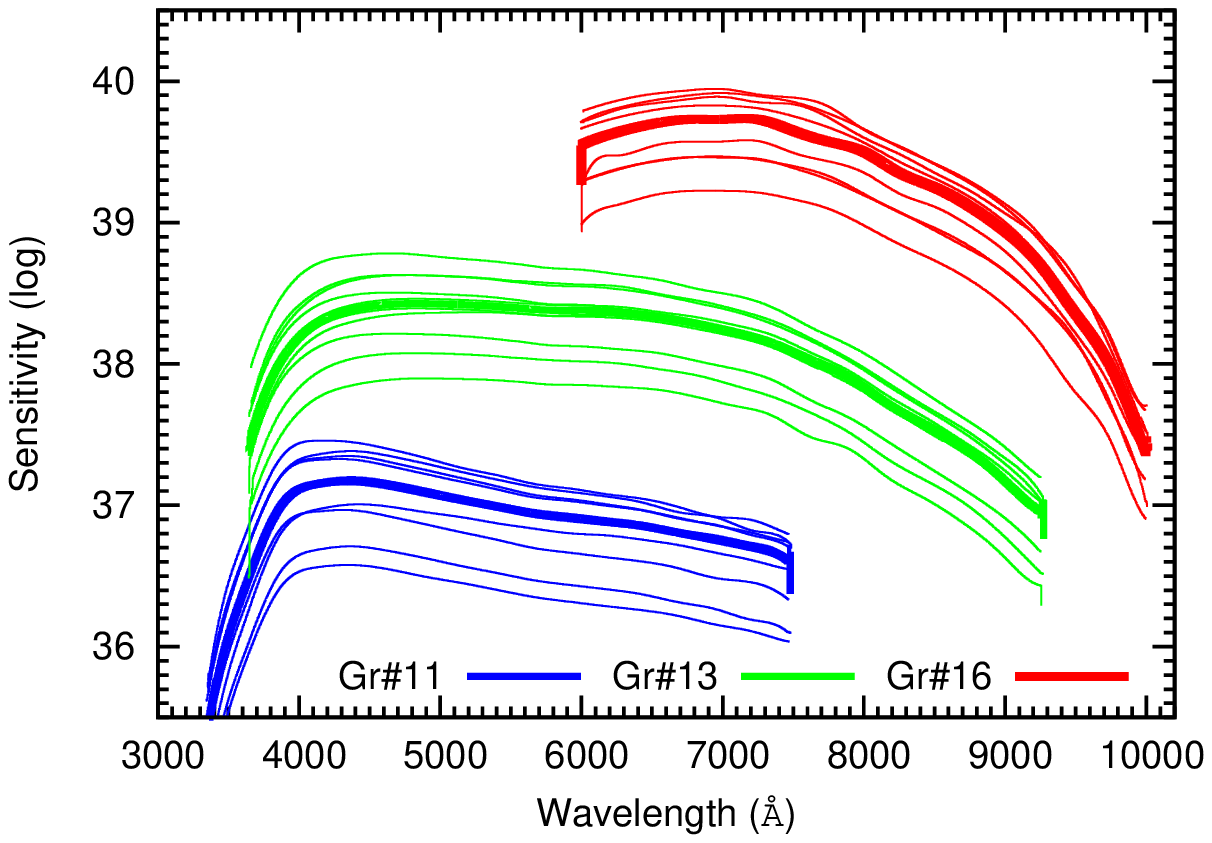}
\includegraphics[scale=0.32,angle=270]{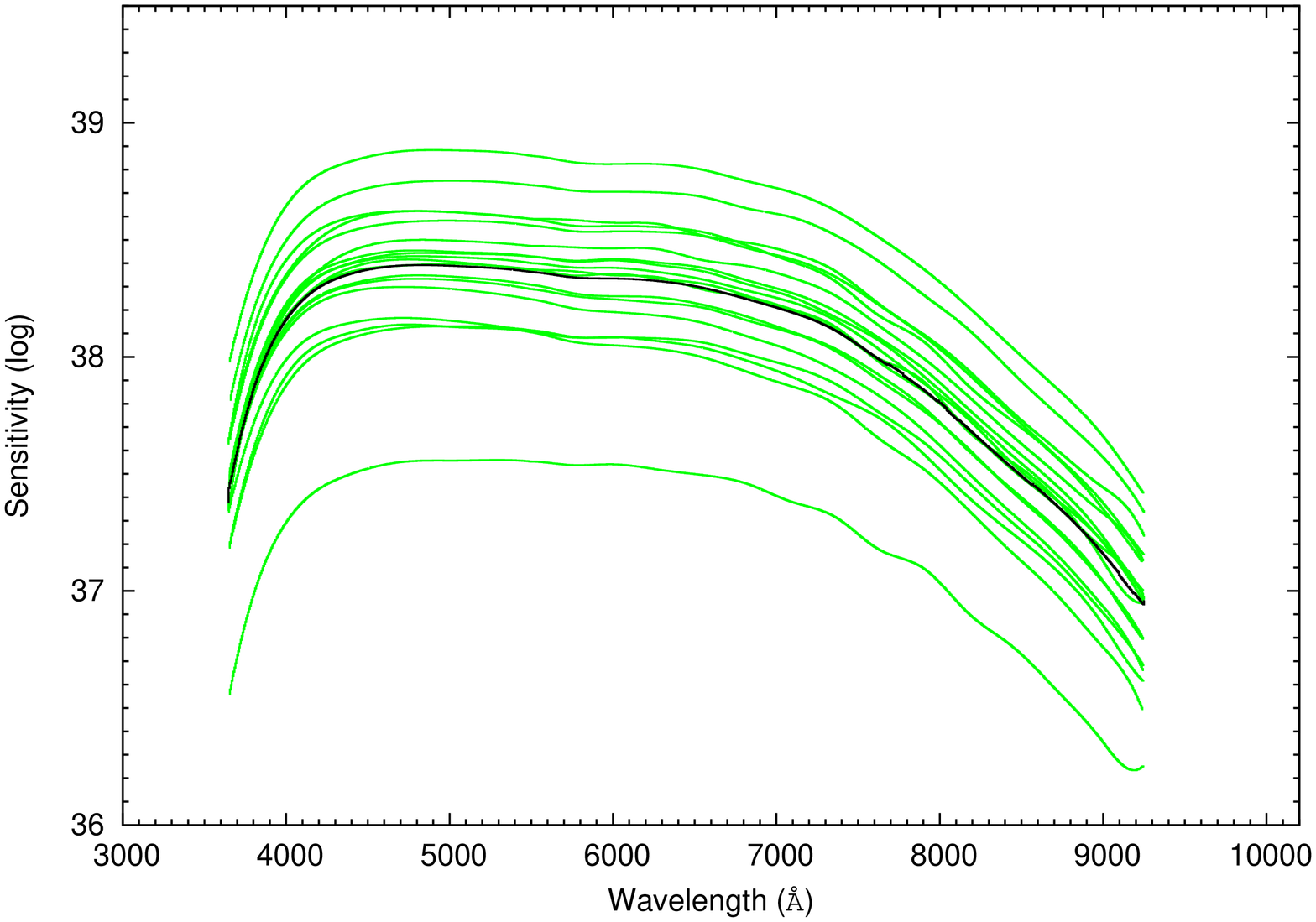}
\caption{{\bf Left: } The average monthly sensitivity curves for  Gr\#11, Gr\#13
  and Gr\#16, derived from spectrophotometric standards taken during
  the first year of PESSTO observations. The thick lines are annual
  sensitivity function. Thin lines are 
monthly averages (Gr\#11 offset by -1, Gr\#16 offset by +1 for legibility)
{\bf Right :} As a comparison, the sensitivity curves 
from individual standard stars over ten nights in March 2013 (for Gr\#13). 
Four different standards were observed, L745a, LTT3864, EG274, GD71. 
The master sensitivity curve for March 2013 with this configuration is shown in black.  
}
\label{fig:sens}
\end{figure*}

A sensitivity function is derived for each EFOSC2 configuration from
the spectrophotometric standards observed. These were averaged to
create a master sensitivity function for each month, which was then
applied to the final reduced spectra. In a few instances, a master
sensitivity curve was not  created for a particular configuration on
a given month, as there were no appropriate standards observed. In
these cases, the sensitivity function from the preceding or following
month was used. In Fig. \ref{fig:sens}, we show the shape of the
sensitivity curves, together with the variation in sensitivity
functions from month to month.
 The sensitivity curves from individual standard stars within 
a particular month are shown for comparison. While there are grey shifts 
between the individual sensitivity curves (as expected for observations taken with a single slit width under 
differing atmospheric conditions), the overall shape of the sensitivity curves are quite similar, 
indicating that the relative flux calibration is reliable.

The
standard method of ensuring spectra are properly flux calibrated is to 
compare synthetic photometry of the science spectra with
contemporaneous calibrated photometry and apply either a constant,
linear or quadratic multiplicative function to the spectra to bring
the synthetic spectra into line with the photometry. For PESSTO SSDR1
this is not yet possible for all spectra since the photometric lightcurves
are not yet finalised for many of the science targets and the
classification spectra do not have a photometric sequence. 
However it is useful to know what the typical uncertainty is in any 
flux calibrated PESSTO spectrum, and this is encoded in the header
keyword \texttt{FLUXERR}. PESSTO observes through non-photometric
nights, and during these nights all targets are still flux
calibrated. Hence the uncertainties in flux calibrations come from 
transparency (clouds), seeing variations that cause mismatches between
sensitivity curves derived using standards with different image
quality, and target slit positioning. Finally, photometric flux is
generally measured with point-spread-function fitting, which inherently
includes an aperture correction to determine the total flux whereas 
spectroscopic flux is typically extracted down to 10 per cent of the
peak  flux (a standard practice in {\sc IRAF}'s {\em apall} task). 
All of this means that large  variations are expected and 
we carried out tests as to how well this method works and what is the
reliability of the absolute flux calibration in the spectra. 

\begin{figure*}
\centering
\includegraphics[scale=1,angle=270]{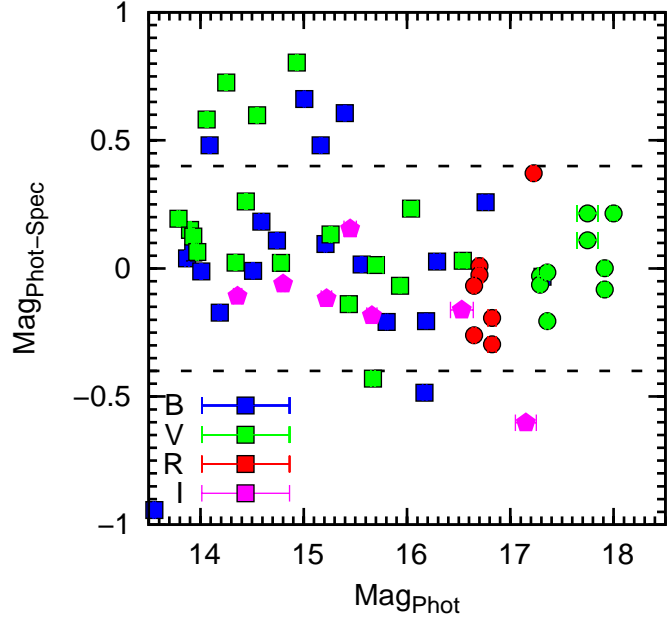}
\caption{Synthetic magnitudes as measured from flux-calibrated spectra
 compared to the photometric magnitude at the same epoch for 
SN2009ip  \citep[][Gr\#11 and Gr\#16]{2013MNRAS.433.1312F}
SN2012fr  \citep[][and Conteras et al., in prep; Gr\#11 and Gr\#16]{2013ApJ...770...29C}
SN2013ai  (Fraser et al. in prep; Gr\#13).  
$Mag_{\rm Phot}$ is the calibrated photometric magnitude and the y-axis is 
the difference between this and the synthetic photometry measured from the
flux calibrated spectra. Colours indicate filters, square symbols are Gr\#11, pentagons are
Gr\#16 and circles are Gr\#13. }
\label{fig:synthphot}
\end{figure*}

We took the Gr\#11, Gr\#16 and Gr\#13 spectra of the three targets for which
a calibrated photometric sequence is either published or has been
measured and is in preparation :
SN2009ip  \citep{2013MNRAS.433.1312F}
SN2012fr  \citep[][and Conteras et al., in prep.]{2013ApJ...770...29C}
SN2013ai  (Fraser et al. in prep).  
Synthetic $BVRI$ photometry on the
PESSTO spectra was calculated using the {\sc synphot} package within {\sc 
iraf} and spectra which covered the entire bandpass of each filter 
were included and the difference between the spectral synthetic magnitudes
and photometric measurements is shown in Fig.\,\ref{fig:synthphot}. 
This illustrates the difficulty and challenges faced in accurately flux
calibrating spectra but also shows promise that in future data
releases we can significantly improve on SSDR1. The standard deviation 
of all points in $\pm$0.31$^{m}$ or $\pm$29\%, but we can 
identify several cases were fairly obvious catastrophic failures have
occurred.  The main bulk of
points lie within  a range of $\pm0.44^{m}$ around zero, with a
formal average offset and standard deviation of $0.042\pm0.164^{m}$. 
The four low lying points have either images of poor seeing, when the
seeing changed rapidly before the slit width was changed or (in the
case of the point at $-$0.9) the object likely was not positioned on
the slit. The points lying above $+0.5$ were found to be owing to a
monthly sensitivity curve for January 2013 which was too low, likely
due to an unusually high frequency of non-photometric cloudy conditions when
the standards were taken. Ignoring these fairly obvious problematic 
cases, the scatter in the flux calibration is $\sim15$\% in the main
locus and is recorded in the headers of all spectra. 

\begin{small}
\begin{verbatim}
FLUXERR = 15.2 / Fractional uncertainty of the flux [%]
\end{verbatim}
\end{small}


Science users should use this as a typical guide, if the seeing (as
can be  measured on the 2D frames and acquisition images)
and night conditions (from the PESSTO wiki night reports and see
Appendix \,\ref{appdx:photomet-nights}) 
 are 
reasonable. Note that we  have recorded the standard deviation formally as 
15.2\% in the file headers but do not attach significance to the decimal digit. 
In future data releases we plan to significantly improve
on the flux calibration scatter and reduce both the failures and
intrinsic scatter. Reviewing the monthly sensitivity curves and
applying photometric calibrations from the $V$-band acquisition images
are the two most promising routes. 
The  $V-$band acquisition images would allow a constant 
offset to be applied in an automated way, but only if all sky catalogues were
available. In the future, the combination of 
Pan-STARRS1 \citep{2013ApJS..205...20M}
and 
SkyMapper \citep{2007PASA...24....1K}
will supplement 
SDSS DR9 \citep{2012ApJS..203...21A}
to provide this all-sky reference catalogue
and would allow an adjustment to the flux to bring down the absolute
flux error to probably a few per cent. PESSTO will pursue this type of 
calibration as far as the reference catalogues will allow in future 
data releases.  

\begin{figure}
\centering
\parbox[t]{8cm}{
\includegraphics[scale=0.7,angle=270]{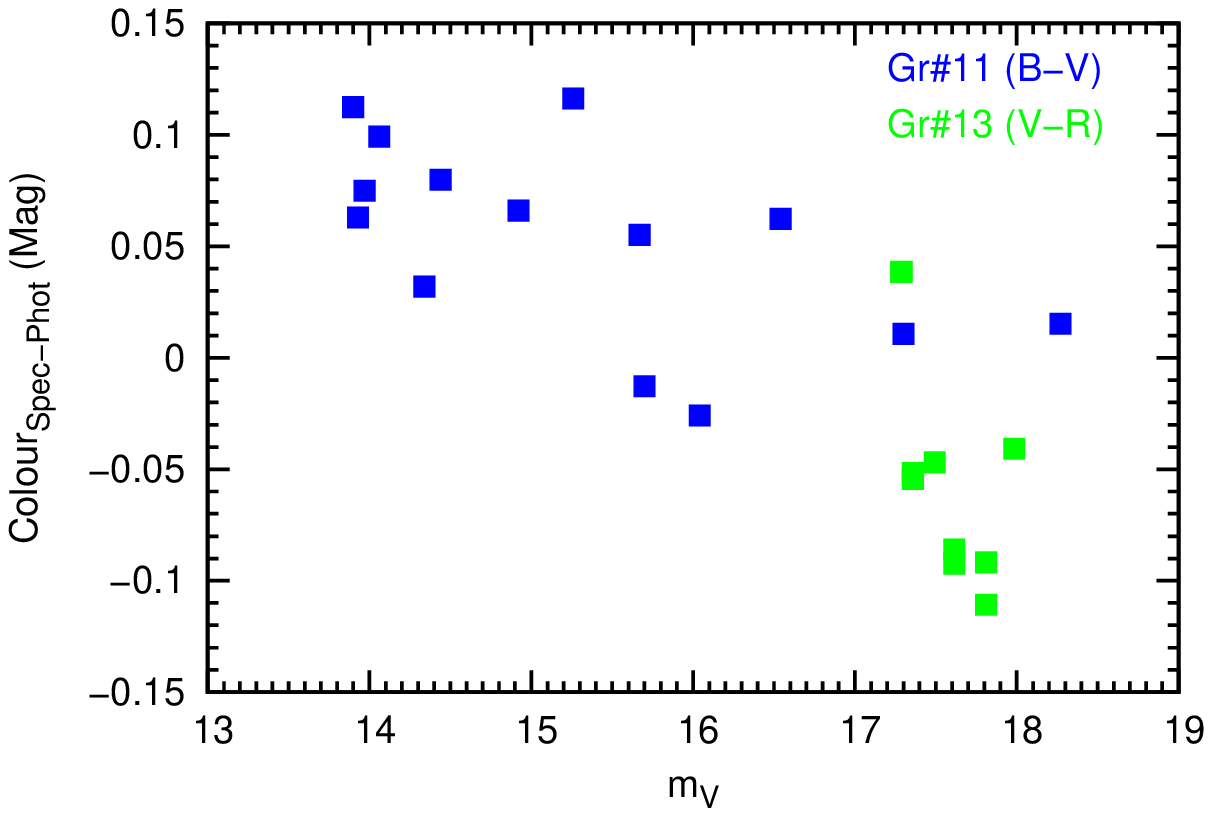}
}
\hfill
\parbox[t]{8cm}{
\includegraphics[scale=0.7, angle=270]{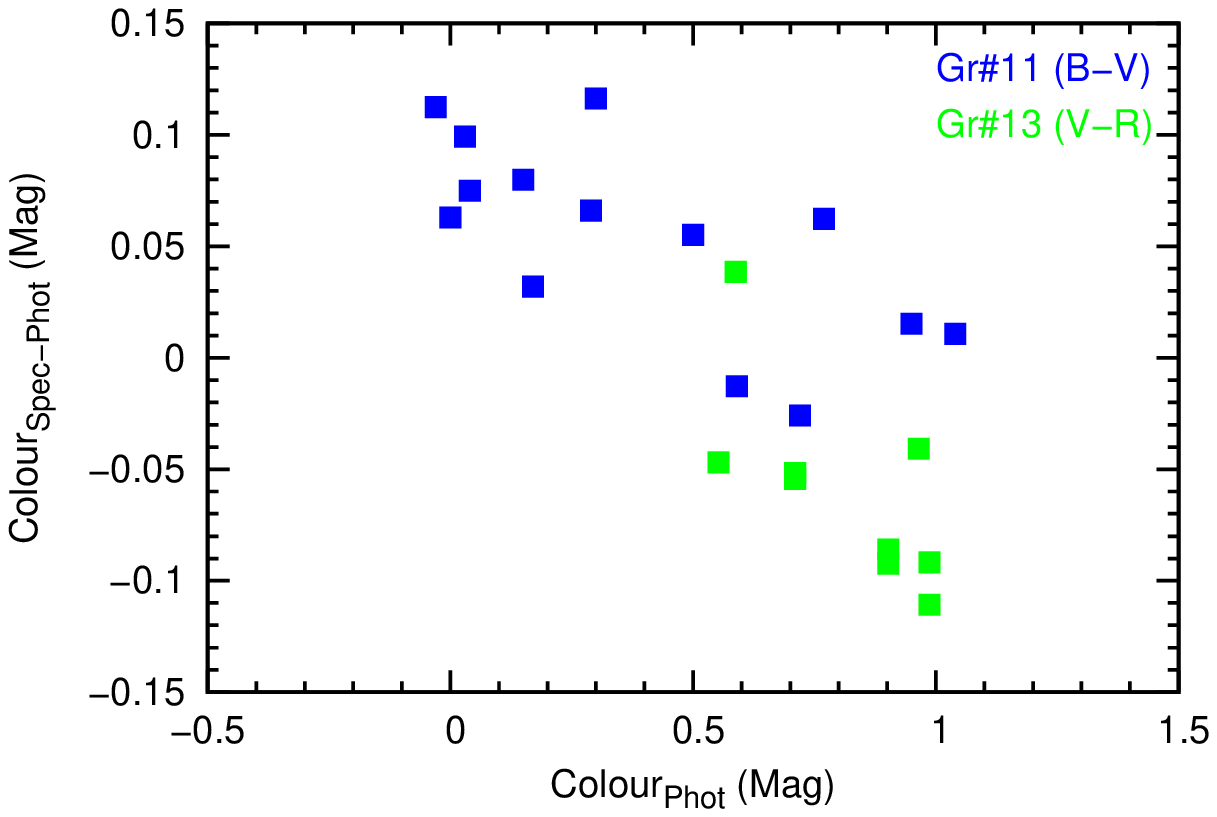}
}
\hfill
\caption{A check on the relative flux calibration of the PESSTO spectra. The difference between the
synthetic photometry colours of SN2009ip (Gr\#11) and SN2013ai (Gr\#13) and photometric 
measurements is plotted on the y-axis. The x-axis is simply the $V$-band photometric magnitude on the left panel and photometric colour (either $B-V$ or $V-R$) on the right. 
}
\label{fig:fluxcal-col}
\end{figure}

We carried out further checks to determine the relative flux calibration across the EFOSC2 
spectra compared to photometric measurements. We employed the 
$BV$ photometry of \cite{2013MNRAS.433.1312F} for SN2009ip and determined synthetic photometry 
from the EFOSC2 Gr\#11spectra, with the results plotted in Fig.\,\ref{fig:fluxcal-col}. 
The average offset of  these 14 spectra gives  $(B-V)_{spec} - (B-V)_{phot} = 0.05 \pm 0.04$\,
 mag
(where the error is the standard deviation of the individual differences). 
For Gr\#13 spectra, we used $VR$ photometry of SN 2013ai from Fraser at al. (in prep.) 
and again the results are shown in Fig.\,\ref{fig:fluxcal-col}. The 10 spectra give an 
average of
$(V-R)_{spec} - (V-R)_{phot} = -0.05 \pm0.05$\, mag.

The comparison plots indicate that there may be a systematic trend. It could 
indicate that the spectra of brighter objects are $\sim$5\% redder than the photometry 
would imply. Or that the Gr\#11 spectra and Gr\#13 spectra have systematic offsets 
of +0.05 mag and $-$0.05 mag in comparison to what the photometry would imply.
However this is not completely clear, since the
average uncertainties in the photometric points are  $\pm$0.04 mag for SN2009ip and $\pm$0.05 mag for SN2013ai 
\citep[][Fraser et al., in prep.]{2013MNRAS.433.1312F}.  We should also 
note that the SN2013ai photometry  comes from SMARTS 1.3m telescope and
 KPNO filters. These are different to Johnson and Bessell filters and one
should ideally do an S-correction on the photometry for consistent comparison. 
These trends will be probed further when we have more calibrated photometry 
and can investigate the trends with better statistics.  From this preliminary 
investigation,  we suggest that the relative flux calibration in the PESSTO spectra
is accurate to around 5\%.

\subsection{Telluric absorption correction} 
\label{sec:efosc-tell}

PESSTO does not specifically observe telluric standards for EFOSC2
such as fast rotating, smooth continuum stars.  Instead, the data
reduction pipeline uses a model of the atmospheric absorption to
correct for the H$_{2}$O and O$_{2}$ absorption.  
The model was computed
by F. Patat using the Line By Line Radiative Transfer Model 
\citep[LBLRTM][]{2005JQSRT..91..233C}.
Details on the model and the parameters used can be found in 
\citep{2011A&A...527A..91P}. 
This is carried out
for all grism set-ups.  The intensities of H$_{2}$O and O$_{2}$
absorptions in the atmospheric absorption model are first Gaussian smoothed to
the nominal resolution of each instrumental set-up, and then rebinned
to the appropriate pixel dispersion.  The pipeline then scales the
model spectrum so that the intensities of H$_{2}$O and O$_{2}$
absorptions match those observed in the spectrophotometric standards, 
hence creating multiple model telluric spectra per night. 
Each science spectrum is then
corrected for telluric absorption, by dividing it by the smoothed,
rebinned, and scaled  absorption model which is most closely matched
 in time i.e. closest match between the standard star observation time
 and the science observation time. 

\begin{table*}
\caption[]{PESSTO spectrophotometric standards. All data in this table
  are taken from Simbad.}
\label{tab:specphotstandards}
\begin{tabular}{lcccccl}
\hline\hline
Standard Name           & RA (FK5, J2000) & DEC (FK5, J2000) & Proper motion (mas/yr)  & V mag & Sp. Type & Instrument  \\\hline
	VMA2         & 00 49 09.902	& $+$05 23 19.01  &	1236.90, $-$2709.19& 	12.374&	DZ8     &   EFOSC2\\	
	GD71         & 05 52 27.614	 &$+$15 53 13.75  &	        85, $-$174  &	13.032	&DA1              &   EFOSC2/SOFI  \\
	L745-46a  & 07 40 20.79	 &$-$17 24 49.1    &	  1129.7, $-$565.7  &	12.98	&DAZ6        &  EFOSC2        \\	
   LTT 3218  & 08 41 32.50     & $-$32 56 34.0    &   $-$1031.7,  1354.3 &  11.85 &  DA5  &  SOFI  \\
	LTT3864   &	10 32 13.603	 &$-$35 37 41.90  &	    $-$263.7, $-$8.0  &	11.84	&Fp...  &     EFOSC2    \\
	GD153      &	12 57 02.337	&$+$22 01 52.68  & $-$46, $-$204  &	13.35	&DA1.5     &       EFOSC2/SOFI     \\
	EG274       &	16 23 33.837	& $-$39 13 46.16  &	     76.19, 0.96  &	11.029	&DA2            &    EFOSC2/SOFI         \\ 
	EG131       &	19 20 34.923	& $-$07 40 00.07  &	 $-$60.87, $-$162.15&  	12.29	&DBQA5 & EFOSC2 \\      
	LTT 7379  &	18 36 25.941	& $-$44 18 36.93  &	 $-$177.05, $-$160.31&  	10.22&	G0   & EFOSC2\\           
   LTT 7989  & 20 11 12.08     & $-$36 06 06.5    &      522, $-$1691   &   11.5  &   M5V    &  SOFI  \\
   Feige110  & 23 19 58.398	& $-$05 09 56.16  &        $-$10.68, 0.31  &	11.5 &	sdO         & EFOSC2/SOFI  \\
\hline\hline
\end{tabular}
\end{table*}

\section{PESSTO  EFOSC2 imaging  observations and calibrations}
\label{sec:efosc-image}

EFOSC2 is used in imaging mode for PESSTO to provide supporting
photometry for some targets. Much of the photometric lightcurve data
is provided by PESSTO scientists through their access to other
facilities such as the SMARTS 1.3m
\citep{2003SPIE.4841..827D}, Liverpool Telescope \citep{2004SPIE.5489..679S}
the LCOGT facilities
\citep{2014SPIE.9149E..1EB}
the SWOPE 1m, 
\citep{2012SPIE.8444E..4HP}
Asiago Telescopes 
\citep{2014AN....335..841T}
and 
PROMPT
\citep{2005NCimC..28..767R}. 
However
EFOSC2 is also used for supporting data, particularly when the targets
are fainter than around 19.5$^{m}$. The detector set-up is exactly the
same as for the spectroscopic observations as described
above in Sect.\,\ref{sec:efosc-ccd}, and during each PESSTO night
the filter wheel is loaded with the filters $U$\#640, $B$\#639,
$V$\#641, $R$\#642, $g$\#782, $r$\#784, $i$\#705, $z$\#623. 
These filters have typically been employed in $UBVRi$ or $Ugriz$
sequences depending on the science target. Additionally, an acquisition image is taken through a $V$-band filter
before every spectroscopic exposure to identify the target and allow
it to be placed on the slit. These are also processed in a similar
manner to the photometric science frames. The data
final products and access are described in  Sect.\ref{sec:efosc-products}.

\begin{figure*}
\centering
\includegraphics[scale=0.65,angle=270]{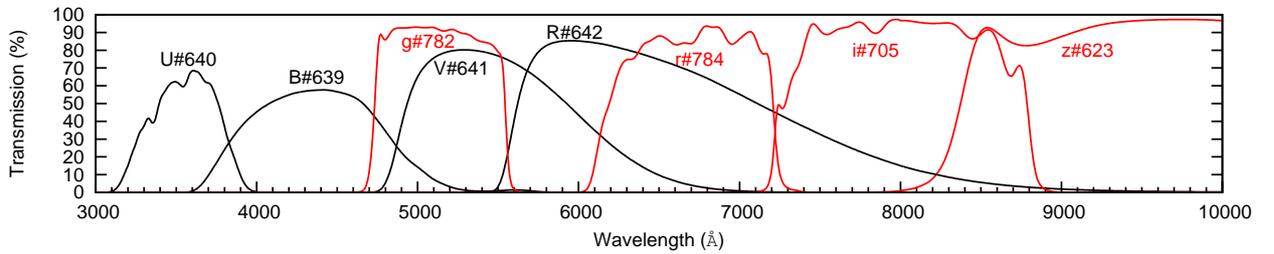}
\caption{Filter functions taken from the ESO database. The black lines are for the Landolt filters, while the red lines are for the Gunn filters. The z filter does not have a cut-off in the red, but is instead limited by the quantum efficiency of the CCD, which drops to 10\% at 1$\mu$m.}
\label{fig:filter}
\end{figure*}

\subsection{EFOSC2 imaging calibration frames and reduction} 
\label{sec:efosc-imagecal}

As the EFOSC2 CCD is read out in the same mode for both imaging and
spectroscopy the CCD characteristics as discussed in Sect.\,\ref{sec:efosc-ccd}
apply and the  bias subtraction calibration is carried out as described in
Sect.\,\ref{sec:efosc-bias}. The filters used for imaging are listed in Table\,\ref{tab:photstd} and their throughputs are illustrated in Fig.\,\ref{fig:filter} (data taken from ESO database). 
Cosmic ray cleaning is generally applied to the full frame imaging data, 
as described in Sect.\,\ref{sec:efosc-lacos}, 
and a header keyword is set to alert the user that this process has been 
run (see Appendix \ref{Appdx:cosmic}).

Twilight sky flatfields for imaging are typically taken once per
sub-run of 3-4N in all of the eight filters (or as many as weather
will allow). A master flat is created and used as close as possible to
the science, or acquisition frames.  The master flat and bias frames used for any particular frame can be found listed by in header keywords. 
The naming nomenclature is similar to that for the
spectroscopic calibration frames but without the grism and slit names. 

\begin{small}
\begin{verbatim}
ZEROCOR = 'bias_20130402_56463.fits' 
FLATCOR = 'flat_20130401_R642_56463.fits' 
\end{verbatim}
\end{small}

In constructing these, the individual flats are checked and those with
a high number of visible stars are rejected and not included in the
masterflat. The masterflats commonly show a feature of apparent
``dots'' in a straight line (along X) in the central pixel area of
[200:700,530:590] in filters $Vgrz$ (it is also faintly visible in
$B$). These are common, but transitory, and it is not clear if they are
illumination ghosts and hence not present in the science frames.
However, the counts level of these patterns differ by only 1\% from
the average level of the masterflat and as a consequence we assume
they do not impinge on science frame calibrations. 
Imaging fringe frames are constructed for the
$i$-band filter from a collection of NTT $i$-band images taken between Jan. 2010 and Apr. 2012.

\begin{figure*}
\includegraphics[width=15cm]{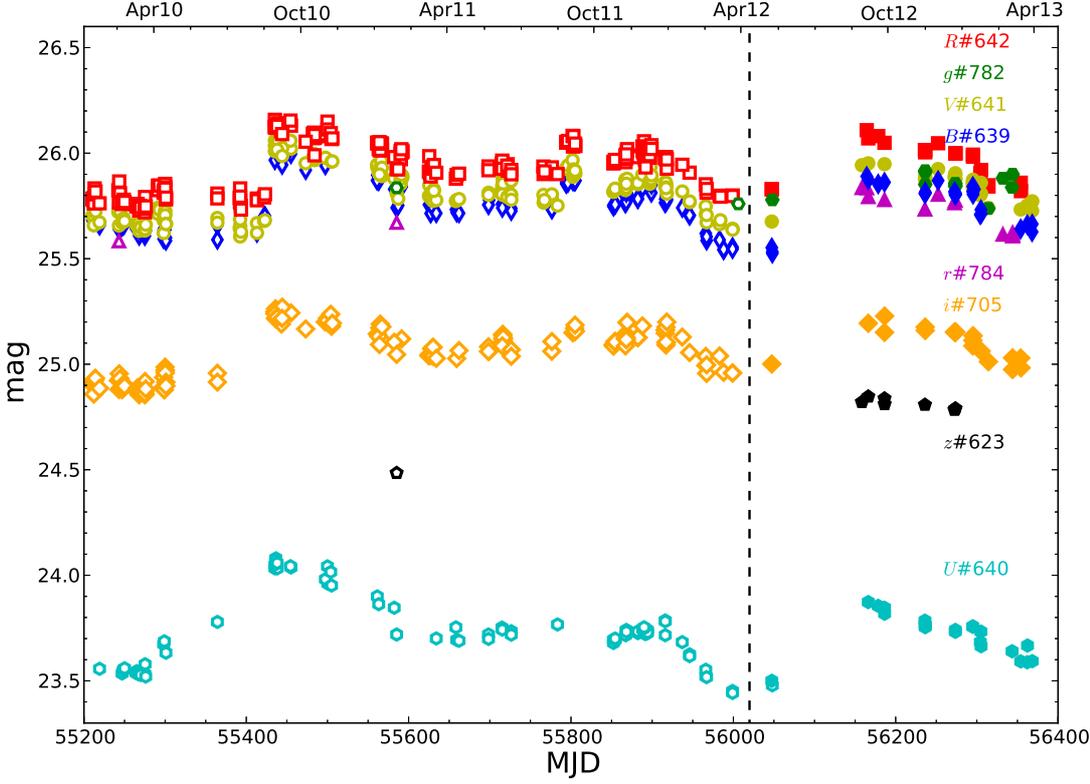}
\caption{Evolution of NTT zero points between 2010 and 2013. PESSTO data are shown by filled symbols,
while the open symbols refer to archival data. The vertical dashed line indicates the first PESSTO night. ZPs were evaluated from observations of PESSTO standard fields using
the PESSTO pipeline.}
\label{fig:zp}
\end{figure*}

PESSTO uses a set of 10 photometric standard fields, six of these 10
fields are photometric standards in both the Landolt and Sloan Digital
Sky Survey (SDSS) systems
(listed in Table\,{\ref{tab:photstd}). If the night appears to be
  photometric to the observers  then a photometric standard field is observed
three times.   As with the spectroscopic standards, at least two of these should be
different fields.  During nights which are clearly non-photometric, PESSTO
does not tend to take standard field calibrations. 
The PESSTO observers record their night reports on the PESSTO public
web pages \footnote{http://wiki.pessto.org/pessto-wiki/home/night-reports}
and record their judgment of whether the night is photometric or
not. This page is publicly available and is a useful
guide when interpreting the flux calibrations and validity of
zeropoints in the FITS headers of the imaging files. The information is
also recorded in the table in Appendix\,\ref{appdx:photomet-nights}.

\begin{table}
\caption{PESSTO photometric standard fields.}
\begin{center}
\begin{tabular}{lccc}
\hline\hline
Standard field & RA (J2000) & DEC (J2000) & Filters\\
\hline
T-Phe & 00 30 14.00 & $-$46 32 00.00 & $UBVRi$\\
PG0231+051 & 02 33 41.00 & 05 18 43.00 & $UBVRgriz$\\
RU149 & 07 24 15.40 & $-$00 32 07.00 & $UBVRgriz$\\
RU152 & 07 29 56.00 & $-$02 05 39.00 & $UBVRgriz$\\
PG1047+003 & 10 50 05.65 & $-$00 01 11.30 & $UBVRgriz$\\
PG1323-085 & 13 25 49.00 & $-$08 50 24.00 & $UBVRgriz$\\
PG1633+099 & 16 35 34.00 & 09 46 17.00 & $UBVRi$\\
PG1657+078 & 16 59 33.00 & 07 42 19.00 & $UBVRi$\\
MarkA & 20 43 58.00 & $-$10 47 11.00 & $UBVRi$\\
PG2336+004 & 23 38 43.00 & 00 42 55.00 & $UBVRgriz$\\
\hline
 \end{tabular}
 \end{center}
 \label{tab:photstd}
\end{table} 

The PESSTO pipeline is constructed to rapidly determine zero points (ZP).
 Instrumental
magnitudes are calculated for standard stars using {\sc daophot}
aperture photometry routines with an aperture set to 3 times the
measured FWHM in the image, which are then compared to catalogue
magnitudes.  We carried out this ZP calculation for all available
EFOSC2 imaging of the PESSTO standard fields for period stretching
back 3 years from April 2013 (along with a few points from
observations of the PG2213-006 standard field which is not a nominated 
PESSTO field).  Many of these data 
come from the Benetti et al. large programme (ESO 184.D-1140) and we built upon the choice of standards
and experience of that. 
The ZP trends are shown for each band in
Fig.~\ref{fig:zp} and in constructing this plot we rejected any night
which had an outlying ZP more than 0.5\,mag away from the average of the
5 ZPs closest in time as these were almost certainly non-photometric
nights. We then rejected measurements when 
the ZPs were greater than 1$\sigma$ from their neighbours within 
$\pm2$ adjacent days. 
 The resultant measurements are likely to be from photometric
nights which is illustrated by the low scatter and long term trends in
Fig.~\ref{fig:zp}. The cyclical long term trends are probably due to
the ZPs being maximised immediately after re-aluminisation of the primary
mirror (annually) and slow degradation afterwards.  An average of ZPs
and colour terms from Aug. 2012 until April 2013 is reported in
Tab.~\ref{table:zp}.   For PESSTO standard fields taken during PESSTO 
time the image products in the archive have zeropoints calculated 
directly with the Landolt or SDSS magnitudes of the stars in the
field, and this is recorded in the header. 
The pipeline also provides ZPs for science
frames if the fields are in the SDSS DR7
footprint and uses reference stars from that catalogue to set the ZPs. 
If the science frame is observed with filters
$g\#642$,$r\#784$ or   $z\#623$ the ZPs are provided in the
SDSS AB system. If the science frame is observed with filters
$U\#640$,$B\#639$,$V\#641$, $R\#642$, or $i\#705$ the magnitudes of the stars in
the SDSS DR7 catalogue are converted to the Landolt system using the
equations from \cite{2005AJ....130..873J} and ZPs are provided in
Landolt system. For these cases when the field is in the DR7
footprint, the ZPs are calculated as follows.   
 Instrumental
magnitudes are calculated for reference stars matched to DR7 stars and 
are reported for an airmass=0 using the extinction coefficient reported in Tab. \ref{table:zp}.  
The ZP is computed as the mean of all ZPs obtained for all the stars
that  have catalogue matches. 
The PESSTO pipeline adds the following keywords which
describe the data product, the measurements of which are described in
full in Appendix\,\ref{appdx:phot}.
}

\begin{scriptsize}
\begin{verbatim}
PSF_FWHM=  1.32371928 / Spatial resolution (arcsec)                    
ELLIPTIC=       0.131 / Average ellipticity of point sources           
PHOTZP  =       25.98 / MAG=-2.5*log(data)+PHOTZP                      
PHOTZPER=         999 / error in PHOTZP   
FLUXCAL = 'ABSOLUTE'  / Certifies the validity of PHOTZP               
PHOTSYS = 'VEGA    '  / Photometric system VEGA or AB                  
ABMAGSAT=    13.34036 / Saturation limit for point sources (AB mags)   
ABMAGLIM=    19.86138 / 5-sigma limiting AB magnitude for point sources
\end{verbatim}
\end{scriptsize}

For images which do not fall in the SDSS DR7 footprint, 
 we generally do not have reference stars in
the EFOSC2 $4.1\times4.1$ arcmin field. Hence we adopt and report the
average  \texttt{PHOTOZP} which we have measured and recorded in
Table\,\ref{table:zp}. If the night was photometric, then the error in the
\texttt{PHOTZP} is recorded as the error reported in
Table\,\ref{table:zp} (\texttt{PHOTZPER})
allowing the user to use the \texttt{PHOTOZP} with some degree of confidence
within the observed spread of the average measurement. If the night
was not photometric, or we are unsure, then \texttt{PHOTZPER} is always set to 999.  
The record of photometric and non-photometric
nights are recorded by the observers on the PESSTO wiki$^{13}$
and in Appendix\,\ref{appdx:photomet-nights}.
In this way the keyword \texttt{FLUXCAL} is always set to ABSOLUTE,
but users should be cautious of the validity. 

Science users can then employ the ZPs to calibrate photometry of stars
in the field using the following equation (and with the calibration
caveats described above) :
\begin{scriptsize}
\begin{equation}
MAG =  - 2.5\times\log_{10}\left({{COUNTS_{ADU}}\over{\texttt{TEXPTIME}}}\right)  + \left(\texttt{AIRMASS} \times K_{filter}\right)  + \texttt{PHOTZP}
\end{equation}
\end{scriptsize}

\noindent
where $COUNTS_{ADU}$ is the measured signal in ADU and
$K_{filter}$ is the average extinction coefficient listed in
for each filter in Table\,\ref{table:zp}.  The other terms are as
defined in the FITS headers. 
Colour terms are not
included, but are listed in Table\,\ref{table:zp} for reference.

\begin{table*}
\caption{Average values of zero points and colour terms from the period Aug 2012 until April 2013, as shown in Fig.\ref{fig:zp}.
The errors are standard deviations of the sample.}
\begin{center}
\begin{tabular}{lccrc}
\hline\hline
Filter & Zero point & Extinction Coefficient & Colour term & \\
\hline
$U$\#640 & $23.655\pm0.090$ &  0.46 $\pm$ 0.09 &   $0.096\pm0.030$ &($U-B$)\\
$B$\#639 & $25.755\pm0.078$ &  0.27 $\pm$ 0.05 &    $0.040\pm0.020$ &($B-V$)\\
$V$\#641 & $25.830\pm0.075$ &  0.12 $\pm$ 0.04&      $0.034\pm0.018$ &($B-V$)\\  
                &                                 & &$0.048\pm0.045$& ($V-R$)\\
$R$\#642 & $25.967\pm0.079$ &  0.09 $\pm$ 0.05 &    $0.031\pm0.042$&
($V-R$)\\ 
              &                          &    & $0.025\pm0.029$ &($R-I$)\\
$g$\#782 & $25.897\pm0.085$ &  0.20 $\pm$ 0.02&      $0.073\pm0.031$ &($g-r$)\\
$r$\#784 & $25.673\pm0.082$ &  0.09 $\pm$ 0.01 &     $0.044\pm0.033$
&($g-r$)\\ 
    &        &        & $0.056\pm0.045$ &($r-i$)\\
 $i$\#705 & $25.112\pm0.081$ & 0.02 $\pm$ 0.01 &     $-0.014\pm0.015$& ($r-i$)\\
 $z$\#623 & $24.777\pm0.081$ & 0.03 $\pm$ 0.01&      $0.126\pm0.042$& ($i-z$)\\
 \hline
 \end{tabular}
 \end{center}
 \label{table:zp}
\end{table*}

The astrometric calibration was derived using the USNO B1 and 2MASS
reference catalogues, 
and a distortion model described by a second-order polynomial.  The
astrometry task within the PESSTO pipeline employs the {\sc images}
package which is part of {\sc pyraf}.  The pipeline makes an initial
estimate for the astrometric solution of the field and iterates at
least three times to reach a confidence level $<$ 2 arcsec in both
$\alpha$ and $\delta$, otherwise it will record a failure to match
catalogued stars. This is recorded in the FITS header with a value of 
 $9999$ for the keyword \texttt{ASTROMET}.  A typical scatter of
0.4-0.5 arcsec was found for the science frames with around 15
stars usually recognised by the catalogue in the EFOSC2 frame.  This
typically   improves to an rms $\sim$0.2-0.3 with $\gtrsim$30
stars. For  standard star 
fields we typically find a scatter of 0.2 arcsec, although the Landolt
fields PG0231 and PG2336 usually produced a higher RMS  of $\sim$0.3-0.5. The
information on the  RMS of  $\alpha$ and $\delta$ and the number of
stars used for the calibration are given by  the header keyword
\texttt{ASTROMET}. Details for the other astrometric keywords are
provided in Appendix\,\ref{appdx:astromet}.

\section{PESSTO EFOSC2 data products}
\label{sec:efosc-products}

\subsection{EFOSC2 Fast Reduced Spectra}
\label{sec:frs}

Since the start of PESSTO survey operations in April 2012, we have
been releasing reduced spectra of all transient targets which have
been classified by PESSTO and announced via the Astronomer's Telegram
system within 24hrs of the data being taken. These spectra are
referred to as PESSTO ``Fast Reduced Spectra'', they are produced
instantly at the telescope by the PESSTO observers. A support team in 
Europe or Chile is always on duty to either re-reduce these, or check
them before they are  made available publicly through
WISeREP\footnote{http://www.weizmann.ac.il/astrophysics/wiserep/}\citep{2012PASP..124..668Y}.
Only EFOSC2 spectra are produced as ``Fast reduced spectra''  (EFOSC2
FRS) as we do not use SOFI for classifications.
These FRS are not ESO Phase 3 compliant and are an intermediate
product to assist the survey and the public with good, but not final, 
data products. They are not sent to the ESO archive (although the raw
data are immediately available in the ESO archive), and are not
as carefully calibrated as the SSDR1 spectra. They are {\em only} ever
made available through WISeREP.  The major differences
are that flat fielding, bias image subtraction, fringe correction, and
telluric  absorption correction are not applied and a library
sensitivity curve is employed for  flux calibration. 
This section describes the data product, 
but we emphasise that these FRS data are now replaced with the fully
reduced SSDR1 spectra in WISeREP and the ESO archive 
only includes the full reduced spectra. 

The PESSTO pipeline has a module to produce the FRS 
from the three fixed EFOSC2 set-ups. These are not flat-fielded
and the bias level is effectively removed during the sky-subtraction
process. 
 Wavelength calibration is achieved by applying a dispersion solution
from an archive arc frame, which is not the one taken during the
previous afternoon's calibrations. However an archival reference night
sky spectrum is cross-correlated with the object frame's sky spectrum
and a linear offset is applied to bring the dispersion solution into
agreement with the observed night sky. As noted above in
Sect.\,\ref{sec:efosc-arc}, the EFOSC2 dispersion solution is stable
over long periods and we find typical observed shifts are 10-30\AA, or
2-6 pixels. The linear shift applied then results in residuals between
the observed sky spectrum and the reference archive spectrum of 
less than a pixel.  An average sensitivity curve for each of the 
EFOSC2 grisms is applied and the PESSTO pipeline hence produces 
wavelength calibrated, and flux calibrated 1D and 2D images. There is 
no correction applied for the telluric absorption lines. 
Bias and flatfields from the night, or indeed the
observing run are {\em not} employed in the FRS.
The PESSTO pipeline then allows the user to interactively select the
object for extraction and set the background regions for background
subtraction within the familiar {\sc IRAF} $apall$ package, and then carry out 
tracing and extraction. The extracted spectra are wavelength calibrated
and then flux calibrated with an archive sensitivity function, after 
correcting for La Silla atmospheric extinction
\citep{2005PASP..117..810S}.  Cosmic ray rejection is generally turned 
on for these FRS, as described in Sect.\ref{sec:efosc-lacos}.

This procedure is carried out by the observers at the NTT, or the 
backup data reduction and analysis team that PESSTO organises each month. 
The backup team can access the raw data at the end of Chilean night, and complete
these reductions. Classifications are then made using one (or a combination)  of the 
SN classification tools 
SNID \citep{2007ApJ...666.1024B}, 
GELATO \citep{2008A&A...488..383H}, 
or SuperFIT \citep{2005ApJ...634.1190H}. 
 These three codes each have different approaches and underlying assumptions. 
Their   application and performance
was discussed recently in \cite{2014AN....335..841T} in 
the context of the Asiago Supernova classification program. The codes each 
have a different set of database spectra for use in the classification algorithm. 
They can provide different answers for ``best classification" depending on the 
signal-to-noise of the spectra and input information such as redshift and reddening. 
SNID is probably the most efficient algorithm, if the redshift of the SN or host galaxy is not 
known, and it also has a library of non-supernova spectra such as M-stars, 
AGN and luminous blue variable stars. 
Conversely, SNID assumes that the input spectrum is purely flux from the supernova, 
while SuperFIT can adjust the fit to include host galaxy contamination. 
GELATO has quite an extended library of spectra which is regularly updated   
and has a web-based interface which has recently undergone some improvements 
including a variable extinction option and smoothing algorithms
 \citep[as described in][]{2014AN....335..841T}. As one might expect, for spectra 
with reasonable signal-to-noise (S/N$\gtrsim$15, as is typical in PESSTO), the 
results from all three are in reasonable agreement. If the continua are featureless, 
or have shallow absorption or weak emission then one needs to be careful, 
irrespective of the code used.  In PESSTO, if the classification is ambiguous then more
than one of the classifiers is always used and the best estimate is provided at the
point of classification.
These classifications, based on the FRS spectra, are posted to the Astronomer's
Telegram website, or occasionally (mostly in the case of amateur and TOCP discoveries) 
to the IAU Central Bureau. These spectra are uploaded to WISeREP and are immediately 
publicly available. The PESSTO target turn around time for this process is 24hrs after the
end of the Chilean night, and to date we have managed 
this on every night, save a very small number of exceptions. 
Some targets have uncertain classifications due to noisy spectra or contamination by host galaxy light. If a reasonable guess at classification cannot be
made, the spectra are anyway made publicly available in WISeREP. In many cases a second attempt is made, 
particularly for those targets that have reasonable signal 
and defy standard classification. The most common 
type that we find are objects with blue featureless continua, which are often classified when more spectra
are taken.

A comparison of the FRS and the SSDR1 spectra for  a PESSTO
classification target (SN 2012fx; also known as PSN J02554120-2725276)
is shown in Fig. \ref{fig:fast_full}. Aside from a uniform scaling in flux, the
overall appearance of the spectrum in the two reductions is very
similar. In the rapid reduction, the uncorrected Telluric B band at
$\sim$6870 \AA\ is apparent (the stronger A band is lost in the deep
O{\sc i} absorption seen in the SN spectrum), while the rapid reduced
spectrum also appears somewhat noisier in the red. In both cases,  SNID finds the same best fitting template (SN 1991bg, z=0.018, age
+1.9 d), giving us confidence that the rapid reduced spectra are adequate for classification purposes.

\subsection{EFOSC2 final data product : SSDR1}
\label{sec:efosc-ssdr1}
The Spectroscopic Survey Data Release 1 (SSDR1) is now available 
through the ESO archive system. This serves the survey data
products which have been through the final data reduction process
via the PESSTO pipeline. The data processing steps that have been
applied are summarised below. 

\begin{enumerate}
\item {\em Bias subtraction:} applied as described in Sect.\ref{sec:efosc-bias}. 
\item {\em Flat fielding:} for Gr\#11 and Gr\#13 flat fields from afternoon dome
flats are applied. For Gr\#16, contemporaneous flat fields taken at
the same instrument and telescope position as the science frames are applied (see Sect.\ref{sec:efosc-flat}). 
For spectrophotometric standards,  daytime dome flats
are used for all grisms. 
\item {\em Wavelength calibration:} the 2D images are calibrated using arc frames as described in 
Sect.\ref{sec:efosc-arc}. 
\item {\em Cosmic Ray cleaning : } the 2D images are cleaned of cosmic rays using
the Laplacian cosmic ray rejection algorithm as discussed in Sect. \ref{sec:efosc-lacos}.
\item  {\em Object extraction and background subtraction :} the PESSTO pipeline implements
the standard {\sc IRAF} task {\em apall} to extract the target and
apply background subtraction. This has been run in interactive mode by
the data reduction team at Queen's University during the preparation
of the SSDR1 data. This process is the most manual and user intensive
in any spectroscopic data reduction process and if the transient
object is on, or close to, a bright host galaxy then the choice of the
background to subtract can be subjective. In all cases the PESSTO data
reduction process has attempted to achieve a clean background
subtraction to provide a target spectrum which is as uncontaminated 
as possible. The SSDR1 also releases the fully calibrated (wavelength
and flux) 2D frames as 
associated products for each 1D spectrum, so that a user can go back
to  this data product and simply re-extract with apertures and
background regions of their choosing. This will provide wavelength and
flux calibrated spectra, without having to go through all the
reduction steps manually. The {\em apall} task has been run in {\em
  pyraf} with the multispec output format with variance weighting
implemented.  Hence each science spectrum
also has an associated error spectrum and sky background spectrum
which are the standard outputs from this process. The error spectrum
produced by {\em apall} is the standard deviation of the variance
weighted science spectrum. 
  \item {\em Flux calibration :} the 1D and 2D frames are flux calibrated as
described in Sect.\ref{sec:efosc-specphot}. 
\item  {\em Telluric absorption correction :} this correction is applied as
detailed in Sec.\ref{sec:efosc-tell}. It is only applied to the 1D
spectra, not the 2D calibrated images released as associated files.
\end{enumerate}

The final step in both the EFOSC2 and SOFI spectral data reduction
processes is to convert the one-dimensional flux calibrated spectrum
images into binary FITS table format as the standard SSDR1 data
products. These conform to the ESO Science Data Products Standard
\citep{p2edpstd}, referred to as the spectrum binary table format.  The
binary table FITS file consists of one primary header (there is no
data in the primary \texttt{HDU} so \texttt{NAXIS=0}), and a single
extension containing a header unit and a \texttt{BINTABLE} with
\texttt{NAXIS=2}. Although the binary FITS table format supports
storing multiple science spectra within a single FITS file, a unique
FITS file is provided for each individual science spectrum. The actual
spectral data is stored within the table as vector arrays in single
cells. As a consequence, there is only one row in the
\texttt{BINTABLE}, which is \texttt{NAXIS2=1}.

Information associated with the science spectrum is also provided
within the same binary table FITS file resulting in a table containing one
row with four data cells. The first cell contains the wavelength array
in angstroms. The other three cells contain the science
spectrum flux array (extracted with variance weighting), its error
array (the standard deviation produced during the extraction
procedure) 
and finally the sky background flux
array. Each flux array is in units of erg\,cm$^{-2}$s$^{-1}$\AA$^{-1}$.
A list of software that can be used to read spectra in binary FITS table is given in Appendix\,\ref{appdx:soft-fitstable}.

The science spectrum has a filename of the following form, 
object name, date of observation, grism, filter, slit width, 
MJD of data reduction date, a numeric counter (beginning at 1) to distinguish
multiple exposures taken on the same night, and a suffix \texttt{\_sb} 
to denote a spectrum in binary table format. 

\begin{small}
\begin{verbatim}
SN2013ak_20130412_Gr11_Free_slit1.0_56448_1_sb.fits
\end{verbatim}
\end{small}

They can be identified as having the data product category
keyword set as 

\begin{small}
\begin{verbatim}
PRODCATG =  SCIENCE.SPECTRUM  / Data product category
\end{verbatim}
\end{small}

 One should note that the ESO Science Archive Facility produces these files with a name which begins ADP and then appended with the date and time the file was created (as is standard ESO policy). The filename described here can always be retrieved from the FITS header with the keyword 
\texttt{ORIGFILE}.

The 2D spectrum images that can be used to re-extract the object as
discussed above are released as associated ancillary data in
SSDR1. They are  associated with the science spectra through the following header
keywords in the science spectra files. The file name is the same as
for the 1D spectrum, but the suffix used is \texttt{\_si} to denote an
image. 

\begin{scriptsize}
\begin{verbatim}
ASSOC1 =  ANCILLARY.2DSPECTRUM        / Category of associated file
ASSON1 = SN2013ak_20130412_Gr11_Free_slit1.0_56448_1_si.fits /Name 
\end{verbatim}
\end{scriptsize}

These 2D files are wavelength and flux calibrated hence a user can
re-extract a region of the data and have a calibrated spectrum
immediately. Users should note the value for \texttt{BUNIT} in these
frames means that the flux should be divided by 10$^{20}$ to provide
the result in  erg\,cm$^{-2}$s$^{-1}$\AA$^{-1}$. 

We are also releasing the reduced acquisition images and reduced multi-colour photometric follow-up frames.
These are reduced as discussed in Sect.\ref{sec:efosc-image} and are currently available directly from the  PESSTO website (www.pessto.org) and will soon be available in the ESO archive. As they don't all 
have reliable absolute photometric zeropoints (as described in Sect\,\ref{sec:efosc-imagecal}) they will be available as 
associated ancillary data from ESO, rather than 
separate science data products. The acquisition 
images may be useful for improving on flux calibration in the future, if reference stars in the field can be accurately calibrated.   Images have the following naming convention for acquisition and science frames 
respectively 

\begin{small}
\begin{verbatim}
acq_SN2012ec_20120907_V641_56462_1.fits 
SN2011hs_20120422_R642_56462_2.fits
\end{verbatim}
\end{small}

This naming scheme is similar to the spectral files: object name, observation date, filter (including ESO number), MJD of date of reduction and a numeric
counter to distinguish multiple exposures from the 
same night. The acquisition images have the 
\texttt{acq\_} prefix.

\section{PESSTO SOFI  spectroscopic observations and calibrations}
\label{sec:sofi-spec}

\begin{figure*}
\centering
\includegraphics[scale=1,angle=270]{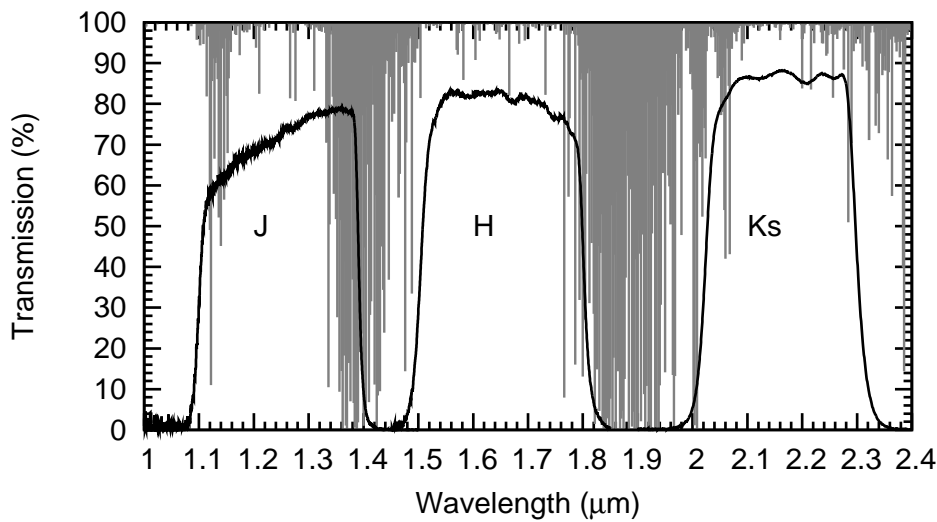}
\caption{Transmission curves for the {\it JHKs} filters used with SOFI. Also plotted in grey is the atmospheric absorption in the NIR \citep[][courtesy of Gemini Observatory]{Lor92}}
\label{fig:sofi_filt}
\end{figure*}

The Son OF ISAAC
(SOFI)\footnote{http://www.eso.org/sci/facilities/lasilla/instruments/sofi.html}
is an infrared spectrograph and imaging camera which is mounted on the
opposite nasmyth platform to EFOSC2 on the NTT (Nasmyth A focus) and
has been installed there since 1997 \citep{1998Msngr..91....9M}.The
instrument has a 1024$\times$1024 Hawaii HgCdTe array with 18.5$\mu$m
pixels. The array sensitivity and range of filters and grisms cover
imaging and spectroscopy between 0.9-2.5$\mu$m. PESSTO operates with
the SOFI imaging and spectroscopy default modes which have pixel
scales of $0\farcs29$\,pix$^{-1}$, and $0\farcs27$\,pix$^{-1}$,
respectively due to the different objectives employed \citep{sofiman}.
The imaging mode provides a FOV of 4.9 arcmins. 
PESSTO uses the long slit spectroscopy mode with the two low-resolution grisms labelled ``Blue'' and ``Red'' and the wavelength
coverage is listed in Table\,\ref{tab:sofispec} and typically only
takes spectra for targets which are in the magnitude range $14 < H <
17$. PESSTO does not use SOFI spectroscopy for any type of
classification, only targets that are picked as PESSTO Key Science
Targets are put forward for SOFI observations and only those bright
enough to give reasonable signal-to-noise (typically $S/N\sim20$ in the
continuum) are spectroscopically observed. PESSTO also uses SOFI in
imaging mode, using the filters $JHK_{\rm s}$, as shown in Fig. \ref{fig:sofi_filt}. The $K-short$ or
$K_{s}$ filter is different to standard $K$ and $K'$ as it transmits
between 2-2.3$\mu$m hence avoiding the 1.9$\mu$m atmospheric
absorption feature and cuts short of the increasing thermal background
beyond 2.3$\mu$m \citep{sofiman}. No other imaging filters are employed
for PESSTO SOFI observations. The amount of SOFI NIR data available
for any PESSTO science target depends critically on the brightness of
the source and the science drivers. Hence, as originally planned in
the survey proposal, SOFI observations make up around 20 per cent of
the total PESSTO time.

\begin{table*}
\caption[]{PESSTO settings for SOFI spectroscopy. 
The order blocking filters used are 0.925$\mu$m (GBF) and 
1.424$\mu$m (GRF) ``cut-on'' filters.
A $1\farcs$ slit projects to 3.4
pixels FWHM, measured from arc lines and the resolution $R$ 
is given at the midpoint of the spectral ranges, as is the 
velocity resolution.   
The column headed Arclines indicates the number of
lines used. The RMS is the typical residual for the wavelength
calibration solution. } 
\label{tab:sofispec}
\begin{tabular}{rlllllllll}
\hline\hline
Grism    &  Wavelength    & Filter         & $n_{\rm pix}$ & Dispersion  & Resolution &  R  & $V$ resolution & Arclines & RMS              \\
              &  ($\mu$m)     & (blocking)  & (pixels) &     (\AA/pixel)  & (\AA) & $\lambda/\Delta\lambda$ & km\,s$^{-1}$ & & \AA \\\hline 
Blue    &  0.935 - 1.645  &   GBF         & 1024    &  6.95  &   23   & 550  &  545 & 12-14   &  0.1-0.2         \\
Red     &  1.497 - 2.536   &  GRF         & 1024   & 10.2  &   33   & 611  &  490  &   7-8	& 0.2-0.5		\\\hline
\end{tabular}
\end{table*}

\subsection{Detector characteristics}
\label{sec:sofi-det}

The detector installed in SOFI is a Rockwell Scientific Hg:Cd:Te
1024x1024 Hawaii array with 18.5$\mu$m pixels and an average quantum
efficiency of 65\% . It has a dark current of typically around 20
e$^{-}$hr$^{-1}$ per pixel, and a documented readout noise of approximately 12e$^{-}$, both of
which are negligible compared to background in PESSTO exposures. The
gain of the array is 5.4e$^{-}$/ADU and well depth around 170,000
electrons (32,000 ADU). The array non-linearity is reported to be less than 1.5\% for a signal up to 10,000 ADU \citep{sofiman}, but the ESO
instrument scientists recommend that exposures keep the background
below 6,000 ADU owing to the bias of the array, which has a complicated
dependence on flux levels.

In imaging mode, we use DCR (double correlated read) mode which results in a readnoise of  around 12e$^{-}$. 
The short noise from the sky (or object if it is bright) dominates and readnoise is negligible for imaging. 
In spectroscopy mode, we always use the NDR (non-destructive read) mode with the settings \texttt{NSAMP=30} and \texttt{NSAMPPIX=4}  \citep[as described in the SOFI manual;][]{sofiman}. This mode is recommended for spectroscopy and the array is read within each \texttt{DIT} a number of times (equal to \texttt{NSAMP}), and for each read-out the signal
is sampled \texttt{NSAMPIX} times. This mode reduces the readnoise  further than for DCR, with read noise values 
typically in the range $2-3$e$^{-}$.

\subsection{SOFI spectroscopic calibration data and reduction}
Similar to PESSTO observations and reductions for EFOSC2, we aim to homogenise the SOFI observations
and calibrations and tie them directly to what is required in the data reduction pipeline. A standard
set of PESSTO OBs for calibrations and science are available on the PESSTO wiki and the 
following sections describe how they are applied in the pipeline reduction process.

\subsubsection{Bias, dark and cross-talk correction}
\label{sec:sofi-bias}

Unlike for EFOSC2, we do not subtract bias (or dark) frames from SOFI images or spectra. The bias level seen in the each image is dependent on the incident flux level, and so it is not practical to correct this with daytime calibration data. Instead, any bias offset or structure is subtracted along with the sky background, as recommended in the SOFI handbook.

The SOFI detector suffers from cross talk, where a bright source on either of the two upper or lower quadrants of the detector will be accompanied by a ``ghost'' on the corresponding row on the opposite two quadrants. This ghost will affect the entire row of the detector, and has a fixed intensity relative to the opposite row. This cross-talk effect is corrected for within the pipeline by summing each row on the detector, scaling by a constant value, and subtracting from the opposite quadrants. 

\subsubsection{Flat field calibration}
\label{sec:sofi-flat}

\begin{figure}
\centering
\includegraphics[scale=0.65,angle=270]{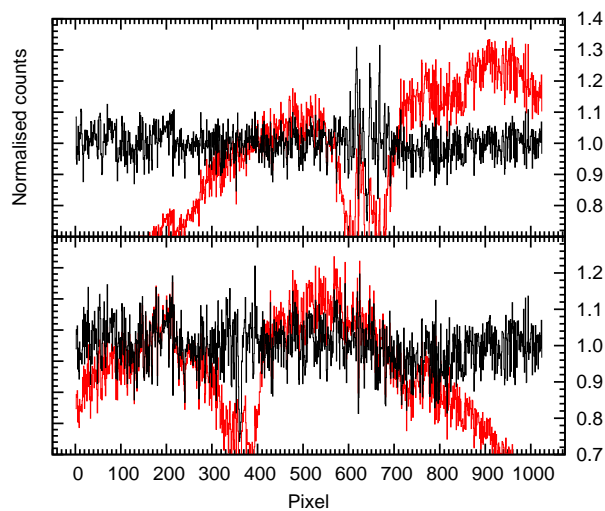}
\caption{Cut across SOFI flat fields along row 512. Blue grism (lower) and red grism (upper panel) are shown, with the normalised flat field shown in black, and the raw flat field (showing the H$_2$O absorption) in red.}
\label{fig:sofi_flat}
\end{figure}

Spectroscopic flats are taken approximately once per month for SOFI; these consist of pairs of flats, taken first with an incandescent lamp illuminating the dome, and then with the dome un-illuminated. The lamp-off flats are subtracted from the lamp-on flats, to remove the thermal background of the system. These subtracted flat fields are then combined and normalised using a high-order (order 80 by default) spline fit; the normalised flat field is used to correct for the pixel to pixel variations in detector sensitivity in the science and standard star frames. Although atmospheric absorption features due to the light path between the dome lamp and the detector can be seen in these flats, the normalisation appears to remove them relatively well. The raw and normalised flat fields for both the blue and red grisms are shown in Fig. \ref{fig:sofi_flat}. 

\begin{figure}
\centering
\includegraphics[scale=0.65,angle=270]{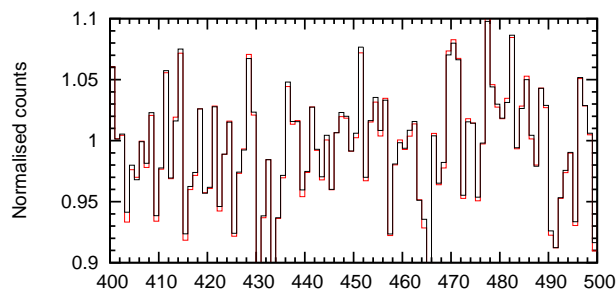}
\caption{Comparison of a cut along column 512 of the normalised flat
  for the red grism taken on 2013 March 3 (black) and the normalised
  flat taken on 2012 October 6 (red) between pixels 400 and 500. The
  two flats are essentially identical, indicating that the ``noise''
  seen in Fig. \ref{fig:sofi_flat} is in fact $\sim$5\% pixel-to-pixel variation in the detector.}
\label{fig:sofi_flat_diff}
\end{figure}

The amplitude of the variability in the flat field is $\sim$4\% for the red grism and $\sim$6\% for the blue grism. The pixel-to-pixel variation in Fig. \ref{fig:sofi_flat} illustrates the real response of the detector, rather than being due to shot noise in the flats.  We verify this in Fig. \ref{fig:sofi_flat_diff}, where we compare a section of two normalised red grism flat fields taken $\sim$5 months apart. Both flats show the same structure, demonstrating that the flat field is stable, and that the use of monthly calibrations is justified.

\subsubsection{Arc frames and wavelength calibrations} 
\label{sec:sofi-arc}

As for the optical spectra, wavelength calibration is performed using
spectra of a Xenon arc lamp. To fit the dispersion solution of the arc
spectra without any systematic residuals requires a fourth-order
polynomial fit. 
As
listed in Table \ref{tab:sofispec}, 7-8 lines were typically used for
the fit in the red grism, and 12-14 lines in the blue grism, giving an
RMS error in the wavelength of around 0.2-0.5 \AA. The dispersion
solution found from the arc frames is then applied to the two
dimensional spectra.  The wavelength calibration is also checked
against the sky lines.  After the 2D science frame is wavelength
  calibrated, the frame is averaged along the spatial axes and
  cross-correlated with sky lines. A linear shift is applied to the
  wavelength calibration and recorded in the header keyword \texttt{SHIFT}. A more
  robust result is obtained if the regions of the spectrum containing strong telluric 
  absorption is removed before the wavelength calibration check
  is performed. As with the EFOSC2 correction, the precision of the
  wavelength correction is limited to 1\AA, owing to the scale of the
shifts in the  library sky spectra  employed. Hence this value of
1\AA, is again  recorded as the systematic error in the wavelength 
calibration (\texttt{SPEC\_SYE}, see Appendix\,\ref{Appdx:wavecal}).

\subsubsection{Sky subtraction and spectral extraction} 
\label{sec:sofi-extract}

A critical part of NIR observations is the bright sky background, which usually has higher flux levels than the target. The sky can vary on timescales of a few minutes, and so must be measured and subtracted at (or close to) the time of the science observations. To accomplish this, SOFI spectra for PESSTO are taken in an ABBA dither pattern. This pattern consists of taking a first (A$_1$) exposure at a position `A', then moving the telescope so that the target is shifted along the slit of SOFI by $\sim$5-10 \arcsec\ to position `B'. Two exposures are taken at `B' (B$_1$ and B$_2$), before the telescope is offset back to `A' where a final exposure (A$_2$) is taken. When reducing the data, the pipeline subtracts each pair of observations (i.e. A$_1$-B$_1$, B$_1$-A$_1$, B$_2$-A$_2$, A$_2$-B$_2$) to give individual bias- and sky-subtracted frames. Next, the PESSTO pipeline attempts to shift these sky-subtracted frames so that the trace of the target is at a constant pixel position, and combine the frames. If the target is relatively faint, and the spectral trace cannot be identified clearly in each frame, this routine in the pipeline will fail, and instead the user will be prompted to interactively align and combine the frames. Finally, the spectrum is optimally extracted in an interactive fashion.

The total on object exposure time of these combined frames is given in the header as \texttt{TEXPTIME}. This is simply a product of the following values, all  found as header keywords :  \texttt{DIT} (the detector integration time), \texttt{NDIT} (the number of DITs), \texttt{NJITTER} (the number of jitters at positions `A' and `B'),  and \texttt{NOFFSETS} (he number of offset positions, which is always 2). Typically \texttt{DIT} is 
kept between 60-240 sec.

\subsection{Telluric absorption correction} 
\label{sec:sofi-tell}

The NIR region covered by SOFI contains multiple strong telluric
absorptions, arising chiefly from water vapour and CO$_2$, and their
absorption strength is a function of both time and airmass. The most
common technique for low-to-medium-resolution spectroscopy is to
observe a star of known spectral type (a ``telluric standard'')
immediately prior to or following the science spectrum, and at a
similar airmass. The spectrum of the telluric standard is then divided
by an appropriate template spectrum of the same spectral type,
yielding an absorption spectrum for the telluric features.  The
absorption spectrum is then divided into the science spectrum to
correct for the telluric absorption.  As part of PESSTO, we observe
either a Vega-like (spectral type A0V) or a Solar analogue (G2V)
telluric standard for each SOFI spectrum. The PESSTO pipeline uses the
closest (in time) observed telluric standard to each science or
standard star spectrum.

\subsection{Spectrophotometric standards and flux calibration} 
\label{sec:sofi-specphot}

\begin{figure*}
\centering
\includegraphics[scale=0.7,angle=270]{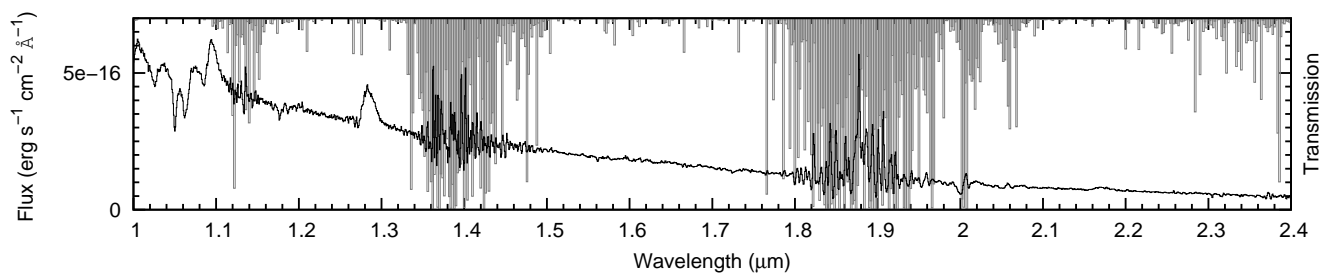}
\caption{Combined blue and red grism SOFI spectra of SN 2012ec taken on 2013 September 24. Overplotted in grey is the atmospheric transmission, showing the correspondence between regions of low transparency and poor S/N in the spectrum.}
\label{fig:sofi_spec}
\end{figure*}

The process for correcting the spectrum for the telluric absorption
also provides a means for flux calibration using 
the Hipparcos $I$ or $V$ photometry of the solar analogues and Vega standards used.
The flux of the observed telluric standard spectrum is scaled to match the tabulated photometry,
with the assumption that the telluric standards have the same colour
(temperature) as Vega or the Sun.
When possible, a second step is performed to flux calibrate the spectra
using a spectrophotometric standard.
The spectrophotometric standard is reduced and corrected for telluric
absorption using a telluric standard, with the same technique as used
for the science targets. This corrected standard spectrum is then
compared with its tabulated flux, and the science frame is then
linearly scaled in flux to correct for any flux discrepancy.
There are only a handful of spectrophotometric standard stars which
have tabulated fluxes extending out as far as the {\it K}-band. We do
observe these standards (listed in Table \ref{tab:specphotstandards}) as far as possible when SOFI
spectra are taken, but nonetheless there are a significant number of
nights where no flux standard was observed in the NIR. For these nights
the spectra will still have an approximate flux calibration performed against the accompanying telluric standard.
An example of a reduced and flux calibrated spectrum is shown in Fig. \ref{fig:sofi_spec}.

All SOFI spectra have the following keyword which denotes which
telluric standard was used for both the telluric correction and the
initial  flux calibration. 

\begin{scriptsize}
\begin{verbatim}
SENSFUN = 'TSTD_Hip109796_20130417_GB_merge_56478_1_ex.fits' / tell 
stand frame 
\end{verbatim}
\end{scriptsize}

If one of the spectrophotometric flux standards from
Table\,\ref{tab:specphotstandards} has been used to additionally scale
the flux then the keyword \texttt{SENSPHOT} is added to the header,
with the spectrum used to apply the flux calibration. This file has
the name of the standard clearly labelled. In this way, users can
distinguish which method has been applied. 

\begin{scriptsize}
\begin{verbatim}
SENSPHOT='sens_GD71_20130417_GB_merge_56478_1_f.fits'/sens for flux cal  
\end{verbatim}
\end{scriptsize}

\begin{figure}
\centering
\includegraphics[scale=0.65,angle=270]{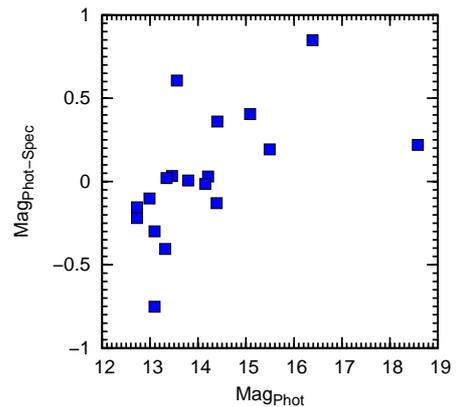}
\caption{Comparison of observed and synthetic {\it JH} magnitudes for
  SN 2012fr. The standard deviation of the distribution is 34.7\%,
  which is a measure of the uncertainty of the absolute
  spectroscopic flux scale of the calibrated SOFI spectra.  }
\label{fig:sofi_flux}
\end{figure}

To check the flux calibration of SOFI spectra, we would ideally have a
large number of targets with both well sampled NIR lightcurves and
SOFI spectra. At this time, the NIR lightcurves for most of the PESSTO
science targets are not complete and not calibrated reliably enough to
allow a large scale comparison. We have used a well observed type Ia
SN \citep[SN 2012fr][]{2013ApJ...770...29C} to determine the accuracy
and reliability of the SSDR1 flux calibrations.  Synthetic {\it
  J}-band photometry was performed on the blue grism spectra, and {\it
  H}-band photometry on the red grism spectra. The difference between
the synthetic magnitudes and the {\it JH} photometry from Conteras et
al., (2014, in prep) is plotted in Fig. \ref{fig:sofi_flux}. Not
surprisingly, a fairly large spread of magnitude offsets is seen, with
the distribution having a mean of 0.04$^{m}$ and a standard deviation of
  0.37$^{m}$. Although this is quite a significant scatter, it 
can be improved upon by users by employing the $JHK_{s}$ imaging
that is normally done when SOFI spectra are taken. Synthetic
photometry will allow more accurate scaling of the absolute flux
levels.  This correction is 
not in SSDR1, but  in future PESSTO data releases, the flux
calibration of SOFI spectra will be cross checked against the
$JHK_{s}$  photometry of the target taken closest to the observations.

\subsection{SOFI  imaging calibration frames and reduction} 
\label{sec:sofi-imagecal}

SOFI imaging is carried out as default when spectroscopy is done,
providing images with a 4.9 arcmin field of view
($0\farcs29$\,pix$^{-1}$). 
The cross talk effect is first corrected as for the spectra and all images are then flat fielded using dome flats.  Dome flats are taken using a screen on the interior of the telescope dome which can be illuminated with a halogen lamp. Pairs of flats are taken with the screen illuminated and un-illuminated; the latter are then subtracted from the former to account for dark current and thermal background. Multiple flats are combined, and then used to reduce the science data. Typically, dome flats are taken once per month with SOFI, although they appear stable over longer periods. 

An illumination correction is also applied, to account for the difference between the illumination pattern of the dome flats and the actual illumination of the night sky. The illumination correction is determined by imaging a bright star at each position in a 4$\times$4 grid on the detector. The intensity of the star is then measured at each position, and a two-dimensional polynomial is fitted. This polynomial is normalised to unity, so that it can be applied to the imaging data as a multiplicative correction. These on-sky calibrations are tested annually within PESSTO.

Sky-subtraction is the most important aspect of NIR imaging and
reductions. For targets that are in relatively uncrowded fields, a
dither pattern is employed where the telescope is moved to four offset
positions on the sky, while keeping the target in the field of view
(``on-source sky subtraction''). To determine the sky background, the
four frames are then median combined {\it without} applying offsets,
rejecting pixels from any individual image which are more than a
certain threshold above the median. This 
\emph{initial}
  sky image is
  subtracted from each individual frame in order to obtain  \emph{initial}
  sky-subtracted images. These frames are used to identify the
  positions of all sources and create a mask frame for each science
  image.  For each set of four images, the frames are then median
  combined {\it again without} applying offsets and using the masks
 created previously to reject all sources and produce the \emph{final}
  sky image.
The \emph{final} sky background image is then subtracted from each of the input frames. The sky-subtracted images are then mosaiced together to create a single image using the {\sc swarp} package \citep{swarp}.

For targets which are in a crowded field, or where there is extended
diffuse emission (such as nearby galaxies), then on-source sky
subtraction is not possible. In these cases, we alternate between
observing the target, and observing an uncrowded off-source field
around $\sim$5 arcmin from the target. We typically observe four
frames on source, then four frames off source, dithering in each case
The off-source frames are then used to compute a sky frame in the same way as for the  ``on-source sky subtraction''.
The off-source sky frame is then subtracted from each of the on-source images of the target, which are then combined to create the final image. 
Since the field of view of SOFI  is rather small (4.9 arcmin) the astrometry is not set for single images. Instead, 
{\sc sextractor} is run to detect sources in individual frames, and to check the nominal dither. The images are 
then mosaiced together using {\sc swarp}. Finally, an astrometric calibration is made, by cross correlating the sources detected by {\sc sextractor} with the 2MASS catalogue, in the same fashion as for the EFOSC2 frames.
 The instrumental aperture magnitudes of the sources in the field as
 measured by {\sc daophot} are then compared to their catalogued 2MASS
 magnitudes to determine the photometric zeropoint, which is recorded
 in the header of the image as   \texttt{PHOTZP}. 

The definition of PHOTZP for SOFI is different to that of EFOSC2. Since the SOFI images
all have astrometric and photometric solutions from 2MASS point source matching, it is 
possible to give a measured zeropoint for all images. The SOFI images are full science
archive products and as such they obey the formal ESO definition of the zeropoint : 

\begin{equation}
MAG =  - 2.5\times\log_{10}\left({COUNTS_{ADU}}\right)    + \texttt{PHOTZP}
\end{equation}

The extinction term is not used since the 2MASS calibration sources are in the same 
image, and the \texttt{EXPTIME} term is incorporated into the \texttt{PHOTZP} value. 
The other photometric keywords are similar to EFOSC2 and are described in
 Appendix\,\ref{appdx:phot}. 

\subsection{PESSTO SOFI data products : SSDR1 }
\label{sec:sofi-products}

PESSTO does not produce fast reduced spectra for SOFI, since the NIR
is never used for classification. Hence only final reduced spectra and
images are described here for SSDR1. The data products for SOFI are
similar to those described in Sect.\ref{sec:efosc-ssdr1} for EFOSC2. 
The spectra are in binary table FITS format, with the same four data
cells corresponding to the wavelength in angstroms, the weighted
science spectrum and its error and the sky background flux array. 
Again, each flux array is in units of
erg\,cm$^{-2}$s$^{-1}$\AA$^{-1}$.
 (see Appendix\,\ref{appdx:soft-fitstable}
 for a list of software that can read binary FITS table format). 
The SSDR1 FITS keywords described
Appendix\ref{Appdx:1} are again applicable here. A typical file name
is 

\begin{small}
\begin{verbatim}
SN2009ip_20130417_GB_merge_56478_1_sb.fits
\end{verbatim}
\end{small}

Where the object name is followed by the date observed, the grism (GB for the blue grism, 
or GR for the red grism), the word ``merge'' to note that that the
individual exposures in the ABBA dither pattern have
been co-added, the MJD date the file was created, a numeric value to
distinguish multiple exposures on the same night and a suffix \texttt{\_sb} 
to denote a spectrum in binary table format.  As
with EFOSC2, this science spectrum can be identified with the label : 

\begin{small}
\begin{verbatim}
PRODCATG =  SCIENCE.SPECTRUM  / Data product category
\end{verbatim}
\end{small}

We also provide the 2D flux calibrated and wavelength calibrated file 
so that users can re-extract their object directly, as described with
EFOSC2.  The identification of the 2D images follow the same
convention as for EFOSC2, with the suffix \texttt{\_si} to denote a spectral
image. 

\begin{scriptsize}
\begin{verbatim}
ASSOC1 =  ANCILLARY.2DSPECTRUM        / Category of associated file
ASSON1 = SN2009ip_20130417_GB_merge_56478_1_si.fits / Name 
of associated file
\end{verbatim}
\end{scriptsize}

We do not reduce and release the SOFI equivalent of the EFOSC2 acquisition images,  
but in nearly all cases where PESSTO 
takes a SOFI spectrum, imaging in $JHK_{\rm s}$ is also taken. 
These images are flux and astrometrically calibrated and released as
science frames  rather than associated files. They are labelled as
follows where \texttt{Ks} labels the filter and the \texttt{merge}
denotes that the dithers have been co-added. 

\begin{small}
\begin{verbatim}
SN2013am_20130417_Ks_merge_56475_1.fits
\end{verbatim}
\end{small}

We also release the image weight map as described in \cite{p2edpstd}.
The definition in this document is  the pixel-to-pixel
variation of the statistical significance of the image array in terms
of a number that is proportional to the inverse variance of the
background, i.e. not including the Poisson noise of sources. This is
labelled as 

\begin{scriptsize}
\begin{verbatim}
ASSOC1 =  ANCILLARY.WEIGHTMAP          / Category of associated file
ASSON1 = SN2013am_20130417_Ks_merge_56475_1.weight.fits / Name of 
associated file
\end{verbatim}
\end{scriptsize}

\begin{figure*}
\centering
\includegraphics[scale=0.6]{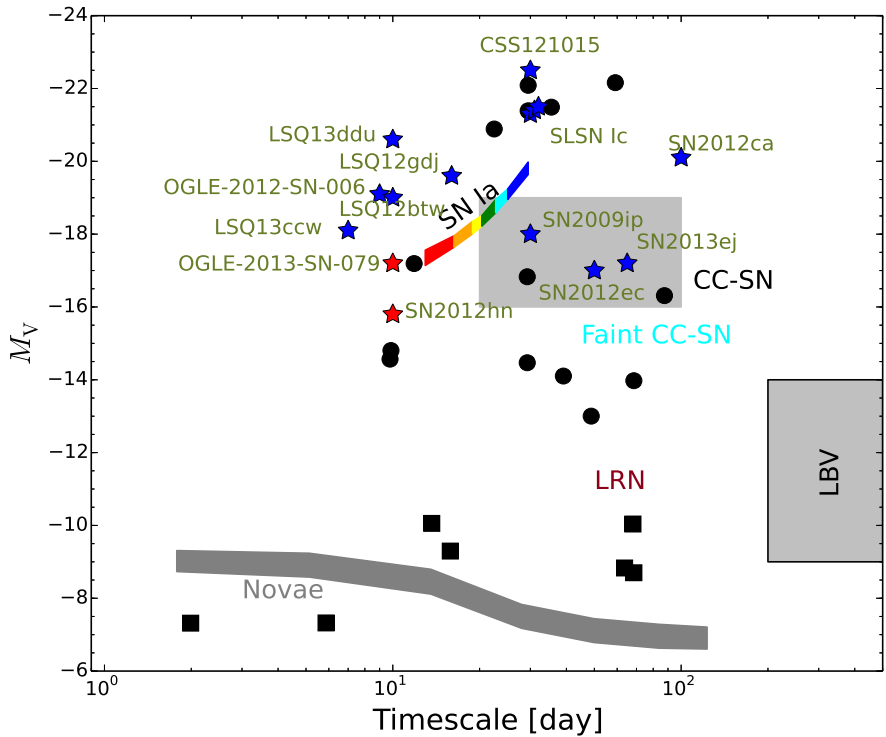}
\caption{PESSTO objects (filled blue stars) in the phase space of cosmic transients as originally developed by \citet{2007Natur.447..458K}.  The grey areas illustrate the known phase space for novae, luminous blue variable 
eruptions and core-collapse supernovae (the common types II and Ibc). The black squares show eruptive transients
lying outside the nova regime, and the black dots are supernovae (normal and superluminous) or extreme erupitve events such as SN2008S \citep[taken from][]{2009aaxo.conf..312K}
The colour for each event represents the colour at peak brightness ($B-V$ or $g-r< 0.2$ in blue; $B-V$ or $g-r> 0.7$ in red).}
\label{fig:kk_diag}
\end{figure*}

\section{Summary and data access}
\label{Conclusions}

This paper describes the processing and calibration of PESSTO
data products that are served by ESO as the Spectroscopic Survey Data
Release 1 (SSDR1).  From this first year of science operations, a 
total of 909 reduced and calibrated spectra from EFOSC2 and SOFI 
for 298 distinct objects have been released along with 234 reduced and calibrated near-infrared
SOFI images for 22 objects.  These spectra and SOFI images are available from the ESO 
archive as Phase 3 compatible data products. In addition we make available the reduced
and calibrated EFOSC2 images now, before they are fully ESO archive compliant. 
As of October 2014, PESSTO has classified around 570
transient objects and is carrying out follow-up campaigns on around 90 of these. All 
information is kept up to date on the PESSTO website to support this public survey and 
classification spectra are released on an ongoing basis via WISeREP \citep{2012PASP..124..668Y}. 
As discussed in Sect.\ref{sec:intro}, one of the major goals of PESSTO is to study the 
extremes of the known transient populations and provide comprehensive datasets to study
the physical mechanisms producing these objects.  An illustrative diagram of  the phase space of 
explosive and eruptive transients was first plotted by 
\cite{2007Natur.447..458K} to show the faint and relatively fast nature of transients in the gap between 
faint supernovae and novae. This was expanded by \cite{2009aaxo.conf..312K}
and \cite{2010ApJ...723L..98K} to higher luminosities and faster declining objects. As an illustration
of PESSTO's science goals, Fig.\ref{fig:kk_diag} shows this Kulkarni \& Kasliwal diagram updated with objects
that PESSTO has classified and is following. The data for these will be released in future public  releases 
via the ESO archive, and this shows the extremes of the transient population that we are now
covering extensively.

We have highlighted  the difficulty in homogenising the
flux calibration of small imaging fields in a public survey, and
providing absolute spectroscopic flux calibration to below 10\% across
many nights which have variable seeing and transparency.  However
methods to improve these for future data releases have been identified
and neither of these greatly affects the 
 science of transient
objects that can be done with PESSTO. Science users have the ability
to adjust the flux measurements since the data releases contain enough
information that improvements to the calibrations can be tailored
for specific objects, with additional manual steps in calibrating. For
example, as all EFOSC2 spectra have an acquisition image in $V$-band,
a calibration of reference stars in the field should allow the
absolute flux to be calibrated on the image to a few per cent. This
has not been possible on a full survey basis for SSDR1 since it would
require re-calibrating several hundred EFOSC2 fields
with reliable photometric measurements, or all 
sky reference catalogues. 
In the
future of all-sky digital surveys such as Pan-STARRS1 and SkyMapper,
the existence of reference stars down to around 20$^{m}$ will provide
this improvement quite easily. We envisage future releases will
improve on this. 

The SSDR1 EFOSC2 and SOFI spectra and the SOFI images are available through the ESO archive server as formal ESO Phase 3 data.  Instructions for accessing these data are available on the PESSTO website \texttt{www.pessto.org}.  The reduced and calibrated EFOSC2 images are available from the PESSTO website, but not yet through the ESO archive. 
All 1D spectra will also available in the Weizmann Interactive Supernova data REPository \citep[WISeREP][]{2012PASP..124..668Y}.

\begin{acknowledgements}
  This work is based on observations collected at the
  European Organisation for Astronomical Research in the Southern
  Hemisphere, Chile as part of PESSTO, (the Public ESO Spectroscopic
  Survey for Transient Objects Survey) ESO programme 188.D-3003, 191.D-0935.
  Research leading to these results has received funding from the
  European Research Council under the European Union's Seventh
  Framework Programme (FP7/2007-2013)/ERC Grant agreement n$^{\rm o}$
  [291222] (PI : S. J. Smartt) and STFC grants ST/I001123/1 and ST/L000709/1.
This research has made use of the SIMBAD database,
operated at CDS, Strasbourg, France.
M.J.C. acknowledges funding from the Australian Research Council Centre of Excellence for All-sky Astrophysics (CAASTRO), through project number CE110001020
MS acknowledges support from the Royal Society and EU/FP7-ERC grant no [615929]. MF acknowledges support by the European Union FP7 programme through ERC grant number 320360. 
SB, EC, AP, LT and MT are partially supported by the PRIN-INAF 2011 with the project Transient Universe: from ESO Large to PESSTO.  We acknowledge CONICYT-Chile grants, Basal-CATA PFB-06/2007 (FEB),
FONDECYT 1141218 (FEB) and 3140534 (SS), 
PCCI 130074 (FEB, SS),  ALMA-CONICYT 31100004 (FEB, CRC), 
"EMBIGGEN" Anillo ACT1101 (FEB),
Project IC120009 "Millennium Institute of Astrophysics (MAS)” of the Iniciativa Cient\'{\i}fica Milenio del Ministerio de Econom\'{\i}a, Fomento y Turismo (FEB, SS, CRC)
AG-Y is supported by the EU/FP7-ERC grant no. [307260], ‘The Quantum Universe' I-Core programme by the Israeli Committee for planning and budgeting and the ISF, the Weizmann-UK making connections
programme, and the Kimmel award.
NER acknowledges support from the European Union Seventh Framework Programme (FP7/2007-2013) under grant agreement n. 267251 ‚"Astronomy Fellowships in Italy"‚ (AstroFIt). 
Support for LG, SG and HK is provided by the Ministry of Economy, Development, and Tourism’s Millennium Science Initiative through grant IC12009, awarded to The Millennium Institute of Astrophysics, MAS. LG, SG and HK acknowledge support by CONICYT through FONDECYT grants 3130680 and 3140563, 3140566

\end{acknowledgements}

\begin{appendix} 
\label{Appdx:1}
\section{SSDR1 FITS Keywords description} 

This section contains details of the some of the more useful PESSTO specific SSDR1 FITS
keywords and their definitions for users. It should be read in
conjunction with the ESO Phase 3 User documentation \citep{p2edpstd}. 
The flux and wavelength related keywords are typically applicable
to both SOFI and EFOSC2 data while some (such as the cosmic ray
rejection flag) are applicable to one or the other only (EFOSC2 in
this case). Their use in the two instruments should be
self-explanatory in the descriptions.

\subsection{Number of exposures}
\noindent All PESSTO EFOSC spectra  are extracted from single epoch exposures, we do not provided merged or co-added spectra in cases where multiple spectra are taken for EFOSC2. This is left to the users to decide. Hence for all EFOSC spectra:

\begin{small}
\begin{verbatim}
SINGLEXP=  T / TRUE if resulting from single exposure
\end{verbatim}
\end{small}

PESSTO SOFI spectra are always taken in an ABBA dither pattern as described in Sect\,\ref{sec:sofi-extract}, and hence:

\begin{small}
\begin{verbatim}
SINGLEXP=  F / TRUE if resulting from single exposure
\end{verbatim}
\end{small}

All PESSTO spectra are taken at a single epoch, hence:

\begin{small}
\begin{verbatim}
M_EPOCH =  F / TRUE if resulting from multiple epochs
\end{verbatim}
\end{small}

\subsection{Cosmic ray rejection}
\label{Appdx:cosmic}

\noindent PESSTO uses the Laplacian cosmic ray rejection algorithm \footnote{http://www.astro.yale.edu/dokkum/lacosmic/}
 of \cite{2001PASP..113.1420V}  for EFOSC2 data (no cleaning is necessary for the SOFI detector). If the Boolean 
 value is set to \texttt{T} as below, then the rejection algorithm has been applied,
 otherwise it has not. Note that in the spectral frames, only the central 200 pixels around the object are cleaned (i.e. central pixel $\pm$100 pixels). Full frame cosmic ray cleaning is generally turned on for acquisition and photometric imaging, and again this is flagged with the following keyword. 
\begin{scriptsize}
\begin{verbatim}
LACOSMIC=  T / TRUE if Laplacian cosmic ray rejection has been applied 
\end{verbatim} 
\end{scriptsize} 

\subsection{Wavelength calibration}
\label{Appdx:wavecal}
\noindent The particular arc frame used for wavelength calibration is always
recorded for information using the \texttt{ARC} keyword. The number of
arc lines used in the fit is given by \texttt{LAMNLIN}, and the root
mean square of the residuals to the fit is listed as
\texttt{LAMRMS}, formally calculated as 

\begin{equation}
LAMRMS = \frac{\sqrt{\sum_{i=1}^N {R_i^2}}}{N}
\end{equation}

\noindent where $R_i$ is the residual of the wavelength fit for the $i$th
arcline and $N$ is the number of arc lines (\texttt{LAMNLIN}). This 
assumes that errors are randomly distributed and without any
systematic errors, which is true as far as we can tell for EFOSC2 and
SOFI. Hence the statistical uncertainty in the wavelength solution at
any point is approximately given by the value   \texttt{SPEC\_ERR},
where 

\begin{equation}
SPEC\_ERR = \frac{LAMRMS}{\sqrt{LAMNLIN}}
\end{equation}

\noindent As described
in Sect.\ref{sec:efosc-arc} the wavelength positions of the skylines in the science  frame (or telluric lines
for bright standard stars) are checked  and a linear shift is
applied. This is listed in the keyword  \texttt{SHIFT}  in
Angstroms. The precision of this is limited to 1\AA\,  and hence we
set the  keyword \texttt{SPEC\_SYE} (the systematic
error in the spectral coordinate system) that is found during the
observation and reduction process to 1\AA. 
After this systematic \texttt{SHIFT}
is applied to correct the skylines to rest, we find no further
systematic effects in EFOSC2 wavelength calibration. 

After the wavelength solution is determined and
the  \texttt{SHIFT} applied, the  following values were  inserted as keywords 

\begin{scriptsize}
\begin{verbatim}
ARC     = 'arc_SN2013XYZ_20130401_Gr11_Free_slit1.0_56448_1.fits'
LAMNLIN =            9.0 / Nb of arc lines used in the fit 
LAMRMS  =         0.0136 / residual RMS [nm]
SPEC_ERR= 0.004533333333 / statistical uncertainty
SHIFT   =               2.0
SPEC_SYE=            1.0 / systematic error
WAVELMIN=    334.3426032 / minimum wavelength [nanometers]
WAVELMAX=    746.9358822 / maximum wavelength [nanometers]
SPEC_BIN=    0.408104133 / average spectral coordinate bin size [nm/pix]
APERTURE=      0.0002778 / [deg] Aperture diameter
SPEC_RES= 432.0955936041 / Spectral resolving power
SPECSYS = 'TOPOCENT'     / Observed frame

\end{verbatim}
\end{scriptsize}

\noindent The dispersion is given by \texttt{SPEC\_BIN}, determined simply from : 

\begin{small}
\begin{equation}
SPEC\_BIN = \frac{WAVELMAX-WAVELMIN}{n_{pix}}
\end{equation}
\end{small}

\noindent where $n_{pix}$ is the number of pixels in the array. 

The slit width is given in degrees as the value \texttt{APERTURE}, and
the resolving power is calculated from the nearest arc calibration
frame (in time) to the science frame.  We do not apply any velocity correction to the spectra, hence
\texttt{SPECSYS} is set to topocentric. 
\subsection{Detector characteristics}

As described in Sect.\,\ref{sec:efosc-ccd} and the read noise and gain
have been remeasured for CCD\#40 on EFOSC2 and the correct values are written into the
header as the following keywords. 

\begin{scriptsize}
\begin{verbatim}
DETRON =                11.6 / Readout noise per output (e-)
GAIN     =              1.18 / Conversion from electrons to ADU
EFFRON  =       13.282436188 / Effective readout noise per output (e-)
\end{verbatim}
\end{scriptsize}

\noindent \texttt{GAIN} is always the same in the EFOSC2 released data products, since they are single images
and not combined. Similarly, the effective readnoise \texttt{EFFRON} is fairly constant since it only relies on the 
the flats and biases used to detrend the data :

\begin{small}
\begin{equation}
EFFRON = DETRON \sqrt{1 + \frac{1}{n_{\rm bias}} + \frac{1}{n_{\rm flat}}}
\end{equation}
\end{small}
where

\noindent$n_{\rm bias} $= number of bias frames making up the masterbias\\
\noindent$n_{\rm flat}$ = number of flat-field frames making up the masterflat\\

\noindent  For SOFI imaging, the dithered images are median combined and hence the values for 
\texttt{EFFRON} and \texttt{GAIN} are calculated appropriately. In general, the images and spectra
are shot noise limited from the high NIR background and readnoise is not a major factor.

\subsection{Instrument set-up and book keeping}

The object name is the primary name used by the supernova and
transient community. Where it exists, an IAU name (e.g. SN2013xy) is
used, otherwise the survey specific names (e.g. LSQ12aaa), or the 
``potential'' SN name from the  CBAT ``Transient Objects Confirmation
Page'' is employed. It is important to note that for all spectral frames the
RA and DEC values refer to those of the target, not the telescope. 
However for   all imaging frames the RA and DEC refer to the telescope pointing position.

\begin{scriptsize}
\begin{verbatim}
OBJECT  = 'PSNJ13540068-0755438' / Original target.
RA      =      208.504924    / 13:54:01.1 RA (J2000) target (deg)
DEC     =        -7.93163    / -07:55:53.8 DEC (J2000) target deg)
\end{verbatim}
\end{scriptsize}

The ESO OB that created the science frames is recorded as OBID1. Since
PESSTO provides the single epoch, individual spectra, there will
always be only one OBID in the header. The title of the dataset is
given as the MJD (of the observations), object name, grism, filter and
slit combinations. In addition, the grism, filter and slit combinations are
listed as below. 

\begin{scriptsize}
\begin{verbatim}
OBID1   =            100324424 / Observation block ID
TITLE   = '56384.305 PSNJ13540068-0755438 Gr11 Free slit1.0'/Dataset title
DISPELEM= 'Gr#11   '           / Dispersive element name
FILTER  = 'Free    '           / Filter name
APERTURE=            0.0002778 / [deg] Aperture diameter
\end{verbatim}
\end{scriptsize}

\noindent The relevant time stamps are listed below and are self-explanatory,
and are as defined in the ESO Science Data Products Standard
\citep{p2edpstd}. We add our own \texttt{AIRMASS} keyword which is the
mean airmass calculated at the midpoint of the exposure. This value 
is the one used in calculations of the sensitivity function, to flux
calibrate
the science spectra and to compute the zeropoints for EFOSC2 imaging. 

\begin{scriptsize}
\begin{verbatim}
TEXPTIME=             900.0006 / Total integ. time of all exposure
TELAPSE =    900.0006001442671 / Total elapsed time [s]
MJD-END =    56384.31092294362 / End of observations (days)
TMID    =    56384.30571460681 /  [d] MJD mid exposure
AIRMASS =                1.148 / mean airmass computed with astcalc
\end{verbatim}
\end{scriptsize}

\noindent The version of the PESSTO pipeline which was used to reduce the data
is recorded using the \texttt{PROCSOFT} keyword. The source code, installation guide, users manual
and tutorial videos are available on the PESSTO
wiki\footnote{http://wiki.pessto.org/pessto-operation-groups/data-reduction-and-quality-control-team}. As
discussed in Sect\,\ref{sec:target}, PESSTO immediately  releases reduced data for all
classification targets via WISeREP \citep{2012PASP..124..668Y} within 24hrs of being taken. We
label these ``Fast'' reductions, while the full reductions for SSDR1 are
given an internal label of ``Final'' to distinguish them. This is
recorded in the header keyword \texttt{QUALITY}.  This publication is
recorded as the primary scientific publication describing the data
content. 

\begin{scriptsize}
\begin{verbatim}
PROCSOFT= 'ntt_2.1.0'          / pipeline version
QUALITY = 'Final   '           / Final or fast reduction
REFERENC= 'Smartt_et_al_2015'       / Bibliographic reference
\end{verbatim}
\end{scriptsize}

\subsection{Flux calibration}
\label{app:efosc-flux}

All objects extracted are by definition point sources hence the
extended object keyword is always set to false. The spectra are flux
calibrated and never
normalised hence \texttt{CONTNORM} is always set to false and the 
\texttt{FLUXCAL} is set to ABSOLUTE. As described in the ESO Science Data Products Standard \citep{p2edpstd}
\texttt{FLUXCAL}  should only be either ABSOLUTE or UNCALIBRATED. 
As PESSTO does not do wide slit observations to ensure that all flux
is captured within the slit, we set \texttt{TOT\_FLUX} to false always. 
The units of the flux calibration are in erg\,cm$^{-2}$\,s$^{-1}$\,\AA$^{-1}$ in the 
FITS binary table spectra. The value for \texttt{FLUXERR} is set to either 
15.2\% for EFOSC2 or 34.7\% for SOFI
as described in Sect.\,\ref{sec:efosc-specphot}
and Sect.\,\ref{sec:sofi-specphot}.

The average signal-noise-ratio (S/N) per pixel is calculated by determining
the S/N in $N$ regions taken at 50\AA\ intervals 
across the spectra and taking the mean. The 
number of regions $N$ is determined simply by
$(WAVELMAX-WAVELMIN)/50$. 

\begin{scriptsize}
\begin{verbatim}
FLUXERR =        15.2 / Fractional uncertainty of the flux [%]
TOT_FLUX=           F / TRUE if phot. cond. and all src flux captured
EXT_OBJ =           F / TRUE if extended
FLUXCAL = 'ABSOLUTE'  / Certifies the validity of PHOTZP
CONTNORM=           F / TRUE if normalised to the continuum
BUNIT   = 'erg/cm2/s/A'/ Physical unit of array values
SNR     =    26.0980159788 / Average signal to noise ratio per pixel
\end{verbatim}
\end{scriptsize}

The associated 2D spectroscopic frame labelled as \texttt{ASSON1} and
is submitted as an ancillary data product. This file is flux calibrated, and wavelength calibrated, and the units for
that are in $10^{-20}$erg\,cm$^{-2}$\,s$^{-1}$\,\AA$^{-1}$ 

\begin{scriptsize}
\begin{verbatim}
ASSON1  = 'SN2009ip_20130419_Gr11_Free_slit1.0_56448_1_si.fits'  
ASSOC1  = 'ANCILLARY.2DSPECTRUM' / Category of associated file
\end{verbatim}
\end{scriptsize}

\subsection{Imaging - photometric calibration }
\label{appdx:phot}

The keywords \texttt{PHOTZP, PHOTZPER, FLUXCAL, PHOTSYS} are
described above in Sect.\ref{sec:efosc-imagecal}. Four other keywords
are used to quantify the data. 

\begin{scriptsize}
\begin{verbatim}
PHOTZP  =           25.98 / MAG=-2.5*log(data)+PHOTZP                      
PHOTZPER=             999 / error in PHOTZP   
FLUXCAL = 'ABSOLUTE'      / Certifies the validity of PHOTZP               
PHOTSYS = 'VEGA    '      / Photometric system VEGA or AB                  
PSF_FWHM=      1.32371928 / Spatial resolution (arcsec)                    
ELLIPTIC=           0.131 / Average ellipticity of point sources   
ABMAGSAT=    13.340369487 / Saturation limit point sources (AB mags)   
ABMAGLIM=    19.861387040 / 5-sigma limiting AB magnitude point sources
\end{verbatim}
\end{scriptsize}

The values for \texttt{PSF\_FWHM}  and \texttt{ELLIPTIC} 
are determined through a {\sc  sextractor} 
\citep{1996A&AS..117..393B}
measurement on the field
which is automatically called within the PESSTO pipeline. 
The 5$\sigma$ limiting magnitude for a point source 
\texttt{ABMAGLIM}  is derived from :

\begin{scriptsize}
\begin{equation}\label{eqn:5sig}
\begin{split}
MAGLIM = PHOTZP  -  2.5\log\bigg(\frac{5}{(GAIN)(EXPTIME)} \times
  (N_{pix})(MBKG \times GAIN) +   \\
 (EFFRON^{2}\times N_{pix})^{1/2}\bigg)  
\end{split}
\end{equation}
\end{scriptsize}

\noindent Where $N_{pix}$ is the number of pixels in an aperture
($N_{pix}  = \pi (PSF\_FWHM/0.24)^2$  for EFOSC2 where the 0.24 scaling
factor is the pixel size in arcseconds; for SOFI this factor is 0.29) and \texttt{MBKG}
is the median background in ADU estimated by {\sc sextractor}. We
ignore extinction as a second-order effect in this calculation. 
 In a small number of cases (around 3\% of the $\sim$2400 EFOSC2 images
images) the images have short exposure times and low background
such that after bias subtraction, the value of  \texttt{MBKG} is
negative. This may be due to bias drift as seen in
Fig.\ref{fig:efgain}, or  a low enough background that read noise
dominates and the overall value is below zero. In these cases Eqn.\ref{eqn:5sig}
is still valid as the read noise will dominate. 

\noindent The magnitude of a point source that will saturate at peak
counts is given by the following equation. This assumes that saturation occurs at 60,000
ADU and that the volume under a 2D Gaussian is $2\pi I_0\sigma^2$
(where $I_0$ is the peak intensity) and that  $FWHM = 2\sqrt{2\ln2}\sigma$, then 

\begin{scriptsize}
\begin{equation}
MAGSAT = PHOTZP - 2.5 \log\left( \frac{\pi}{4\ln2} (60000 - MBKG) (PSF\_FWHM/0.24)^2 \right),
\end{equation}
\end{scriptsize}

\noindent The saturation value of 60,000 is assumed for EFOSC2 and for SOFI we assume
32000 ADU \citep[from the SOFI manual;][]{sofiman}
The $MBKG$ value is simply the median background sky in
ADU  (the bias level has a negligible effect since it is 0.3\% of the
ADU 16-bit saturation level) and is recorded in the headers as such. 
The short exposure time problem, where $MBKG$ may go negative, is 
not significant for this calculation. Again, the scaling factor of 0.24 is
simply the pixel size for EFOSC2 and for SOFI it is 0.29. 

Although the header keywords are always listed as \texttt{ABMAGSAT}
and \texttt{ABMAGLIM}, they should be interpreted in the photometric
system given by \texttt{PHOTSYS} and not always assumed to be in the
AB system.

\subsection{Imaging - astrometric calibration }
\label{appdx:astromet}

As described in  Sect. \ref{sec:efosc-imagecal}, the keyword
\texttt{ASTROMET} provides the number of catalogued stars used for the
astrometric calibration and the RMS of  $\alpha$ and $\delta$ in
arcseconds.  The other keywords are mandatory  ESO Science Data
Products Standard keywords
\citep{p2edpstd}.  
\texttt{CSYER1} and \texttt{CSYER2} should specify the contribution to
the uncertainty of the astrometric calibration due to systematic
errors intrinsic to the registration process. In our case this is
dominated by the 
uncertainty intrinsic to the astrometric reference catalogues
used. For data registered to the 2MASS point source catalogue we list
the uncertainty in each axis as 100 milli-arcseconds 
\citep{2006AJ....131.1163S}
or 2.78E-05 degrees and 
and for USNO  B1 it is 200 milli-arcseconds
\citep{2003AJ....125..984M} 
or 5.55E-05 degrees. 

\begin{scriptsize}
\begin{verbatim}
ASTROMET= '0.372 0.507 10'     / rmsx rmsy nstars                               
CUNIT1  = 'deg     '           / unit of the coord. trans.                      
CRDER1  = 7.30677007226099E-05 / Random error (degree)                          
CSYER1  =             2.78E-05 / Systematic error (RA_m - Ra_ref)   
CUNIT2  = 'deg     '           / unit of the coord. trans.                      
CRDER2  = 9.95842050171054E-05 / Random error (degree)                          
CSYER2  =             2.78E-05 / Systematic error (DEC_m - DEC_ref)             
\end{verbatim}
\end{scriptsize}

\section{Photometric nights} 
\label{appdx:photomet-nights}

The PESSTO observers record the night conditions in a night report
which is publicly available on the PESSTO web pages (via the PESSTO
wiki $^{13}$. The following table summarises that information. Where
the conditions are labelled with ``photometric'', then the night was considered
photometric in that there were no visible clouds at dusk or dawn and
no obvious signs of clouds or transparency problems during the night. 
A ``non-photometric'' label means that the night was definitely not photometric, and a $?$
means that there were no obvious transparency issues but with the
information available we cannot be  completely certain that it was
photometric. 

\begin{table}
\caption[]{List of records on photometric nights} 
\label{tab:phot-record}
\begin{tabular}{llll}
\hline\hline
Night    &  La Silla Conditions   &                Night & La Silla Conditions \\   \hline       
2013 Apr 19   &       photometric      &               2012 Dec 11   &      photometric    \\                                      
2013 Apr 18   &      photometric      &                2012 Dec 06   &      ?        \\                                            
2013 Apr 17   &      photometric      &                2012 Dec 05   &      ?        \\                                            
2013 Apr 13   &      photometric      &                2012 Dec 04   &      ?        \\                                            
2013 Apr 12   &       non-photometric      &       2012 Dec 03   &      ?        \\                                            
2013 Apr 11   &      photometric      &                2012 Nov 22   &      ?        \\                                            
2013 Apr 05   &      photometric      &                2012 Nov 21   &      ?        \\                                            
2013 Apr 04   &      photometric      &                2012 Nov 20   &      ?        \\                                            
2013 Apr 03   &      photometric      &                2012 Nov 14   &      ?        \\                                            
2013 Apr 02   &      non-photometric      &        2012 Nov 13   &      ?        \\                                            
2013 Apr 01   &      non-photometric      &        2012 Nov 12   &      non-photometric    \\                                  
2013 Mar 18   &      non-photometric      &        2012 Nov 07   &      non-photometric    \\                                  
2013 Mar 17   &      ?          &                              2012 Nov 06   &      photometric    \\                                      
2013 Mar 16   &      photometric      &                2012 Nov 05   &      photometric    \\                                      
2013 Mar 12   &      photometric      &                2012 Nov 04   &      photometric    \\                                      
2013 Mar 11   &      photometric      &                2012 Oct 22   &      ?        \\                                            
2013 Mar 10   &      photometric      &                2012 Oct 21   &      non-photometric    \\                                  
2013 Mar 05   &      photometric      &                2012 Oct 20   &      ?        \\                                            
2013 Mar 04   &      photometric      &                2012 Oct 16   &      non-photometric    \\                                  
2013 Mar 03   &      non-photometric      &        2012 Oct 15   &      ?        \\                                            
2013 Mar 02   &      non-photometric      &        2012 Oct 14   &      ?        \\                                            
2013 Mar 01   &      non-photometric      &        2012 Oct 09   &      non-photometric    \\                                  
2013 Feb 21   &      photometric      &                2012 Oct 08   &      ?        \\                                            
2013 Feb 20   &      photometric      &                2012 Oct 07   &      non-photometric    \\                                  
2013 Feb 19   &      non-photometric      &        2012 Oct 06   &      non-photometric    \\                                  
2013 Feb 08   &      non-photometric      &        2012 Sep 25   &      non-photometric    \\                                  
2013 Feb 07   &      non-photometric      &        2012 Sep 24   &      photometric    \\                                      
2013 Feb 06   &      non-photometric      &        2012 Sep 23   &      non-photometric    \\                                  
2013 Jan 30   &      non-photometric      &        2012 Sep 22   &      non-photometric    \\                                  
2013 Jan 29   &      ?          &                              2012 Sep 17   &      non-photometric    \\                                  
2013 Jan 28   &      ?          &                              2012 Sep 16   &      non-photometric    \\                                  
2013 Jan 27   &      ?          &                              2012 Sep 15   &      non-photometric    \\                                  
2013 Jan 21   &      non-photometric      &        2012 Sep 09   &      photometric    \\                                      
2013 Jan 20   &      non-photometric      &        2012 Sep 08   &      photometric    \\                                      
2013 Jan 19   &      non-photometric      &        2012 Sep 07   &         non-photometric    \\                               
2013 Jan 13   &      non-photometric      &        2012 Sep 06   &         non-photometric    \\                               
2013 Jan 12   &      ?          &                              2012 Aug 26   &         non-photometric    \\                               
2013 Jan 11   &      ?          &                              2012 Aug 25   &         non-photometric    \\                               
2013 Jan 04   &      ?          &                              2012 Aug 24   &        photometric    \\                                    
2013 Jan 03   &      ?          &                              2012 Aug 18   &        photometric    \\                                    
2013 Jan 02   &      ?          &                              2012 Aug 17   &         non-photometric    \\                               
2013 Jan 01   &      ?          &                              2012 Aug 16   &         non-photometric    \\                               
2012 Dec 22   &      ?          &                              2012 Aug 10   &         photometric    \\                                   
2012 Dec 21   &      ?          &                              2012 Aug 09   &         non-photometric    \\                               
2012 Dec 20   &      ?          &                              2012 Aug 08   &         non-photometric    \\                               
2012 Dec 13   &      non-photometric      &        2012 Aug 07   &         photometric    \\                                   
2012 Dec 12   &      photometric      &     \\
\hline 
 \end{tabular}
 \end{table} 

\section{Software for  reading FITS binary tables} 
\label{appdx:soft-fitstable}
As described in Section\,\ref{sec:efosc-ssdr1}
and Section\,\ref{sec:sofi-products}, 
the PESSTO spectra from the ESO Science Archive Facility are in FITS binary table format. Not all astronomical software routines can read this format easily. Listed here are some examples of software that can be used. This information is  linked from the PESSTO survey home page and there are links there to follow to get the relevant software. 
\begin{enumerate}
\item  The new  {\sc IRAF} external package  {\sc SPTABLE}  is able to read, display, and analyse (via the onedspec and rv packages). 
\item {\em Fv} is a graphical program for viewing and editing any FITS format image or table available from NASA's  High Energy Astrophysics Science Archive Research Center (HEASARC). 
\item {\em VOSpec} is a multi-wavelength spectral analysis tool from the ESA Virtual Observatory team.
\item {\em SPLAT-VO} is a Virtual Observatory enabled package that originated in {\sc STARLINK} and is
now released as part of the 
German Astrophysical Virtual Observatory (GAVO)
\item IDL and {\em python} can also read FITS binary tables through the 
IDL  Astronomy User's Library
at Goddard and through 
{\em pyfits} respectively. 
\item Further details on all the above are linked from the PESSTO website and are available at 
\begin{small}
\verb+http://archive.eso.org/cms/eso-data/help/1dspectra.html+
\end{small}

\end{enumerate}

\end{appendix} 


\begin{thebibliography}{78}
\expandafter\ifx\csname natexlab\endcsname\relax\def\natexlab#1{#1}\fi

\bibitem[{{Ahn} {et~al.}(2012){Ahn}, {Alexandroff}, {Allende Prieto},
  {Anderson}, {Anderton}, {Andrews}, {Aubourg}, {Bailey}, {Balbinot}, {Barnes},
  \& et~al.}]{2012ApJS..203...21A}
{Ahn}, C.~P., {Alexandroff}, R., {Allende Prieto}, C., {et~al.} 2012, \apjs,
  203, 21

\bibitem[{{Baltay} {et~al.}(2012){Baltay}, {Rabinowitz}, {Hadjiyska},
  {Schwamb}, {Ellman}, {Zinn}, {Tourtellotte}, {McKinnon}, {Horowitz},
  {Effron}, \& {Nugent}}]{2012Msngr.150...34B}
{Baltay}, C., {Rabinowitz}, D., {Hadjiyska}, E., {et~al.} 2012, The Messenger,
  150, 34

\bibitem[{{Baltay} {et~al.}(2013){Baltay}, {Rabinowitz}, {Hadjiyska}, {Walker},
  {Nugent}, {Coppi}, {Ellman}, {Feindt}, {McKinnon}, {Horowitz}, \&
  {Effron}}]{2013PASP..125..683B}
{Baltay}, C., {Rabinowitz}, D., {Hadjiyska}, E., {et~al.} 2013, \pasp, 125, 683

\bibitem[{{Benetti} {et~al.}(2014){Benetti}, {Nicholl}, {Cappellaro},
  {Pastorello}, {Smartt}, {Elias-Rosa}, {Drake}, {Tomasella}, {Turatto},
  {Harutyunyan}, {Taubenberger}, {Hachinger}, {Morales-Garoffolo}, {Chen},
  {Djorgovski}, {Fraser}, {Gal-Yam}, {Inserra}, {Mazzali}, {Pumo}, {Sollerman},
  {Valenti}, {Young}, {Dennefeld}, {Le Guillou}, {Fleury}, \&
  {L{\'e}get}}]{2014MNRAS.441..289B}
{Benetti}, S., {Nicholl}, M., {Cappellaro}, E., {et~al.} 2014, \mnras, 441, 289

\bibitem[{{Berger} {et~al.}(2012){Berger}, {Chornock}, {Lunnan}, {Foley},
  {Czekala}, {Rest}, {Leibler}, {Soderberg}, {Roth}, {Narayan}, {Huber},
  {Milisavljevic}, {Sanders}, {Drout}, {Margutti}, {Kirshner}, {Marion},
  {Challis}, {Riess}, {Smartt}, {Burgett}, {Hodapp}, {Heasley}, {Kaiser},
  {Kudritzki}, {Magnier}, {McCrum}, {Price}, {Smith}, {Tonry}, \&
  {Wainscoat}}]{2012ApJ...755L..29B}
{Berger}, E., {Chornock}, R., {Lunnan}, R., {et~al.} 2012, \apjl, 755, L29

\bibitem[{{Bertin} \& {Arnouts}(1996)}]{1996A&AS..117..393B}
{Bertin}, E. \& {Arnouts}, S. 1996, \aaps, 117, 393

\bibitem[{{Bertin} {et~al.}(2002){Bertin}, {Mellier}, {Radovich}, {Missonnier},
  {Didelon}, \& {Morin}}]{swarp}
{Bertin}, E., {Mellier}, Y., {Radovich}, M., {et~al.} 2002, in Astronomical
  Society of the Pacific Conference Series, Vol. 281, Astronomical Data
  Analysis Software and Systems XI, ed. D.~A. {Bohlender}, D.~{Durand}, \&
  T.~H. {Handley}, 228

\bibitem[{{Blondin} \& {Tonry}(2007)}]{2007ApJ...666.1024B}
{Blondin}, S. \& {Tonry}, J.~L. 2007, \apj, 666, 1024

\bibitem[{{Boroson} {et~al.}(2014){Boroson}, {Brown}, {Hjelstrom}, {Howell},
  {Lister}, {Pickles}, {Rosing}, {Saunders}, {Street}, \&
  {Walker}}]{2014SPIE.9149E..1EB}
{Boroson}, T., {Brown}, T., {Hjelstrom}, A., {et~al.} 2014, in Society of
  Photo-Optical Instrumentation Engineers (SPIE) Conference Series, Vol. 9149,
  Society of Photo-Optical Instrumentation Engineers (SPIE) Conference Series

\bibitem[{{Botticella} {et~al.}(2010){Botticella}, {Trundle}, {Pastorello},
  {Rodney}, {Rest}, {Gezari}, {Smartt}, {Narayan}, {Huber}, {Tonry}, {Young},
  {Smith}, {Bresolin}, {Valenti}, {Kotak}, {Mattila}, {Kankare}, {Wood-Vasey},
  {Riess}, {Neill}, {Forster}, {Martin}, {Stubbs}, {Burgett}, {Chambers},
  {Dombeck}, {Flewelling}, {Grav}, {Heasley}, {Hodapp}, {Kaiser}, {Kudritzki},
  {Luppino}, {Lupton}, {Magnier}, {Monet}, {Morgan}, {Onaka}, {Price},
  {Rhoads}, {Siegmund}, {Sweeney}, {Wainscoat}, {Waters}, {Waterson}, \&
  {Wynn-Williams}}]{2010ApJ...717L..52B}
{Botticella}, M.~T., {Trundle}, C., {Pastorello}, A., {et~al.} 2010, \apjl,
  717, L52

\bibitem[{{Cao} {et~al.}(2013){Cao}, {Kasliwal}, {Arcavi}, {Horesh}, {Hancock},
  {Valenti}, {Cenko}, {Kulkarni}, {Gal-Yam}, {Gorbikov}, {Ofek}, {Sand},
  {Yaron}, {Graham}, {Silverman}, {Wheeler}, {Marion}, {Walker}, {Mazzali},
  {Howell}, {Li}, {Kong}, {Bloom}, {Nugent}, {Surace}, {Masci}, {Carpenter},
  {Degenaar}, \& {Gelino}}]{2013ApJ...775L...7C}
{Cao}, Y., {Kasliwal}, M.~M., {Arcavi}, I., {et~al.} 2013, \apjl, 775, L7

\bibitem[{{Childress} {et~al.}(2013){Childress}, {Scalzo}, {Sim}, {Tucker},
  {Yuan}, {Schmidt}, {Cenko}, {Silverman}, {Contreras}, {Hsiao}, {Phillips},
  {Morrell}, {Jha}, {McCully}, {Filippenko}, {Anderson}, {Benetti}, {Bufano},
  {de Jaeger}, {Forster}, {Gal-Yam}, {Le Guillou}, {Maguire}, {Maund},
  {Mazzali}, {Pignata}, {Smartt}, {Spyromilio}, {Sullivan}, {Taddia},
  {Valenti}, {Bayliss}, {Bessell}, {Blanc}, {Carson}, {Clubb}, {de Burgh-Day},
  {Desjardins}, {Fang}, {Fox}, {Gates}, {Ho}, {Keller}, {Kelly}, {Lidman},
  {Loaring}, {Mould}, {Owers}, {Ozbilgen}, {Pei}, {Pickering}, {Pracy}, {Rich},
  {Schaefer}, {Scott}, {Stritzinger}, {Vogt}, \& {Zhou}}]{2013ApJ...770...29C}
{Childress}, M.~J., {Scalzo}, R.~A., {Sim}, S.~A., {et~al.} 2013, \apj, 770, 29

\bibitem[{{Chomiuk} {et~al.}(2011){Chomiuk}, {Chornock}, {Soderberg}, {Berger},
  {Chevalier}, {Foley}, {Huber}, {Narayan}, {Rest}, {Gezari}, {Kirshner},
  {Riess}, {Rodney}, {Smartt}, {Stubbs}, {Tonry}, {Wood-Vasey}, {Burgett},
  {Chambers}, {Czekala}, {Flewelling}, {Forster}, {Kaiser}, {Kudritzki},
  {Magnier}, {Martin}, {Morgan}, {Neill}, {Price}, {Roth}, {Sanders}, \&
  {Wainscoat}}]{2011ApJ...743..114C}
{Chomiuk}, L., {Chornock}, R., {Soderberg}, A.~M., {et~al.} 2011, \apj, 743,
  114

\bibitem[{{Clough} {et~al.}(2005){Clough}, {Shephard}, {Mlawer}, {Delamere},
  {Iacono}, {Cady-Pereira}, {Boukabara}, \& {Brown}}]{2005JQSRT..91..233C}
{Clough}, S.~A., {Shephard}, M.~W., {Mlawer}, E.~J., {et~al.} 2005, \jqsrt, 91,
  233

\bibitem[{{DePoy} {et~al.}(2003){DePoy}, {Atwood}, {Belville}, {Brewer},
  {Byard}, {Gould}, {Mason}, {O'Brien}, {Pappalardo}, {Pogge}, {Steinbrecher},
  \& {Teiga}}]{2003SPIE.4841..827D}
{DePoy}, D.~L., {Atwood}, B., {Belville}, S.~R., {et~al.} 2003, in Society of
  Photo-Optical Instrumentation Engineers (SPIE) Conference Series, Vol. 4841,
  Instrument Design and Performance for Optical/Infrared Ground-based
  Telescopes, ed. M.~{Iye} \& A.~F.~M. {Moorwood}, 827--838

\bibitem[{{Drake} {et~al.}(2011){Drake}, {Djorgovski}, {Mahabal}, {Anderson},
  {Roy}, {Mohan}, {Ravindranath}, {Frail}, {Gezari}, {Neill}, {Ho}, {Prieto},
  {Thompson}, {Thorstensen}, {Wagner}, {Kowalski}, {Chiang}, {Grove},
  {Schinzel}, {Wood}, {Carrasco}, {Recillas}, {Kewley}, {Archana}, {Basu},
  {Wadadekar}, {Kumar}, {Myers}, {Phinney}, {Williams}, {Graham}, {Catelan},
  {Beshore}, {Larson}, \& {Christensen}}]{2011ApJ...735..106D}
{Drake}, A.~J., {Djorgovski}, S.~G., {Mahabal}, A., {et~al.} 2011, \apj, 735,
  106

\bibitem[{{Drake} {et~al.}(2009){Drake}, {Djorgovski}, {Mahabal}, {Beshore},
  {Larson}, {Graham}, {Williams}, {Christensen}, {Catelan}, {Boattini},
  {Gibbs}, {Hill}, \& {Kowalski}}]{2009ApJ...696..870D}
{Drake}, A.~J., {Djorgovski}, S.~G., {Mahabal}, A., {et~al.} 2009, \apj, 696,
  870

\bibitem[{{Drake} {et~al.}(2010){Drake}, {Djorgovski}, {Prieto}, {Mahabal},
  {Balam}, {Williams}, {Graham}, {Catelan}, {Beshore}, \&
  {Larson}}]{2010ApJ...718L.127D}
{Drake}, A.~J., {Djorgovski}, S.~G., {Prieto}, J.~L., {et~al.} 2010, \apjl,
  718, L127

\bibitem[{{Ellis} {et~al.}(2008){Ellis}, {Sullivan}, {Nugent}, {Howell},
  {Gal-Yam}, {Astier}, {Balam}, {Balland}, {Basa}, {Carlberg}, {Conley},
  {Fouchez}, {Guy}, {Hardin}, {Hook}, {Pain}, {Perrett}, {Pritchet}, \&
  {Regnault}}]{2008ApJ...674...51E}
{Ellis}, R.~S., {Sullivan}, M., {Nugent}, P.~E., {et~al.} 2008, \apj, 674, 51

\bibitem[{{Fraser} {et~al.}(2013){Fraser}, {Inserra}, {Jerkstrand}, {Kotak},
  {Pignata}, {Benetti}, {Botticella}, {Bufano}, {Childress}, {Mattila},
  {Pastorello}, {Smartt}, {Turatto}, {Yuan}, {Anderson}, {Bayliss}, {Bauer},
  {Chen}, {F{\"o}rster Bur{\'o}n}, {Gal-Yam}, {Haislip}, {Knapic}, {Le
  Guillou}, {Marchi}, {Mazzali}, {Molinaro}, {Moore}, {Reichart}, {Smareglia},
  {Smith}, {Sternberg}, {Sullivan}, {Tak{\'a}ts}, {Tucker}, {Valenti}, {Yaron},
  {Young}, \& {Zhou}}]{2013MNRAS.433.1312F}
{Fraser}, M., {Inserra}, C., {Jerkstrand}, A., {et~al.} 2013, \mnras, 433, 1312

\bibitem[{{Gal-Yam} {et~al.}(2014){Gal-Yam}, {Arcavi}, {Ofek}, {Ben-Ami},
  {Cenko}, {Kasliwal}, {Cao}, {Yaron}, {Tal}, {Silverman}, {Horesh}, {De Cia},
  {Taddia}, {Sollerman}, {Perley}, {Vreeswijk}, {Kulkarni}, {Nugent},
  {Filippenko}, \& {Wheeler}}]{2014Natur.509..471G}
{Gal-Yam}, A., {Arcavi}, I., {Ofek}, E.~O., {et~al.} 2014, \nat, 509, 471

\bibitem[{{Gal-Yam} {et~al.}(2011){Gal-Yam}, {Kasliwal}, {Arcavi}, {Green},
  {Yaron}, {Ben-Ami}, {Xu}, {Sternberg}, {Quimby}, {Kulkarni}, {Ofek},
  {Walters}, {Nugent}, {Poznanski}, {Bloom}, {Cenko}, {Filippenko}, {Li},
  {Silverman}, {Walker}, {Sullivan}, {Maguire}, {Howell}, {Mazzali}, {Frail},
  {Bersier}, {James}, {Akerlof}, {Yuan}, {Law}, {Fox}, \&
  {Gehrels}}]{2011ApJ...736..159G}
{Gal-Yam}, A., {Kasliwal}, M.~M., {Arcavi}, I., {et~al.} 2011, \apj, 736, 159

\bibitem[{{Gezari} {et~al.}(2012){Gezari}, {Chornock}, {Rest}, {Huber},
  {Forster}, {Berger}, {Challis}, {Neill}, {Martin}, {Heckman}, {Lawrence},
  {Norman}, {Narayan}, {Foley}, {Marion}, {Scolnic}, {Chomiuk}, {Soderberg},
  {Smith}, {Kirshner}, {Riess}, {Smartt}, {Stubbs}, {Tonry}, {Wood-Vasey},
  {Burgett}, {Chambers}, {Grav}, {Heasley}, {Kaiser}, {Kudritzki}, {Magnier},
  {Morgan}, \& {Price}}]{2012Natur.485..217G}
{Gezari}, S., {Chornock}, R., {Rest}, A., {et~al.} 2012, \nat, 485, 217

\bibitem[{{Harutyunyan} {et~al.}(2008){Harutyunyan}, {Pfahler}, {Pastorello},
  {Taubenberger}, {Turatto}, {Cappellaro}, {Benetti}, {Elias-Rosa},
  {Navasardyan}, {Valenti}, {Stanishev}, {Patat}, {Riello}, {Pignata}, \&
  {Hillebrandt}}]{2008A&A...488..383H}
{Harutyunyan}, A.~H., {Pfahler}, P., {Pastorello}, A., {et~al.} 2008, \aap,
  488, 383

\bibitem[{{Hodgkin} {et~al.}(2013){Hodgkin}, {Wyrzykowski}, {Blagorodnova}, \&
  {Koposov}}]{2013RSPTA.37120239H}
{Hodgkin}, S.~T., {Wyrzykowski}, L., {Blagorodnova}, N., \& {Koposov}, S. 2013,
  Royal Society of London Philosophical Transactions Series A, 371, 20239

\bibitem[{{Howell} {et~al.}(2005){Howell}, {Sullivan}, {Perrett}, {Bronder},
  {Hook}, {Astier}, {Aubourg}, {Balam}, {Basa}, {Carlberg}, {Fabbro},
  {Fouchez}, {Guy}, {Lafoux}, {Neill}, {Pain}, {Palanque-Delabrouille},
  {Pritchet}, {Regnault}, {Rich}, {Taillet}, {Knop}, {McMahon}, {Perlmutter},
  \& {Walton}}]{2005ApJ...634.1190H}
{Howell}, D.~A., {Sullivan}, M., {Perrett}, K., {et~al.} 2005, \apj, 634, 1190

\bibitem[{{Inserra} {et~al.}(2013){Inserra}, {Smartt}, {Jerkstrand}, {Valenti},
  {Fraser}, {Wright}, {Smith}, {Chen}, {Kotak}, {Pastorello}, {Nicholl},
  {Bresolin}, {Kudritzki}, {Benetti}, {Botticella}, {Burgett}, {Chambers},
  {Ergon}, {Flewelling}, {Fynbo}, {Geier}, {Hodapp}, {Howell}, {Huber},
  {Kaiser}, {Leloudas}, {Magill}, {Magnier}, {McCrum}, {Metcalfe}, {Price},
  {Rest}, {Sollerman}, {Sweeney}, {Taddia}, {Taubenberger}, {Tonry},
  {Wainscoat}, {Waters}, \& {Young}}]{2013ApJ...770..128I}
{Inserra}, C., {Smartt}, S.~J., {Jerkstrand}, A., {et~al.} 2013, \apj, 770, 128

\bibitem[{{Inserra} {et~al.}(2014){Inserra}, {Smartt}, {Scalzo}, {Fraser},
  {Pastorello}, {Childress}, {Pignata}, {Jerkstrand}, {Kotak}, {Benetti},
  {Della Valle}, {Gal-Yam}, {Mazzali}, {Smith}, {Sullivan}, {Valenti}, {Yaron},
  {Young}, \& {Reichart}}]{2014MNRAS.437L..51I}
{Inserra}, C., {Smartt}, S.~J., {Scalzo}, R., {et~al.} 2014, \mnras, 437, L51

\bibitem[{{Jester} {et~al.}(2005){Jester}, {Schneider}, {Richards}, {Green},
  {Schmidt}, {Hall}, {Strauss}, {Vanden Berk}, {Stoughton}, {Gunn},
  {Brinkmann}, {Kent}, {Smith}, {Tucker}, \& {Yanny}}]{2005AJ....130..873J}
{Jester}, S., {Schneider}, D.~P., {Richards}, G.~T., {et~al.} 2005, \aj, 130,
  873

\bibitem[{{Kaiser} {et~al.}(2010){Kaiser}, {Burgett}, {Chambers}, {Denneau},
  {Heasley}, {Jedicke}, {Magnier}, {Morgan}, {Onaka}, \&
  {Tonry}}]{2010SPIE.7733E..12K}
{Kaiser}, N., {Burgett}, W., {Chambers}, K., {et~al.} 2010, in Society of
  Photo-Optical Instrumentation Engineers (SPIE) Conference Series, Vol. 7733,
  Society of Photo-Optical Instrumentation Engineers (SPIE) Conference Series

\bibitem[{{Kasliwal} {et~al.}(2010){Kasliwal}, {Kulkarni}, {Gal-Yam}, {Yaron},
  {Quimby}, {Ofek}, {Nugent}, {Poznanski}, {Jacobsen}, {Sternberg}, {Arcavi},
  {Howell}, {Sullivan}, {Rich}, {Burke}, {Brimacombe}, {Milisavljevic},
  {Fesen}, {Bildsten}, {Shen}, {Cenko}, {Bloom}, {Hsiao}, {Law}, {Gehrels},
  {Immler}, {Dekany}, {Rahmer}, {Hale}, {Smith}, {Zolkower}, {Velur},
  {Walters}, {Henning}, {Bui}, \& {McKenna}}]{2010ApJ...723L..98K}
{Kasliwal}, M.~M., {Kulkarni}, S.~R., {Gal-Yam}, A., {et~al.} 2010, \apjl, 723,
  L98

\bibitem[{{Keller} {et~al.}(2007){Keller}, {Schmidt}, {Bessell}, {Conroy},
  {Francis}, {Granlund}, {Kowald}, {Oates}, {Martin-Jones}, {Preston},
  {Tisserand}, {Vaccarella}, \& {Waterson}}]{2007PASA...24....1K}
{Keller}, S.~C., {Schmidt}, B.~P., {Bessell}, M.~S., {et~al.} 2007, \pasa, 24,
  1

\bibitem[{{Klotz} {et~al.}(2008){Klotz}, {Vachier}, \&
  {Bo{\"e}r}}]{2008AN....329..275K}
{Klotz}, A., {Vachier}, F., \& {Bo{\"e}r}, M. 2008, Astronomische Nachrichten,
  329, 275

\bibitem[{{Koz{\l}owski} {et~al.}(2013){Koz{\l}owski}, {Udalski},
  {Wyrzykowski}, {Poleski}, {Pietrukowicz}, {Skowron}, {Szyma{\'n}ski},
  {Kubiak}, {Pietrzy{\'n}ski}, {Soszy{\'n}ski}, \&
  {Ulaczyk}}]{2013AcA....63....1K}
{Koz{\l}owski}, S., {Udalski}, A., {Wyrzykowski}, {\L}., {et~al.} 2013, \actaa,
  63, 1

\bibitem[{{Kulkarni} \& {Kasliwal}(2009)}]{2009aaxo.conf..312K}
{Kulkarni}, S. \& {Kasliwal}, M.~M. 2009, in Astrophysics with All-Sky X-Ray
  Observations, ed. N.~{Kawai}, T.~{Mihara}, M.~{Kohama}, \& M.~{Suzuki}, 312

\bibitem[{{Kulkarni} {et~al.}(2007){Kulkarni}, {Ofek}, {Rau}, {Cenko},
  {Soderberg}, {Fox}, {Gal-Yam}, {Capak}, {Moon}, {Li}, {Filippenko}, {Egami},
  {Kartaltepe}, \& {Sanders}}]{2007Natur.447..458K}
{Kulkarni}, S.~R., {Ofek}, E.~O., {Rau}, A., {et~al.} 2007, \nat, 447, 458

\bibitem[{{Law} {et~al.}(2009){Law}, {Kulkarni}, {Dekany}, {Ofek}, {Quimby},
  {Nugent}, {Surace}, {Grillmair}, {Bloom}, {Kasliwal}, {Bildsten}, {Brown},
  {Cenko}, {Ciardi}, {Croner}, {Djorgovski}, {van Eyken}, {Filippenko}, {Fox},
  {Gal-Yam}, {Hale}, {Hamam}, {Helou}, {Henning}, {Howell}, {Jacobsen},
  {Laher}, {Mattingly}, {McKenna}, {Pickles}, {Poznanski}, {Rahmer}, {Rau},
  {Rosing}, {Shara}, {Smith}, {Starr}, {Sullivan}, {Velur}, {Walters}, \&
  {Zolkower}}]{2009PASP..121.1395L}
{Law}, N.~M., {Kulkarni}, S.~R., {Dekany}, R.~G., {et~al.} 2009, \pasp, 121,
  1395

\bibitem[{{Leaman} {et~al.}(2011){Leaman}, {Li}, {Chornock}, \&
  {Filippenko}}]{2011MNRAS.412.1419L}
{Leaman}, J., {Li}, W., {Chornock}, R., \& {Filippenko}, A.~V. 2011, \mnras,
  412, 1419

\bibitem[{{Li} {et~al.}(2011){Li}, {Leaman}, {Chornock}, {Filippenko},
  {Poznanski}, {Ganeshalingam}, {Wang}, {Modjaz}, {Jha}, {Foley}, \&
  {Smith}}]{2011MNRAS.412.1441L}
{Li}, W., {Leaman}, J., {Chornock}, R., {et~al.} 2011, \mnras, 412, 1441

\bibitem[{{Lidman,} {et~al.}(2012){Lidman,}, {Cuby,}, {Vanzi}, {Billeres},
  {Ivanov}, \& {Saviane}}]{sofiman}
{Lidman,}, C., {Cuby,}, J.-G., {Vanzi}, L., {et~al.} 2012,
  LSO-MAN-ESO-40100-0004, Issue 2.3

\bibitem[{{Lipunov} {et~al.}(2010){Lipunov}, {Kornilov}, {Gorbovskoy},
  {Shatskij}, {Kuvshinov}, {Tyurina}, {Belinski}, {Krylov}, {Balanutsa},
  {Chazov}, {Kuznetsov}, {Kortunov}, {Sankovich}, {Tlatov}, {Parkhomenko},
  {Krushinsky}, {Zalozhnyh}, {Popov}, {Kopytova}, {Ivanov}, {Yazev}, \&
  {Yurkov}}]{2010AdAst2010E..30L}
{Lipunov}, V., {Kornilov}, V., {Gorbovskoy}, E., {et~al.} 2010, Advances in
  Astronomy, 2010

\bibitem[{{Lord}(1992)}]{Lor92}
{Lord}, S. 1992, NASA Technical Memorandum, 103957

\bibitem[{{Magnier} {et~al.}(2013){Magnier}, {Schlafly}, {Finkbeiner}, {Juric},
  {Tonry}, {Burgett}, {Chambers}, {Flewelling}, {Kaiser}, {Kudritzki},
  {Morgan}, {Price}, {Sweeney}, \& {Stubbs}}]{2013ApJS..205...20M}
{Magnier}, E.~A., {Schlafly}, E., {Finkbeiner}, D., {et~al.} 2013, \apjs, 205,
  20

\bibitem[{{Maguire} {et~al.}(2014){Maguire}, {Sullivan}, {Pan}, {Gal-Yam},
  {Hook}, {Howell}, {Nugent}, {Mazzali}, {Chotard}, {Clubb}, {Filippenko},
  {Kasliwal}, {Kandrashoff}, {Poznanski}, {Saunders}, {Silverman}, {Walker}, \&
  {Xu}}]{2014MNRAS.444.3258M}
{Maguire}, K., {Sullivan}, M., {Pan}, Y.-C., {et~al.} 2014, \mnras, 444, 3258

\bibitem[{{Maguire} {et~al.}(2013){Maguire}, {Sullivan}, {Patat}, {Gal-Yam},
  {Hook}, {Dhawan}, {Howell}, {Mazzali}, {Nugent}, {Pan}, {Podsiadlowski},
  {Simon}, {Sternberg}, {Valenti}, {Baltay}, {Bersier}, {Blagorodnova}, {Chen},
  {Ellman}, {Feindt}, {F{\"o}rster}, {Fraser}, {Gonz{\'a}lez-Gait{\'a}n},
  {Graham}, {Guti{\'e}rrez}, {Hachinger}, {Hadjiyska}, {Inserra}, {Knapic},
  {Laher}, {Leloudas}, {Margheim}, {McKinnon}, {Molinaro}, {Morrell}, {Ofek},
  {Rabinowitz}, {Rest}, {Sand}, {Smareglia}, {Smartt}, {Taddia}, {Walker},
  {Walton}, \& {Young}}]{2013MNRAS.436..222M}
{Maguire}, K., {Sullivan}, M., {Patat}, F., {et~al.} 2013, \mnras, 436, 222

\bibitem[{{Maund} {et~al.}(2013){Maund}, {Fraser}, {Smartt}, {Botticella},
  {Barbarino}, {Childress}, {Gal-Yam}, {Inserra}, {Pignata}, {Reichart},
  {Schmidt}, {Sollerman}, {Taddia}, {Tomasella}, {Valenti}, \&
  {Yaron}}]{2013MNRAS.431L.102M}
{Maund}, J.~R., {Fraser}, M., {Smartt}, S.~J., {et~al.} 2013, \mnras, 431, L102

\bibitem[{{Monaco} {et~al.}(2012){Monaco}, {Snodgrass}, {Schmidtobreick}, \&
  {Saviane}}]{efosc2manual}
{Monaco}, L., {Snodgrass}, C., {Schmidtobreick}, L., \& {Saviane}, I. 2012,
  LSO-MAN-ESO-36100-0004, Issue 3.6

\bibitem[{{Monet} {et~al.}(2003){Monet}, {Levine}, {Canzian}, {Ables}, {Bird},
  {Dahn}, {Guetter}, {Harris}, {Henden}, {Leggett}, {Levison}, {Luginbuhl},
  {Martini}, {Monet}, {Munn}, {Pier}, {Rhodes}, {Riepe}, {Sell}, {Stone},
  {Vrba}, {Walker}, {Westerhout}, {Brucato}, {Reid}, {Schoening}, {Hartley},
  {Read}, \& {Tritton}}]{2003AJ....125..984M}
{Monet}, D.~G., {Levine}, S.~E., {Canzian}, B., {et~al.} 2003, \aj, 125, 984

\bibitem[{{Moorwood} {et~al.}(1998){Moorwood}, {Cuby}, \&
  {Lidman}}]{1998Msngr..91....9M}
{Moorwood}, A., {Cuby}, J.-G., \& {Lidman}, C. 1998, The Messenger, 91, 9

\bibitem[{{Nicholl} {et~al.}(2014){Nicholl}, {Smartt}, {Jerkstrand}, {Inserra},
  {Anderson}, {Baltay}, {Benetti}, {Chen}, {Elias-Rosa}, {Feindt}, {Fraser},
  {Gal-Yam}, {Hadjiyska}, {Howell}, {Kotak}, {Lawrence}, {Leloudas},
  {Margheim}, {Mattila}, {McCrum}, {McKinnon}, {Mead}, {Nugent}, {Rabinowitz},
  {Rest}, {Smith}, {Sollerman}, {Sullivan}, {Taddia}, {Valenti}, {Walker}, \&
  {Young}}]{2014MNRAS.444.2096N}
{Nicholl}, M., {Smartt}, S.~J., {Jerkstrand}, A., {et~al.} 2014, \mnras, 444,
  2096

\bibitem[{{Nugent} {et~al.}(2011){Nugent}, {Sullivan}, {Cenko}, {Thomas},
  {Kasen}, {Howell}, {Bersier}, {Bloom}, {Kulkarni}, {Kandrashoff},
  {Filippenko}, {Silverman}, {Marcy}, {Howard}, {Isaacson}, {Maguire},
  {Suzuki}, {Tarlton}, {Pan}, {Bildsten}, {Fulton}, {Parrent}, {Sand},
  {Podsiadlowski}, {Bianco}, {Dilday}, {Graham}, {Lyman}, {James}, {Kasliwal},
  {Law}, {Quimby}, {Hook}, {Walker}, {Mazzali}, {Pian}, {Ofek}, {Gal-Yam}, \&
  {Poznanski}}]{2011Natur.480..344N}
{Nugent}, P.~E., {Sullivan}, M., {Cenko}, S.~B., {et~al.} 2011, \nat, 480, 344

\bibitem[{{O'Brien} \& {Smartt}(2013)}]{2013RSPTA.37120498O}
{O'Brien}, P.~T. \& {Smartt}, S.~J. 2013, Royal Society of London Philosophical
  Transactions Series A, 371, 20498

\bibitem[{{Ofek} {et~al.}(2013){Ofek}, {Sullivan}, {Cenko}, {Kasliwal},
  {Gal-Yam}, {Kulkarni}, {Arcavi}, {Bildsten}, {Bloom}, {Horesh}, {Howell},
  {Filippenko}, {Laher}, {Murray}, {Nakar}, {Nugent}, {Silverman}, {Shaviv},
  {Surace}, \& {Yaron}}]{2013Natur.494...65O}
{Ofek}, E.~O., {Sullivan}, M., {Cenko}, S.~B., {et~al.} 2013, \nat, 494, 65

\bibitem[{{Pastorello} {et~al.}(2013){Pastorello}, {Cappellaro}, {Inserra},
  {Smartt}, {Pignata}, {Benetti}, {Valenti}, {Fraser}, {Tak{\'a}ts}, {Benitez},
  {Botticella}, {Brimacombe}, {Bufano}, {Cellier-Holzem}, {Costado}, {Cupani},
  {Curtis}, {Elias-Rosa}, {Ergon}, {Fynbo}, {Hambsch}, {Hamuy}, {Harutyunyan},
  {Ivarson}, {Kankare}, {Martin}, {Kotak}, {LaCluyze}, {Maguire}, {Mattila},
  {Maza}, {McCrum}, {Miluzio}, {Norgaard-Nielsen}, {Nysewander}, {Ochner},
  {Pan}, {Pumo}, {Reichart}, {Tan}, {Taubenberger}, {Tomasella}, {Turatto}, \&
  {Wright}}]{2013ApJ...767....1P}
{Pastorello}, A., {Cappellaro}, E., {Inserra}, C., {et~al.} 2013, \apj, 767, 1

\bibitem[{{Pastorello} {et~al.}(2010){Pastorello}, {Smartt}, {Botticella},
  {Maguire}, {Fraser}, {Smith}, {Kotak}, {Magill}, {Valenti}, {Young},
  {Gezari}, {Bresolin}, {Kudritzki}, {Howell}, {Rest}, {Metcalfe}, {Mattila},
  {Kankare}, {Huang}, {Urata}, {Burgett}, {Chambers}, {Dombeck}, {Flewelling},
  {Grav}, {Heasley}, {Hodapp}, {Kaiser}, {Luppino}, {Lupton}, {Magnier},
  {Monet}, {Morgan}, {Onaka}, {Price}, {Rhoads}, {Siegmund}, {Stubbs},
  {Sweeney}, {Tonry}, {Wainscoat}, {Waterson}, {Waters}, \&
  {Wynn-Williams}}]{2010ApJ...724L..16P}
{Pastorello}, A., {Smartt}, S.~J., {Botticella}, M.~T., {et~al.} 2010, \apjl,
  724, L16

\bibitem[{{Pastorello} {et~al.}(2007){Pastorello}, {Smartt}, {Mattila},
  {Eldridge}, {Young}, {Itagaki}, {Yamaoka}, {Navasardyan}, {Valenti}, {Patat},
  {Agnoletto}, {Augusteijn}, {Benetti}, {Cappellaro}, {Boles}, {Bonnet-Bidaud},
  {Botticella}, {Bufano}, {Cao}, {Deng}, {Dennefeld}, {Elias-Rosa},
  {Harutyunyan}, {Keenan}, {Iijima}, {Lorenzi}, {Mazzali}, {Meng}, {Nakano},
  {Nielsen}, {Smoker}, {Stanishev}, {Turatto}, {Xu}, \&
  {Zampieri}}]{2007Natur.447..829P}
{Pastorello}, A., {Smartt}, S.~J., {Mattila}, S., {et~al.} 2007, \nat, 447, 829

\bibitem[{{Patat} {et~al.}(2011){Patat}, {Moehler}, {O'Brien}, {Pompei},
  {Bensby}, {Carraro}, {de Ugarte Postigo}, {Fox}, {Gavignaud}, {James},
  {Korhonen}, {Ledoux}, {Randall}, {Sana}, {Smoker}, {Stefl}, \&
  {Szeifert}}]{2011A&A...527A..91P}
{Patat}, F., {Moehler}, S., {O'Brien}, K., {et~al.} 2011, \aap, 527, A91

\bibitem[{{Perez} {et~al.}(2012){Perez}, {Bagish}, {Bredthauer}, {Espoz},
  {Jones}, \& {Pinto}}]{2012SPIE.8444E..4HP}
{Perez}, F., {Bagish}, A., {Bredthauer}, G., {et~al.} 2012, in Society of
  Photo-Optical Instrumentation Engineers (SPIE) Conference Series, Vol. 8444,
  Society of Photo-Optical Instrumentation Engineers (SPIE) Conference Series

\bibitem[{{Pignata} {et~al.}(2008){Pignata}, {Maza}, J.~{Hamuy}, {Other1},
  {Other2}, \& {Other3}}]{arxiv.org.0812.4923}
{Pignata}, G., {Maza}, J.~{Hamuy}, M., {et~al.} 2008, arXiv:0812.4923

\bibitem[{{Rau} {et~al.}(2009){Rau}, {Kulkarni}, {Law}, {Bloom}, {Ciardi},
  {Djorgovski}, {Fox}, {Gal-Yam}, {Grillmair}, {Kasliwal}, {Nugent}, {Ofek},
  {Quimby}, {Reach}, {Shara}, {Bildsten}, {Cenko}, {Drake}, {Filippenko},
  {Helfand}, {Helou}, {Howell}, {Poznanski}, \&
  {Sullivan}}]{2009PASP..121.1334R}
{Rau}, A., {Kulkarni}, S.~R., {Law}, N.~M., {et~al.} 2009, \pasp, 121, 1334

\bibitem[{{Reichart} {et~al.}(2005){Reichart}, {Nysewander}, {Moran},
  {Bartelme}, {Bayliss}, {Foster}, {Clemens}, {Price}, {Evans}, {Salmonson},
  {Trammell}, {Carney}, {Keohane}, \& {Gotwals}}]{2005NCimC..28..767R}
{Reichart}, D., {Nysewander}, M., {Moran}, J., {et~al.} 2005, Nuovo Cimento C
  Geophysics Space Physics C, 28, 767

\bibitem[{{Retzlaff} {et~al.}(2013){Retzlaff}, {Delmotte}, {Arnaboldi}, \&
  {Romaniello}}]{p2edpstd}
{Retzlaff}, J., {Delmotte}, N., {Arnaboldi}, M., \& {Romaniello}, M. 2013,
  GEN-SPE-ESO-33000-5335, Issue 5

\bibitem[{{Scalzo} {et~al.}(2014){Scalzo}, {Childress}, {Tucker}, {Yuan},
  {Schmidt}, {Brown}, {Contreras}, {Morrell}, {Hsiao}, {Burns}, {Phillips},
  {Campillay}, {Gonzalez}, {Krisciunas}, {Stritzinger}, {Graham}, {Parrent},
  {Valenti}, {Lidman}, {Schaefer}, {Scott}, {Fraser}, {Gal-Yam}, {Inserra},
  {Maguire}, {Smartt}, {Sollerman}, {Sullivan}, {Taddia}, {Yaron}, {Young},
  {Taubenberger}, {Baltay}, {Ellman}, {Feindt}, {Hadjiyska}, {McKinnon},
  {Nugent}, {Rabinowitz}, \& {Walker}}]{2014MNRAS.445...30S}
{Scalzo}, R.~A., {Childress}, M., {Tucker}, B., {et~al.} 2014, \mnras, 445, 30

\bibitem[{{Shappee} {et~al.}(2014){Shappee}, {Prieto}, {Grupe}, {Kochanek},
  {Stanek}, {De Rosa}, {Mathur}, {Zu}, {Peterson}, {Pogge}, {Komossa}, {Im},
  {Jencson}, {Holoien}, {Basu}, {Beacom}, {Szczygie{\l}}, {Brimacombe},
  {Adams}, {Campillay}, {Choi}, {Contreras}, {Dietrich}, {Dubberley},
  {Elphick}, {Foale}, {Giustini}, {Gonzalez}, {Hawkins}, {Howell}, {Hsiao},
  {Koss}, {Leighly}, {Morrell}, {Mudd}, {Mullins}, {Nugent}, {Parrent},
  {Phillips}, {Pojmanski}, {Rosing}, {Ross}, {Sand}, {Terndrup}, {Valenti},
  {Walker}, \& {Yoon}}]{2014ApJ...788...48S}
{Shappee}, B.~J., {Prieto}, J.~L., {Grupe}, D., {et~al.} 2014, \apj, 788, 48

\bibitem[{{Skrutskie} {et~al.}(2006){Skrutskie}, {Cutri}, {Stiening},
  {Weinberg}, {Schneider}, {Carpenter}, {Beichman}, {Capps}, {Chester},
  {Elias}, {Huchra}, {Liebert}, {Lonsdale}, {Monet}, {Price}, {Seitzer},
  {Jarrett}, {Kirkpatrick}, {Gizis}, {Howard}, {Evans}, {Fowler}, {Fullmer},
  {Hurt}, {Light}, {Kopan}, {Marsh}, {McCallon}, {Tam}, {Van Dyk}, \&
  {Wheelock}}]{2006AJ....131.1163S}
{Skrutskie}, M.~F., {Cutri}, R.~M., {Stiening}, R., {et~al.} 2006, \aj, 131,
  1163

\bibitem[{{Smartt}(2009)}]{2009ARA&A..47...63S}
{Smartt}, S.~J. 2009, \araa, 47, 63

\bibitem[{{Smartt} {et~al.}(2014){Smartt}, {Smith}, {Wright}, {Young}, {Kotak},
  {Nicholl}, {Polshaw}, {Inserra}, {Chen}, {Terreran}, {Gall}, {Fraser},
  {McCrum}, {Valenti}, {Foley}, {Lawrence}, {Gezari}, {Burgett}, {Chambers},
  {Huber}, {Kudritzki}, {Magnier}, {Morgan}, {Tonry}, {Sweeney}, {Stubbs},
  {Kirshner}, {Metcalfe}, \& {Rest}}]{2014ATel.5850....1S}
{Smartt}, S.~J., {Smith}, K.~W., {Wright}, D., {et~al.} 2014, The Astronomer's
  Telegram, 5850, 1

\bibitem[{{Soderberg} {et~al.}(2008){Soderberg}, {Berger}, {Page}, {Schady},
  {Parrent}, {Pooley}, {Wang}, {Ofek}, {Cucchiara}, {Rau}, {Waxman}, {Simon},
  {Bock}, {Milne}, {Page}, {Barentine}, {Barthelmy}, {Beardmore}, {Bietenholz},
  {Brown}, {Burrows}, {Burrows}, {Byrngelson}, {Cenko}, {Chandra}, {Cummings},
  {Fox}, {Gal-Yam}, {Gehrels}, {Immler}, {Kasliwal}, {Kong}, {Krimm},
  {Kulkarni}, {Maccarone}, {M{\'e}sz{\'a}ros}, {Nakar}, {O'Brien}, {Overzier},
  {de Pasquale}, {Racusin}, {Rea}, \& {York}}]{2008Natur.453..469S}
{Soderberg}, A.~M., {Berger}, E., {Page}, K.~L., {et~al.} 2008, \nat, 453, 469

\bibitem[{{Steele} {et~al.}(2004){Steele}, {Smith}, {Rees}, {Baker}, {Bates},
  {Bode}, {Bowman}, {Carter}, {Etherton}, {Ford}, {Fraser}, {Gomboc}, {Lett},
  {Mansfield}, {Marchant}, {Medrano-Cerda}, {Mottram}, {Raback}, {Scott},
  {Tomlinson}, \& {Zamanov}}]{2004SPIE.5489..679S}
{Steele}, I.~A., {Smith}, R.~J., {Rees}, P.~C., {et~al.} 2004, in Society of
  Photo-Optical Instrumentation Engineers (SPIE) Conference Series, Vol. 5489,
  Ground-based Telescopes, ed. J.~M. {Oschmann}, Jr., 679--692

\bibitem[{{Stritzinger} {et~al.}(2005){Stritzinger}, {Suntzeff}, {Hamuy},
  {Challis}, {Demarco}, {Germany}, \& {Soderberg}}]{2005PASP..117..810S}
{Stritzinger}, M., {Suntzeff}, N.~B., {Hamuy}, M., {et~al.} 2005, \pasp, 117,
  810

\bibitem[{{Sullivan}(2013)}]{2013A&G....54f6.17S}
{Sullivan}, M. 2013, Astronomy and Geophysics, 54, 060006

\bibitem[{{Tomasella } {et~al.}(2014){Tomasella }, {Benetti}, {Cappellaro},
  {Pastorello}, {Turatto}, {Barbon}, {Elias-Rosa}, {Harutyunyan}, {Ochner},
  {Tartaglia}, \& {Valenti}}]{2014AN....335..841T}
{Tomasella }, L., {Benetti}, S., {Cappellaro}, E., {et~al.} 2014, Astronomische
  Nachrichten, 335, 841

\bibitem[{{Valenti} {et~al.}(2009){Valenti}, {Pastorello}, {Cappellaro},
  {Benetti}, {Mazzali}, {Manteca}, {Taubenberger}, {Elias-Rosa}, {Ferrando},
  {Harutyunyan}, {Hentunen}, {Nissinen}, {Pian}, {Turatto}, {Zampieri}, \&
  {Smartt}}]{2009Natur.459..674V}
{Valenti}, S., {Pastorello}, A., {Cappellaro}, E., {et~al.} 2009, \nat, 459,
  674

\bibitem[{{Valenti} {et~al.}(2014{\natexlab{a}}){Valenti}, {Sand},
  {Pastorello}, {Graham}, {Howell}, {Parrent}, {Tomasella}, {Ochner}, {Fraser},
  {Benetti}, {Yuan}, {Smartt}, {Maund}, {Arcavi}, {Gal-Yam}, {Inserra}, \&
  {Young}}]{2014MNRAS.438L.101V}
{Valenti}, S., {Sand}, D., {Pastorello}, A., {et~al.} 2014{\natexlab{a}},
  \mnras, 438, L101

\bibitem[{{Valenti} {et~al.}(2014{\natexlab{b}}){Valenti}, {Yuan},
  {Taubenberger}, {Maguire}, {Pastorello}, {Benetti}, {Smartt}, {Cappellaro},
  {Howell}, {Bildsten}, {Moore}, {Stritzinger}, {Anderson}, {Benitez-Herrera},
  {Bufano}, {Gonzalez-Gaitan}, {McCrum}, {Pignata}, {Fraser}, {Gal-Yam}, {Le
  Guillou}, {Inserra}, {Reichart}, {Scalzo}, {Sullivan}, {Yaron}, \&
  {Young}}]{2014MNRAS.437.1519V}
{Valenti}, S., {Yuan}, F., {Taubenberger}, S., {et~al.} 2014{\natexlab{b}},
  \mnras, 437, 1519

\bibitem[{{van Dokkum}(2001)}]{2001PASP..113.1420V}
{van Dokkum}, P.~G. 2001, \pasp, 113, 1420

\bibitem[{{Wyrzykowski} {et~al.}(2014){Wyrzykowski}, {Kostrzewa-Rutkowska},
  {Koz{\l}owski}, {Udalski}, {Poleski}, {Skowron}, {Blagorodnova}, {Kubiak},
  {Szyma{\'n}ski}, {Pietrzy{\'n}ski}, {Soszy{\'n}ski}, {Ulaczyk},
  {Pietrukowicz}, \& {Mr{\'o}z}}]{2014AcA....64..197W}
{Wyrzykowski}, {\L}., {Kostrzewa-Rutkowska}, Z., {Koz{\l}owski}, S., {et~al.}
  2014, \actaa, 64, 197

\bibitem[{{Yaron} \& {Gal-Yam}(2012)}]{2012PASP..124..668Y}
{Yaron}, O. \& {Gal-Yam}, A. 2012, \pasp, 124, 668

\end{thebibliography}
\end{document}